%% file: PaperV3.tex
\documentclass{iopart}

\usepackage[english]{babel}
\usepackage[latin1]{inputenc}
\usepackage{latexsym}
\usepackage{amsmath}
\usepackage{amsthm}
\usepackage{amssymb}
\usepackage{graphics}
\usepackage{graphicx}
\usepackage{epsfig}
\usepackage{color}
\usepackage{feynmp}
\usepackage{bbm}
\usepackage[normalem]{ulem}
\usepackage{cancel}
\usepackage{mathrsfs}
\usepackage{hyperref}
\hypersetup{
colorlinks=true,
citecolor=blue,
linkcolor=red,
urlcolor=black
}

\include{./TexFiles/journalsVaps}

\include{./TexFiles/Notations}

\begin{document}

\unitlength = 0.008\linewidth

\title{Quantum transport of atomic matterwaves in anisotropic 2D and 3D disorder}

\author{M~Piraud$^{1}$, L~Pezz\'e$^{1,2}$ and L~Sanchez-Palencia$^{1}$}
\address{
$^1$Laboratoire Charles Fabry,
Institut d'Optique, CNRS, Univ Paris Sud,
2 avenue Augustin Fresnel,
F-91127 Palaiseau cedex, France\\
$^2$INO-CNR and LENS,
Largo Enrico Fermi 6,
I-50125 Firenze, Italy}
\ead{lsp@institutoptique.fr}

\begin{abstract}
The macroscopic transport properties in a disordered potential, namely diffusion and weak/strong localization, closely depend on the microscopic and statistical properties of the disorder itself.
This dependence is rich of counter-intuitive consequences.
It can be particularly exploited in matter wave experiments, where the disordered potential can be tailored and controlled, and anisotropies are naturally present.
In this work, we apply a perturbative microscopic transport theory and the self-consistent theory of Anderson localization to study the transport properties of ultracold atoms in anisotropic 2D and 3D speckle potentials.
In particular, we discuss the anisotropy of single-scattering, diffusion and localization.
We also calculate a disorder-induced shift of the energy states
and propose a method to include it, which amounts to renormalize energies in the standard on-shell approximation.
We show that the renormalization of energies strongly affects the prediction for the 3D localization threshold (mobility edge).
We illustrate the theoretical findings with examples which are revelant for current matter wave experiments, where the disorder is created with a laser speckle.
This paper provides a guideline for future experiments aiming at the precise location of the 3D mobility edge and study of anisotropic diffusion and localization effects in 2D and 3D.
\end{abstract}

\date{\today}


\maketitle

\tableofcontents

\input{TexFiles/Introduction.07}

\section{Matter waves in disordered media \label{part:quantum-transport}}
\input{TexFiles/Basics-Transport.07}

\input{TexFiles/Propagator-Dyson.06}
\input{TexFiles/SpectralFunction.08}

\input{TexFiles/BSE.07}

\input{TexFiles/Conductivity.06}

\input{TexFiles/CorrF.07}

\section{Single-scattering \label{par:scat}}
We now focus on the first time scale introduced in Sec.~\ref{par:basics}:
The scattering mean free time.
\input{TexFiles/Scattering.06}
\input{TexFiles/ScatteringExample2DGauss.06}

\section{Boltzmann diffusion \label{par:diff}}
We now turn to the behaviour of the spatial density in the incoherent diffusive regime, which is characterized by the Boltzmann diffusion tensor $\DiffTensB(E)$.
We first give an explicit formula for the diffusion tensor, in the framework of the usual on-shell approximation, and then apply it to 2D disorder (3D cases are discussed in Sec.~\ref{part:diff-ex3D}).
\input{TexFiles/Diffusion.06}
\input{TexFiles/DiffusionExample2DGauss.06}

\section{Weak and strong localization \label{par:loc}}
We now consider interference effects, which lead to weak and strong localization.
We first describe the quantum corrections (Sec.~\ref{part:weakloc}), then the self-consistent theory (Sec.~\ref{part:strongloc}), and apply it to the 2D speckle potential (Sec.~\ref{part:locex2D}).
The 3D case, which follows the same route, is discussed in Sec.~\ref{part:loc-ex3D}.
\input{TexFiles/Localization.05}

\input{TexFiles/LocalizationExample2DGauss.05}

\section{Three-dimensional anisotropic disorder \label{par:3Dexamples}}
In this section we apply the formalism introduced in Secs.~\ref{par:scat} to \ref{par:loc} to the 3D speckle potential of Sec.~\ref{part:correl-single}.
We discuss single-scattering (Sec.~\ref{part:scatt-3D}), Boltzmann diffusion (Sec.~\ref{part:diff-ex3D}) and localization (Sec.~\ref{part:loc-ex3D}) properties and the position of the mobility edge (Sec.~\ref{part:mobedge}). 
\input{TexFiles/ScatteringExample3D.04}

\input{TexFiles/DiffusionExample3D.04}

\input{TexFiles/LocalizationExample3D.05}

\subsection{About the 3D mobility edge \label{part:mobedge}}
\input{TexFiles/MobilityEdge.07}

\section{Conclusions \label{part:concl}}
\input{TexFiles/Conclusion.04}

\ack
We thank B.~van~Tiggelen and P. W\"olfle for enlightening discussions.
This research was supported by
the European Research Council (FP7/2007-2013 Grant Agreement No.\ 256294),
the CoopIntEER CNRS-CNR joint project "AtoFerTwoD" (No.\ EDC25123),
the Minist\`ere de l'Enseignement Sup\'erieur et de la Recherche,
and
the Institut Francilien de Recherche sur les Atomes Froids (IFRAF).
We acknowledge the use of the computing facility cluster GMPCS of the
LUMAT federation (FR LUMAT 2764). 
\begin{appendix}
\input{TexFiles/Appendix-BSE.03}

\input{TexFiles/Appendix-iso.02}

\section{Conductivity}
\input{TexFiles/Appendix-Einstein.02}
\input{TexFiles/Appendix-Vrenorm.01}

\input{TexFiles/Appendix-CorrCond.02}
\end{appendix}


\section*{References}
%

\end{document}

%% file: TexFiles/journalsVaps.tex
\newcommand{\Jnature}{Nature (London)}
\newcommand{\Jnatphys}{Nat. Phys.}

\newcommand{\Jscience}{Science}

\newcommand{\Jprl}{Phys. Rev. Lett.}
\newcommand{\Jpr}{Phys. Rev.}
\newcommand{\Jpra}{Phys. Rev. A}
\newcommand{\Jprb}{Phys. Rev. B}

\newcommand{\Jpre}{Phys. Rev. E}
\newcommand{\Jrmp}{Rev. Mod. Phys.}

\newcommand{\JplA}{Phys. Lett. A}

\newcommand{\Jepl}{Europhys. Lett.}
\newcommand{\Jnjp}{New J. Phys.}
\newcommand{\Jepjb}{Eur. Phys. J. B}
\newcommand{\Jepjd}{Eur. Phys. J. D}
\newcommand{\Jepjst}{Eur. Phys. J. Special Topics}

\newcommand{\JApplPhysLett}{Appl. Phys. Lett.}

\newcommand{\Jprocroysoc}{Proc. Roy. Soc. A: Math. Phys. Eng. Sci.}

\newcommand{\Jphysrep}{Phys. Rep.}
\newcommand{\JRepProgPhys}{Rep. Prog. Phys.}

\newcommand{\JjphysA}{J. Phys. A: Math. Theor.}

\newcommand{\JjphysC}{J. Phys. C: Solid State Phys.}

\newcommand{\Jphystoday}{Phys. Today}

\newcommand{\Jadvphys}{Adv. Phys.}

\newcommand{\Jadvatmoloptphys}{Adv. At. Mol. Opt. Phys.}

%% file: TexFiles/Notations.tex
\newcommand{\new}[1]{#1}

\newcommand{\ie}{{i.e.}}
\newcommand{\eg}{{e.g.}}
\newcommand{\ud}{\mathrm{d}}
\newcommand{\vect}[1]{\mathbf{#1}}

\newcommand{\av}[1]{\overline{#1}}
\newcommand{\geomav}[1]{#1^{\textrm{av}}}
\newcommand{\Vr}{V_\textrm{\tiny R}}
\newcommand{\sigmaOrth}{\sigma_{\perp}}
\newcommand{\sigmaOrthx}{{\sigmaOrth}_{x}}
\newcommand{\sigmaOrthy}{{\sigmaOrth}_{y}}
\newcommand{\sigmaPara}{\sigma_{\parallel}}
\newcommand{\sigmaParax}{{\sigmaPara}_{x}}
\newcommand{\sigmaParay}{{\sigmaPara}_{y}}
\newcommand{\sigmar}{\sigma}

\newcommand{\ls}{l_\textrm{\tiny s}}
\newcommand{\smft}{\tau_{\textrm{\tiny s}}}
\newcommand{\DB}{D_\textrm{\tiny B}}
\newcommand{\UB}{U_\textrm{\tiny B}}
\newcommand{\lB}{l_\textrm{\tiny B}}
\newcommand{\Lloc}{L_\textrm{\tiny loc}}

\newcommand{\Cor}{C}
\newcommand{\cor}{c}
\newcommand{\TFCor}{\tilde{C}}
\newcommand{\TFcor}{\tilde{c}}
\newcommand{\kE}{k_\textrm{\tiny \textit E}}

\newcommand{\selfE}{\Sigma}
\newcommand{\selfEa}{\Sigma^{\dagger}}

\newcommand{\Diff}{\vect{D}}
\newcommand{\DiffTensB}{\vect{D}_{\textrm{B}}}
\newcommand{\DiffTens}{\vect{D}_{*}}
\newcommand{\DiffTensCor}{\Delta\vect{D}}

\newcommand{\LocTens}{\vect{L}_\textrm{loc}}
\newcommand{\LocTensTwo}{\vect{\Lambda}}

\newcommand{\Emob}{E_{\textrm{c}}}

\newcommand{\EmobNoShift}{E^\prime_{\textrm{c}}}

\newcommand{\distE}{\mathcal{D}_{\textrm{\tiny E}}}
\newcommand{\distk}{\mathcal{D}_{\textrm{\tiny k}}}
\newcommand{\distO}{\mathcal{D}_0}
\newcommand{\dens}{n}
\newcommand{\vecr}{\textbf{r}}
\newcommand{\vecR}{\textbf{R}}
\newcommand{\veck}{\textbf{k}}
\newcommand{\vecq}{\textbf{q}}

\newcommand{\vecQ}{\textbf{Q}}
\newcommand{\vecJ}{\textbf{J}}
\newcommand{\kOrth}{k_{\perp}}
\newcommand{\veckOrth}{{\textbf{k}}_{\perp}}
\newcommand{\uveck}{\hat{\textbf{k}}}

\newcommand{\uvecu}{\hat{\textbf{u}}}
\newcommand{\be}{\begin{equation}}
\newcommand{\beq}{\begin{eqnarray}}
\newcommand{\ee}{\end{equation}}
\newcommand{\eeq}{\end{eqnarray}}
\newcommand{\Ham}{H}  
\newcommand{\Gr}{G}
\newcommand{\Ga}{G^{\dagger}}
\newcommand{\eps}[1]{\epsilon(#1)}

\newcommand{\U}{\mathrm{U}}
\newcommand{\Oplamb}{\Lambda}

\newcommand{\vv}[1]{\mathbf{#1}}
\newcommand{\sigtens}{\boldsymbol \sigma}
\newcommand{\sigtensB}{{\boldsymbol \sigma}_{\textrm{\tiny B}}}
\newcommand{\sig}{\sigma}

\newcommand{\Oo}{\textrm{O}}

\newcommand{\Id}{I_\textrm{\tiny D}}
\newcommand{\Iscreen}{\mathcal{I}}
\newcommand{\Vopt}{V}
\newcommand{\anifact}{\xi}

\newcommand{\lambdaL}{\lambda_\textrm{\tiny L}}
\newcommand{\kL}{k_\textrm{\tiny L}}

%% file: TexFiles/Introduction.07.tex
\section{Introduction}

Transport in disordered media is a fascinatingly rich field, which sparks a broad range of phenomena such as
  Brownian motion~\cite{risken1989},
  electronic conductivity~\cite{ashcroft1976,mott1990},
  superconductivity~\cite{degennes1995},
  superfluid flows of $^4$He on Vycor substrates~\cite{crowell1995},
  as well as localization of
  classical (electromagnetic or sound) waves in dense media~\cite{akkermans2006,lagendijk2009}
  and of ultracold atoms in controlled disorder~\cite{fallani2008,aspect2009,lsp2010,modugno2010,shapiro2012}.
In the case of a matter particle
for instance, two regimes should be distinguished.
In the classical regime, where the de~Broglie wavelength is vanishingly small,
transport leads to normal or anomalous diffusion~\cite{bouchaud1990,metzler2000}.
The dynamics is characterized by the appearance of a percolation transition, which separates
  a trapping regime -- where the particle is bound in deep potential wells --
  from a diffusion regime -- where the particle trajectory is spatially unbounded~\cite{zallen1971,isichenko1992}.
In the quantum regime, the wave nature of the particle determines its transport properties, in close analogy with those of a classical wave~\cite{john1984,lagendijk1996}.
In this case, interference effects can survive disorder averaging, 
leading to striking effects
such as weak localization~\cite{akkermans2006}, the related coherent back-scattering effect~\cite{akkermans1986},
and strong (Anderson) localization~\cite{anderson1958,lee1985,janssen1998}.

Localization shows a widely universal behaviour~\cite{abrahams1979},
but observable features significantly depend on the details of the system.
It shows a renewed interest in the context of ultracold matter waves~\cite{fallani2008,aspect2009,lsp2010,modugno2010,shapiro2012}.
On the one hand, the microscopic parameters in these systems are precisely known and, in many cases, tunable,
which paves the way to unprecedented direct comparison between experiments and theory~\cite{lsp2007,piraud2011a}.
This is a great advantage of ultracold atoms, compared to traditional condensed-matter systems.
On the other hand, these systems offer new situations, which can induce original effects~\cite{pezze2011a}
and provide new test-grounds in non-standard disorder~\cite{gurevich2009,lugan2009,plodzien2011,piraud2011b,piraud2012c}.
Major advances in this field were the observation of one-dimensional (1D) Anderson localization of matterwaves~\cite{billy2008,roati2008}
and studies of the effects of 
weak~\cite{paul2007,paul2009,lugan2007a,lugan2007b,lugan2011,pikovsky2008,kopidakis2008,flach2009,deissler2010,aleiner2010}
and strong~\cite{damski2003,fallani2007,white2009,pasienski2010}
interactions in disordered gases.
Presently, a major challenge is the study of quantum transport in dimensions higher than one. While localization is the dominant effect in one dimension~\cite{mott1961,borland1963}, higher dimensions show a richer phenomenology where regimes of diffusion, weak localization and Anderson localization can appear~\cite{abrahams1979}.
Recent experiments reported the
  observation of an Anderson transition in momentum space using cold-atom kick-rotor setups~\cite{chabe2008,lemarie2010,lopez2012},
  study of classical diffusion in two-dimensional (2D) speckle potentials~\cite{mrsv2010,pezze2011b},
  coherent back-scattering~\cite{labeyrie2012,jendrzejewski2012},
and
  evidence of Anderson localization in noninteracting Fermi~\cite{kondov2011} and Bose~\cite{jendrzejewski2011} gases in three-dimensional (3D) speckle potentials.

From a theoretical viewpoint, diffusion and localization of noninteracting matter waves have been thoroughly studied for disordered potentials
with zero-range correlations~\cite{shapiro2007,skipetrov2008}
and isotropic correlation functions~\cite{kuhn2005,kuhn2007,miniatura2009,yedjour2010,beilin2010,cherroret2012}.
However, transport experiments in dimensions higher than one are most often performed with speckle potentials which are anisotropic,
either effectively in 2D setups~\cite{mrsv2010,pezze2011b},
or for fundamental optical constraints in 3D~\cite{kondov2011,jendrzejewski2011}.
Moreover, correlations in speckle potentials can be tailored in a broad range of configurations~\cite{clement2006},
which offers scope for investigation of localization in nonstandard models of disorder~\cite{plodzien2011,piraud2011b}.
Taking into account anisotropic effects is of fundamental importance because they can strongly affect coherent transport and localization properties.
This was demonstrated in various stretched media \cite{bishop1984,pine1988,nickell2000,kao1996,wiersma1999,johnson2002,gurioli2005,woelfe1984,kaas2008,tiggelen1996,stark1997}.
Optical disorder, relevant to ultracold-atom experiments~\mbox{\cite{kondov2011,jendrzejewski2011}}, can show significantly more complex anisotropic correlation functions,
the effect of which has been addressed only recently~\cite{piraud2012a}.

In this paper, we study quantum transport and Anderson localization of matter waves in 2D and 3D anisotropic speckle potentials.
We first introduce the basics of quantum transport of matter waves in disordered media (Sec.~\ref{part:quantum-transport}) and the models of disorder we focus on in 2D and 3D (Sec.~\ref{part:correl}).
We then present a detailed description of the theoretical framework pioneered in Refs.~\cite{woelfe1984,vollhardt1992}, which intends to be pedagogical.
We study
  single-scattering (Sec.~\ref{par:scat}), 
  Boltzmann diffusion (Sec.~\ref{par:diff}),
  and localization (Sec.~\ref{par:loc}),
as a function of the particle energy, and discuss in particular the different anisotropies of these quantities.
From a technical viewpoint, while the scattering allows for analytic expressions as for isotropic models of disorder~\cite{kuhn2007}, diffusion and localization are more involved and require in general numerical  diagonalization of a certain operator.
Some analytic expressions are however found in some limits for anisotropic disorder.
In Secs.~\ref{par:scat}, \ref{par:diff} and \ref{par:loc}, we focus on the 2D case, which contains most of the anisotropy effects discussed in the paper.
The 3D cases are discussed in the next sections, where we study the same quantities as above (Sec.~\ref{par:3Dexamples}).
We also show that energy-dependent quantities calculated in the usual on-shell approximation should be renormalized in strong disorder, and propose a method to do it.
It does not strongly alter the overall energy-dependence of the quantities calculated in the previous sections, and in particular their anisotropies.
However, it may be important when comparing to energy-resolved experimental measurements.
Most importantly, we show that it strongly affects the calculation of the 3D mobility edge.
Finally, we summarize our results and discuss their impact on recent and future experiments on ultra-cold atoms in speckle potentials in the conclusion (Sec.~\ref{part:concl}).

%% file: TexFiles/Basics-Transport.07.tex
\subsection{Basics of quantum transport\label{par:basics}}

\begin{figure}[!t] 
\begin{center}
\includegraphics[width=0.7\textwidth]{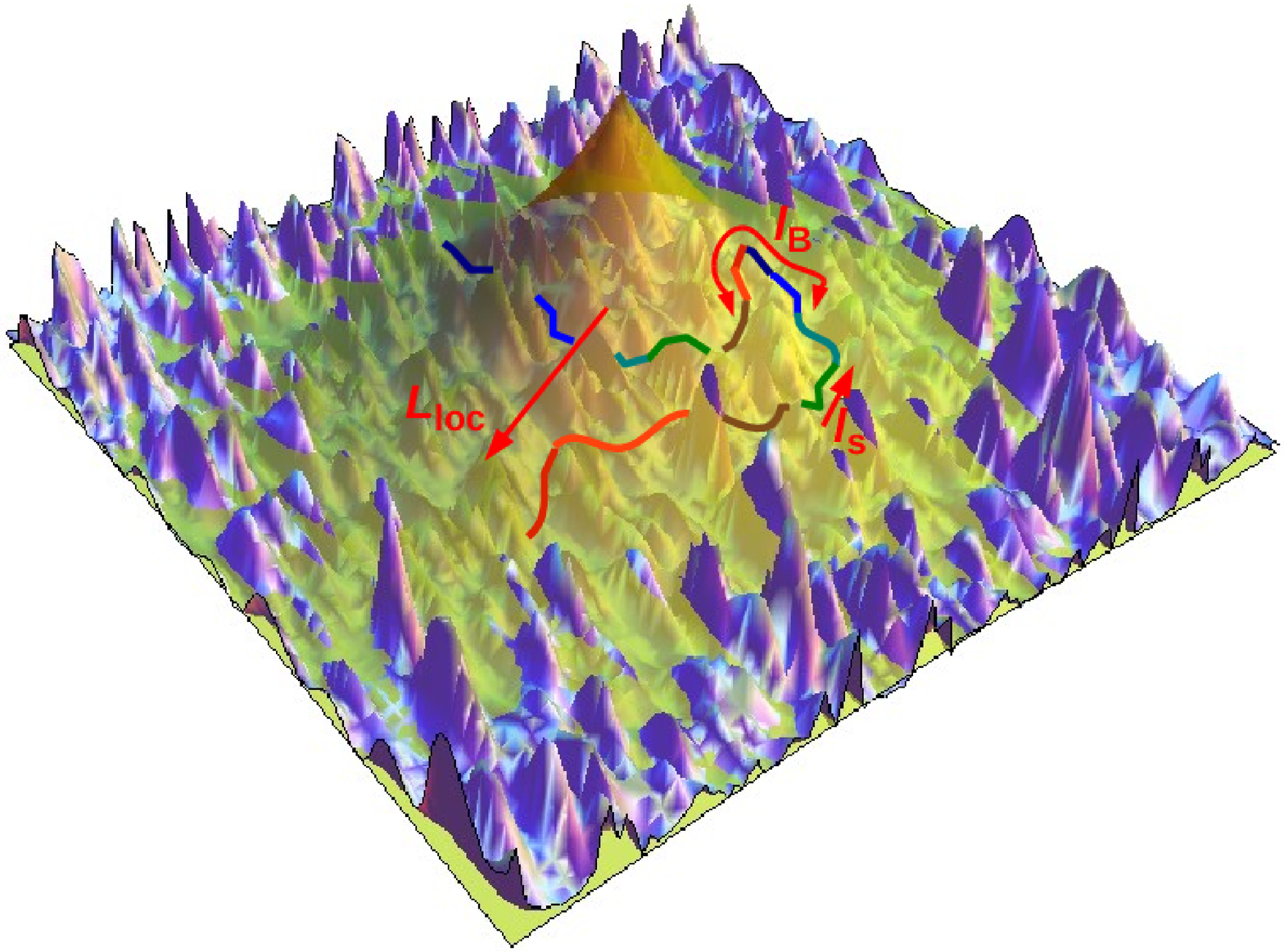}
\end{center} 
\caption{(Color online)
Schematic view of the coherent transport of a matter wave in a disordered medium, with special emphasis on the characteristic length scales.
The figure shows a trajectory of a particle (solid multicolor line) in a two-dimensional disordered landscape (blue surface).
Along its trajectory, the wave loses the memory of its phase (encoded in the various colors along the trajectory) on the characteristic length $\ls$ (scattering mean-free path).
Multiple scattering then deflects the trajectory and the wave loses the memory of its direction on the characteristic length $\lB$ (transport mean-free path).
Interference between the multiple-scattering paths can finally cancel diffusion (strong or Anderson localization). The wave then acquires an exponentially decaying probability profile (orange-green surface) of characteristic length $\Lloc$ (localization length).}
\label{fig:lengths}
\end{figure}

Before turning to a more formal description, it is worth recalling the basic picture of coherent transport in a disordered medium, which is genuinely understood in a microscopic approach~\cite{rammer1998,vollhardt1992}.
Consider a wave of momentum $\veck$ and velocity ${\boldsymbol \upsilon}=\vect{\nabla}_{\vect{k}}\epsilon/\hbar$ [$\eps{\vect{k}}$ is the dispersion relation] propagating in a disordered medium.
We assume for the moment that the medium is isotropic and will drop this assumption in the following sections.
The wave propagation is governed by scattering from the random impurities.
Three typical energy-dependent length scales can be identified,
which characterize three basic effects induced by the disorder
(see Fig.~\ref{fig:lengths}).
%
First, single scattering from impurities depletes the $\veck$-wave states,
which can be seen as quasiparticles in the disordered medium, with a finite life-time
$\smft(\veck)$.
Single scattering hence defines the first length scale, namely the \textit{scattering mean-free path},
$\ls=\upsilon \smft$,
which characterizes the typical length travelled by the wave before it
loses the memory of its initial state,
and primarily the memory of its initial phase.
Then, multiple scattering defines the second length scale, namely the \textit{transport (Boltzmann) mean-free path}, $\lB$, which characterizes the typical length travelled by the wave before it loses the memory of its initial direction.
In general, several scattering events are necessary to significantly deflect the trajectories so that
$\lB \geq \ls$. The two length scales are found to be equal only in the white-noise limit (if it exists), where the wavelength is smaller than the typical size of the impurities.
In this case the scattering is isotropic and the wave loses the memory of its phase and initial propagation direction at the same time.
Within the distance $\lB$, the transport crosses over from ballistic to diffusive.
The average squared size of the wavepacket increases linearly in time, $\av{\vect{r}^2} \sim 2d \DB t$
with $\DB = \upsilon \lB/d$ the Boltzmann diffusion constant ($d$ is the space dimension)~\cite{ashcroft1976,mott1990}.
Finally, diffusive transport allows the wave to return to its initial position via loop paths, and interference effects enter the game.
Each loop can be traveled in one way or the other, which generates two multiple-scattering paths along which exactly the same phase is accumulated during the successive scattering events.
This coherent effect holds for any specific realization of the disordered potential and thus survives disorder averaging.
Moreover, since these two paths are in phase, 
it gives rise to a constructive interference of the matter wave, which significantly enhances its return probability.
This effect induces coherent back-scattering and weak localization,
which leads to diffusive transport with a reduced diffusion coefficient,
$D_* < \DB$~\cite{akkermans2006}.
For strong enough disorder, the diffusion can completely cancel, an effect known as strong, or Anderson, localization~\cite{janssen1998}. Then, the probability distribution of the wave decays exponentially in space, hence defining the third characteristic length, $\Lloc$, the so-called localization length.

The picture above shows that localization relies on two characteristics of the medium: coherence along the multiple-scattering paths and return probablity to the origin. One then understands that the strength of localization should be governed by the interference parameter $k\lB$~\cite{ioffe1960} (since the more the coherence length exceeds the typical length of a loop path, the more significant interference terms are) and by the dimension of space $d$ (since the return probability decreases when $d$ increases).
As a matter of fact, in 1D and 2D, any state is localized, although disorder correlations may lead to strong energy-dependence of the localization length~\cite{izrailev1999,tessieri2002,lugan2009,gurevich2009}.
In 1D, one finds that $\Lloc \sim \lB$ so that diffusion is strictly absent.
In 2D, one finds $\lB<\Lloc$, and diffusion shows up at intermediate distances and times.
In 3D, the return probability is finite and localization appears only for sufficiently low values of $k\lB$.
A mobility edge shows up for $k\lB \sim 1$, which separates localized states (for $k\lB \lesssim 1$) from diffusive states (for $k\lB \gtrsim 1$)~\cite{abrahams1979,edwards1972}.

The microscopic description outlined above offers a comprehensive picture of transport and localization effects for coherent waves in disordered media.
The next subsections give mathematical support to this picture within a formalism adapted to anisotropic disorder.

%% file: TexFiles/Propagator-Dyson.06.tex
\subsection{Green functions}

Consider a quantum particle in a given homogeneous underlying medium and subjected to some static randomness.
Its dynamics is governed by the single-particle Hamiltonian $\Ham = \Ham_0 + V(\vecr)$,
where $\Ham_0$ is the disorder-free, translation-invariant, Hamiltonian of the underlying medium, and $V(\vecr)$ is the time-independent (conservative) disordered potential.
Without loss of generality, the disordered potential can be assumed to be of zero \new{statistical average}\footnote{\new{Here we choose the zero of energies such that the disordered potential is of zero average, \ie\ $\av{V}=0$.
For any other choice of the energy reference all energies appearing below should be shifted by $\av{V}$, \ie\ replace $E$ by $ E-\av{V}$.}}, $\av{V}=0$.
The evolution of the wave function between $t_0$ and $t>t_0$ is determined by the retarded single-particle propagator $\Gr(t,t_0) \equiv  \exp [-i \Ham (t-t_0) /\hbar] \, \Theta(t-t_0)$,
where the Heaviside step function $\Theta(t-t_0)$ accounts for temporal ordering.
In the energy domain\footnote{Here, we use $\Gr(E) \equiv \frac{-i}{\hbar} \int \ud\tau \Gr(\tau) \exp[i E \tau/\hbar]$.}, $\Gr$ is the retarded Green operator
\be
\Gr(E) = \left( E - \Ham + i 0^{+} \right)^{-1}, 
\label{eq:GreenH}
\ee
where $E$ is the particle energy.
It is the solution of the equation
\be
\Gr(E) =\Gr_0(E) + \Gr_0(E) \, V \, \Gr(E),
\label{eq:Gpertub}
\ee
where $\Gr_0 =  \left( E - \Ham_0 + i 0^{+} \right)^{-1}$ is the disorder-free retarded Green function associated to the unperturbed Hamiltonian $\Ham_0$.

%% file: TexFiles/SpectralFunction.08.tex
\subsection{Properties of the disordered medium \label{par:spectral}}

In a disordered medium, meaningful observable quantities correspond to statistical averages over realizations of the disorder.
When averaging over disorder realizations, some quantities can be written in terms of the average Green function $\av{\Gr}(E)$, for instance the spectral function (see below).
The Born series of Eq.~(\ref{eq:Gpertub}), averaged over the disorder, reads
\be
\av{\Gr} = \Gr_0 + \Gr_0 \av{V \Gr_0 V} \Gr_0 + \Gr_0 \av{V \Gr_0 V \Gr_0 V} \Gr_0 + ...
\label{eq:Born-series}
\ee
since the first order term, $\Gr_0 \av{V} \Gr_0$, vanishes
It is convenient to represent this equation diagrammatically:
\input{Diagrammes/diag-born}
where a plain line is a Green function (grey for $\Gr_0$ and black for $\av{\Gr}$), the vertices (black dots) are scattering events and the dashed lines recall that they are correlated.
The Dyson equation~\cite{mahan2000}
\be
\av{\Gr} = \Gr_0 + \Gr_0 \selfE \av{\Gr}, 
\label{eq:Dyson}
\ee  
with $\selfE(E)$ the self energy, can be developped in powers of $V$ thanks to Eq.~(\ref{eq:Born-series}) so as to determine $\selfE$.
The average Green function then reads
\be
\av{\Gr} = \left(\Gr_0^{-1}- \selfE \right)^{-1}.
\label{eq:sol-dyson}
\ee
If the disorder is homogeneous, \ie\ if its statistical properties are translation-invariant~\cite{lifshits1988}, then the disorder-averaged Green function is diagonal in \veck-space\footnote{Here $| \veck \rangle$ is normalized so that $\int \frac{\ud \veck}{(2\pi)^d} \, | \veck \rangle \langle \veck | = 1$.}:
\beq
\langle \veck | \av{\Gr}(E) | \veck' \rangle & \equiv & (2 \pi)^d \delta(\veck- \veck') \av{\Gr}(E,\veck) \nonumber \\
 & = &
\frac{(2 \pi)^d \delta(\veck- \veck')}{E-\eps{\veck} - \selfE(E,\veck) + i 0^{+}},
\label{eq:GreenS}
\eeq
where $\eps{\veck}$ is the dispersion relation associated to $\Ham_0$
and $d$ the space dimension.
In addition, if the statistical properties of the disorder are isotropic, then $\av{\Gr}(E,\veck) \equiv \av{\Gr}(E,|\veck|)$.

This features an effective homogeneous (\ie\ translation-invariant) medium, which contains all necessary information to determine the disorder average of any quantity linear in $\Gr$.
It is the case of the spectral function $A(E,\veck)$ defined by~\cite{rammer1998}:
\be
2\pi \langle \veck | \av{\delta(E-\Ham)} | \veck' \rangle \equiv (2 \pi)^d \delta(\veck- \veck') A(E,\veck).
\label{eq:AkEdef}
\ee
It contains all the information about the spectrum of the disordered medium.
Using Eq.~(\ref{eq:GreenH}), it yields
\be
A(E,\veck) = -2\Im\left[ \av{\Gr}(E,\veck) \right].
\label{eq:AkE0}
\ee
The spectral function can be interpreted (up to a numerical factor) as the (normalized) probability density for an excitation of momentum $\veck$ to have energy E and $\int \frac{\ud E}{2 \pi} \, A(E,\veck) = 1$.
It is also the unnormalized probability, per unit energy, to find a particle of energy $E$ with momentum $\veck$ and $\int \frac{\ud \veck}{(2 \pi)^d} \, A(E,\veck) = 2 \pi N(E)$,
where $N(E)$ is the density of states per unit volume.
For a particle in disorder-free space, it is given by $A_0(E,\veck)= 2 \pi \delta\left[E - \eps{\veck}\right]$.
In the presence of disorder, Eqs.~(\ref{eq:GreenS}) and (\ref{eq:AkE0}) yield
\be
A(E,\veck) = 
\frac{-2 \selfE''(E,\veck)}{\big(E- \eps{\veck} -\selfE'(E,\veck)\big)^2+ 
\selfE''(E,\veck)^2},
\label{eq:AkE1}
\ee
with $\selfE'$ and $\selfE''$ the real and imaginary parts of $\selfE$, respectively.
\begin{figure}[!t] 
\begin{center}  
\includegraphics[width=0.6\textwidth]{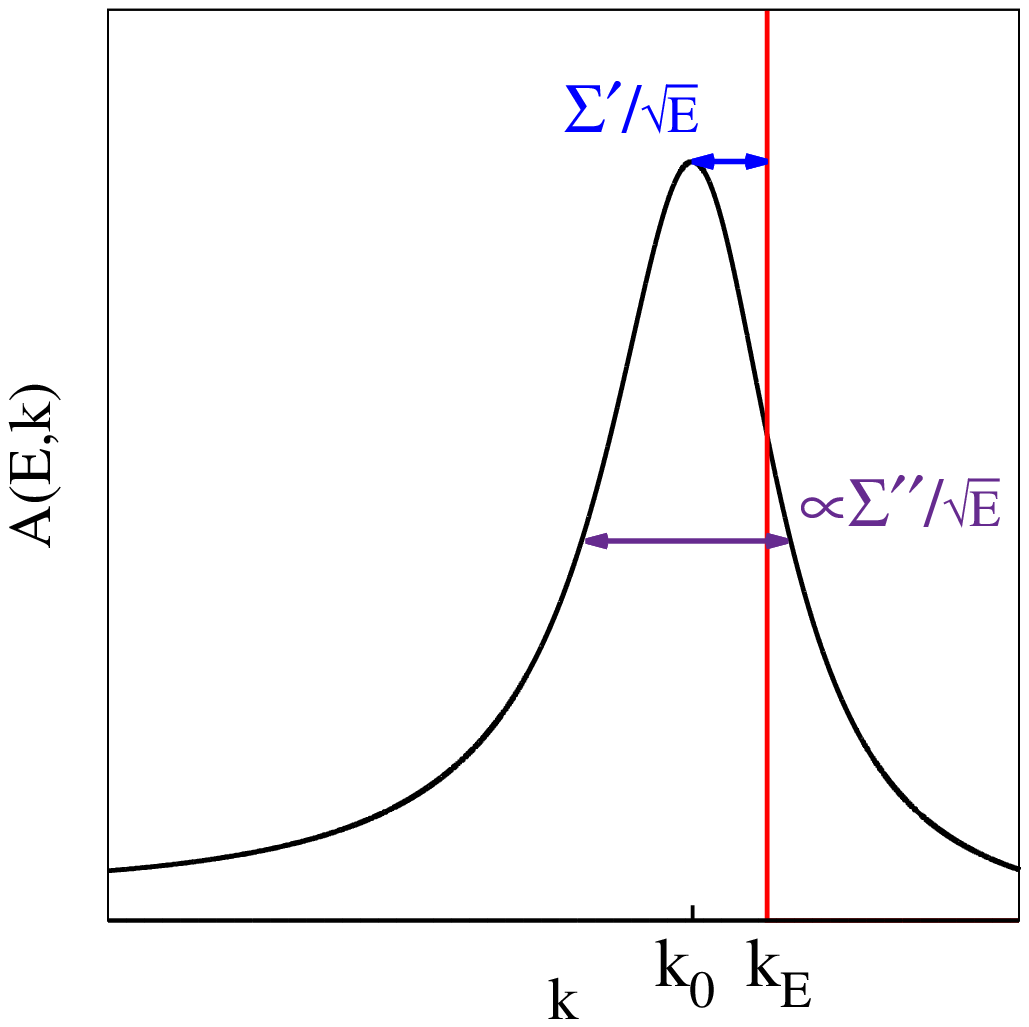}
\end{center} 
\caption{\small{(Color online) 
Schematic representation of the spectral function $A(E,\veck)$ of a particle of energy $E=\hbar^2 \kE^2/2m$, as a function of the particle momentum $\veck$.
The vertical red line is the spectral function for the disorder free particle $A_0(E,\veck)= 2 \pi \delta\left[E - \eps{\veck}\right]$ with $\eps{\veck}=\hbar^2 k^2/2m$.
In the presence of disorder the spectral function is shifted and broadened (black line).
}}
\label{fig:SpecFunction}
\end{figure} 
As represented schematically in Fig.~\ref{fig:SpecFunction}, for a particle in free space [$\eps{\veck}=\hbar^2 k^2/2m$, where $m$ is the mass of the particle] with a weak disordered potential, the spectral function has a Lorentzian-like shape as a function of $\veck$.
It is centered in $\veck_0$, solution of
$E- \eps{\veck_0} -\selfE'(E,\veck_0)=0$.
The quantity $\selfE'(E,\veck_0)$ thus describes the shift in energy of the free-particle modes when they are dressed by the disorder.
The quantity $\selfE''(E,\veck)$ is the energy width of the spectral function, which defines the scattering mean free time
\be
\smft(E,\veck) = -\frac{\hbar}{2 \selfE''(E,\veck)},
\label{eq:tau_s-def}
\ee
or equivalently the scattering mean free path $\ls(E,\veck) = |{\boldsymbol  \upsilon}| \smft(E,\veck)$.
It accounts for the depletion of the free particle mode at $E=\eps{\veck}$ due to scattering from the disordered medium.

The spectral function, which contains all the information about the relative weight, the energy, and the life time of the quasi-particles, will be the key ingredient in the following calculations.
In addition, in ultracold atomic systems, a broad range of energies are involved, but only the momentum distribution is usually measured by time-of-flight techniques.
The spectral function
relates the energy distribution ($\distE$) and the momentum distribution ($\distk$) of the stationary particles in the disorder via
\be
\distE(E)=\int\frac{\ud\veck}{(2\pi)^d}\, A(E,\veck) \distk(\veck),
\ee
which is normalized by $\int\frac{\ud E}{2\pi}\,\distE(E)=1$.
The exact calculation of the spectral function requires the knowledge of the real and imaginary parts of the self energy $\selfE$ [see Eq.~(\ref{eq:AkE1})], or, according to Eq.~(\ref{eq:AkEdef}), the direct diagonalization of the disordered Hamiltonian and an average over disorder realizations.
This is, in general, a complicated task, especially in dimensions larger than one and for anisotropic disorder.
In Secs.~\ref{par:diff} to \ref{par:3Dexamples}, we work within the usual \emph{on-shell approximation}~\cite{vollhardt1992}, in which one neglects the real-part of the self energy $\selfE'(E,\veck)$ and the structure of the spectral function (see schematic dashed blue line in Fig.~\ref{fig:SpecFunction}).
In Sec.~\ref{part:mobedge}, we describe a method to go beyond the on-shell approximation, which amounts to renormalizing the energies in a self-consistent way~\cite{piraud2012a}.

%% file: Diagrammes/diag-born.tex
\begin{fmffile}{diag-born-fmf}
\be
\parbox{0.1\linewidth}{
	    \begin{fmfgraph*}(10,7)
		\fmfleft{i1}
		\fmfright{o1}
		\fmf{plain_arrow,width=2}{i1,o1}
	    \end{fmfgraph*}
	    }
\, = \, \parbox{0.1\linewidth}{
	    \begin{fmfgraph*}(10,7)
		\fmfleft{i1}
		\fmfright{o1}
		\fmf{plain_arrow,foreground=0.5white,width=2}{i1,o1}
	    \end{fmfgraph*}
	    }
\, + \,
\parbox{0.2\linewidth}{
	    \begin{fmfgraph*}(20,7)
		\fmfleft{i1}
		\fmfright{o1}
		\fmf{plain_arrow,foreground=0.5white,width=2}{i1,v1}
		\fmf{plain,foreground=0.5white,width=2}{v1,v2}
		\fmf{plain_arrow,foreground=0.5white,width=2}{v2,o1}
		\fmffreeze
		\fmf{dashes,left=0.5}{v1,v2}
		\fmfdot{v1,v2}
	    \end{fmfgraph*}
}
\, + \,
\parbox{0.3\linewidth}{
	    \begin{fmfgraph*}(30,7)
		\fmfleft{i1}
		\fmfright{o1}
		\fmftop{t1}
		\fmf{plain_arrow,foreground=0.5white,width=2}{i1,v1}
		\fmf{plain,foreground=0.5white,width=2}{v1,v2}
		\fmf{plain,foreground=0.5white,width=2}{v2,v3}
		\fmf{plain_arrow,foreground=0.5white,width=2}{v3,o1}
		\fmffreeze
		\fmf{dashes,left=0.2}{v1,t1}
		\fmf{dashes,left=0.2}{t1,v3}
		\fmf{dashes}{v2,t1}
		\fmfdot{v1,v2,v3}
	    \end{fmfgraph*}
}
\, + \, ...
\label{Diag:Born-series}
\ee
\end{fmffile}\\

%% file: TexFiles/BSE.07.tex
\subsection{Propagation of the Wigner function \label{part:BSE}}

Some quantities are not simply related to the averaged Green function $\av{\Gr}$ and require a more elaborate treatment.
It is for instance the case of the spatial density and the momentum distribution.
More generally, consider the time evolution of the one-body density matrix $\rho(t)$~\cite{rammer1998} or equivalently of the Wigner function~\cite{hillery1984}
\be \label{Wigner}
W(\vecr,\veck,t) \equiv \int \frac{\ud \vecq}{(2\pi)^d} \, e^{i \vecq \cdot \vecr} \,
\left\langle \veck + \frac{\vecq}{2} \right\vert \rho(t) \left\vert \veck - \frac{\vecq}{2} \right\rangle.
\ee
The spatial density probability is given by $\dens(\vecr,t)=\int \frac{\ud\veck}{(2\pi)^d}\, W(\vecr,\veck,t)$ and the momentum distribution by $\distk(\veck,t)=\int \ud\vecr\, W(\vecr,\veck,t)$.
It is fruitful to rewrite Eq.~(\ref{Wigner}) in a form indicating explicitly the initial conditions, using the relation $\rho(t)=\Theta(t-t_0) e^{-i\Ham (t-t_0)/\hbar} \rho(t_0) e^{+i\Ham (t-t_0)/\hbar}$.
When averaging over the disorder, if there is no correlations between the initial state and the disorder, one finds~\cite{miniatura2009}
\be
\av{W}(\vecr,\veck,t) = \int \ud \vecr' \, \int \frac{\ud \veck'}{(2\pi)^d} \, 
W_0(\vecr',\veck') \, F_{\veck,\veck'}(\vecr-\vecr'; t-t_0),
\label{av_Wign}
\ee
where $W_0(\vecr,\veck) \equiv W(\vecr,\veck,t_0)$ is the initial Wigner function and $F_{\veck,\veck'}(\vect{R}; t)$ is the phase-space propagation kernel, defined by (if $t>0$)
\be
F_{\veck,\veck'}(\vect{R}; t) \equiv
\int \frac{\ud E}{2\pi} \,
\int \frac{\ud \vecq}{(2\pi)^d} \, 
\int \frac{\ud \hbar \omega}{2\pi} \, 
 e^{i \vecq \cdot \vect{R}}\, e^{-i \omega t} \, \Phi_{\veck,\veck'}(\vecq,\omega,E),
\label{Fkernel}
\ee
and
\be \label{eq:phi-bb}
\av{ \langle \veck_+ \vert \Gr(E_+) \vert \veck'_+ \rangle 
\langle \veck'_- \vert \Ga(E_-) \vert \veck_- \rangle}
\equiv (2\pi)^d \delta (\vecq-\vecq') \Phi_{\veck,\veck'}(\vecq,\omega,E),
\ee
with $\veck_\pm \equiv \veck \pm \vecq/2$, $\veck_\pm' \equiv \veck' \pm \vecq'/2$, $E_\pm \equiv E \pm \hbar\omega/2$, and $(\vecq$, $\omega)$ the Fourier conjugates of the space and time variables\footnote{We use the Fourier transform $\tilde{f}(\vecq,\omega) \equiv \int \ud\vecr \ud t\ f(\vecr,t) \exp[-i (\vecq\cdot\vecr - \omega t)]$.}.
As discussed above, disorder averaging features a translational invariance in space so that Eq.~(\ref{Fkernel}) depends only on the difference $\vect{R}=\vecr-\vecr'$.
For the same reason, translational invariance, or equivalently momentum conservation, imposes that the sum of the in-going wavevectors ($\veck_+$ and $\veck_-'$) on one hand, and out-going wavevectors ($\veck_+'$ and $\veck_-$) on the other hand, are equal.
It leads to the condition on momentum transfer, $\vecq=\vecq'$, in Eq.~(\ref{eq:phi-bb}).

As can be seen in Eqs.~(\ref{av_Wign}) and (\ref{Fkernel}), the building block to describe wave propagation in random media is the density propagator $\Phi$, which can be represented as a four-point vertex with $\veck_\pm$ and $\veck_\pm'$ the left and right entries [see left-hand side of Eq.~(\ref{Diag:BSE})].
The skeleton of this vertex is made by retarded and advanced Green functions (respectively $\Gr$, represented by the top line, and $\Ga$, represented by the bottom line). It contains all possible correlations between the scattering events of these Green functions.
Following the same approach as used for the average field propagator $\av{\Gr}$ [leading to the Dyson equation~(\ref{eq:Dyson})], the vertex $\Phi=\av{\Gr \otimes \Ga}$ is formally constructed from the uncorrelated-average vertex $\av{\Gr} \otimes \av{\Ga}$.
Without any approximation, $\Phi$ is then governed by the so-called Bethe-Salpeter equation (BSE)~\cite{rammer1998}
\be \label{eq:BSE}
\Phi = \av{\Gr} \otimes \av{\Ga} + \av{\Gr} \otimes \av{\Ga} \, \U \, \Phi,
\ee
represented diagrammatically as
\vspace{0.02\linewidth}
\input{Diagrammes/diag-BSE}
where $\U$ is the vertex function including all irreducible four-point scattering diagrams:
\vspace{0.05\linewidth}
\input{Diagrammes/diag-U-vert}
The first term in the BSE~(\ref{eq:BSE})-(\ref{Diag:BSE}) describes uncorrelated propagation of the field and its conjugate in the effective medium. The second term accounts for all correlations in the density propagation.

Analogously to Eq.~(\ref{eq:sol-dyson}), the solution of the BSE~(\ref{eq:BSE})-(\ref{Diag:BSE}) can be formally obtained from the inverse, if it exists, of the four-point operator $\Oplamb \equiv 1 - \av{\Gr} \otimes \av{\Ga} \,\U$~\footnote{In this context, the inverse of an operator $\Oplamb$ is defined by
$\int \frac{\ud \veck_1}{(2\pi)^d} \Oplamb_{\veck,\veck_1}(\vecq,\omega,E) \Oplamb^{-1}_{\veck_1,\veck'}(\vecq,\omega,E) = (2\pi)^d
\delta(\veck-\veck')$.}~\cite{mackintosh1989}:
\be
\Phi=\Oplamb^{-1}\,\av{\Gr} \otimes \av{\Ga}.
\label{eq:sol-BSE}
\ee
More explicitly, the $(\veck,\veck')$ component of a four-point vertex $\Oplamb$ which fulfills momentum conservation is $\Oplamb_{\veck,\veck'} (\vecq,\omega,E)$, such that
$\langle \veck_+, \veck_-' | \Oplamb | \veck_+',\veck_- \rangle \equiv (2\pi)^d \delta(\vecq-\vecq') \Oplamb_{\veck,\veck'} (\vecq,\omega,E)$,
and
\be
\Oplamb_{\veck,\veck'}(\vecq,\omega,E) = (2\pi)^d \delta(\veck-\veck') 
- f_\veck(\vecq,\omega,E) U_{\veck,\veck'}(\vecq,\omega,E),
\ee
and
\be
f_\veck(\vecq,\omega,E) \equiv  \av{\Gr}(E_+,\veck_+)\av{\Ga}(E_-,\veck_-).
\label{eq:fkE}
\ee 
Therefore Eq.~(\ref{eq:sol-BSE}) reads
\be
\Phi_{\veck,\veck'}(\vecq,\omega,E) = \Oplamb^{-1}_{\veck,\veck'}(\vecq,\omega,E) f_{\veck'}(\vecq,\omega,E),
\label{eq:sol-BSE2}
\ee
and can be expressed as a geometric series 
\beq
\Phi_{\veck,\veck'}(\vecq,\omega,E) &=& (2\pi)^d \delta(\veck-\veck') f_\veck(\vecq,\omega,E) \\
&+& f_\veck(\vecq,\omega,E) U_{\veck,\veck'} (\vecq,\omega,E) f_{\veck'}(\vecq,\omega,E) \nonumber \\
&+& \int \frac{\ud \veck_1}{(2 \pi)^d} \, f_\veck(\vecq,\omega,E)  U_{\veck,\veck_1} (\vecq,\omega,E) f_{\veck_1}(\vecq,\omega,E) \nonumber \\
&\times& U_{\veck_1,\veck'} (\vecq,\omega,E) f_{\veck'}(\vecq,\omega,E) + ... \nonumber
\eeq
The operator $\Oplamb^{-1}(\omega,E)$ can be expressed in terms of 
the eigenvectors and associated eigenvalues of the operator 
$\Oplamb(\omega,E)$ which was used in Refs.~\cite{woelfe1984,bhatt1985} to solve the BSE.
It then gives access, via Eq.~(\ref{eq:sol-BSE2}) to $\Phi$, which is the quantity of interest [see Eqs.~(\ref{Wigner}) to (\ref{Fkernel})].

In the following we will see that the intensity kernel $\Phi$ has a diffusion pole, which takes the form
\be
\Phi_{\veck,\veck'}(\vecq,\omega,E) = \frac{1}{2\pi N(E)}\frac{A(E,\veck) A(E,\veck')}{i \hbar \omega - \hbar \vecq \! \cdot \! \Diff(\omega,E) \! \cdot \! \vecq}
\label{Kernel-usual}
\ee
where $\Diff$ is the so-called dynamic diffusion tensor.
The average spatial density distribution is then given by
\beq
\av{\dens}(\vecr,t)&=&\int \frac{\ud\veck}{(2\pi)^d}\, \av{W}(\vecr,\veck,t) \nonumber\\
&=&\int \frac{\ud E}{2\pi} \, \int \ud \vecr' \, \distO(\vecr',E) P(\vecr-\vecr',t-t_0|E)
\eeq
where $\distO(\vecr',E)=\int \frac{\ud\veck'}{(2\pi)^d} \, A(E,\veck') W_0(\vecr',\veck')$ represents the initial joint position-energy density and $P(\vecr-\vecr',t-t_0|E)$ is the probability of quantum transport, \ie\ the probability for a particle of energy $E$ originating from point $\vecr'$ at time $t_0$ to be in $\vecr$ at $t$.
It can be expressed thanks to Eqs.~(\ref{av_Wign}), (\ref{Fkernel}) and (\ref{Kernel-usual}) as the space-time Fourier Transform of the diffusion pole $1/[i \hbar \omega - \hbar \vecq \! \cdot \! \Diff(\omega,E) \! \cdot \! \vecq]$.
In the long-time limit, we will encounter two different situations.
First, if $\lim_{\omega \rightarrow 0} \Diff(\omega,E)= \Diff(E)$ is a real definite positive tensor, the diffusion pole of the intensity kernel~(\ref{Kernel-usual}) describes normal diffusion with the anisotropic diffusion tensor $\Diff(E)$, and the probability of quantum transport reads
\be
P(\vecR,t\rightarrow \infty|E) = \frac{e^{-\vecR \cdot \Diff^{-1}(E) \cdot \vecR/4t}}{\sqrt{(4\pi t)^d \det \left\{ \Diff(E) \right\}}} \Theta(t).
\label{eq:prop-kernel-diff}
\ee
Second, if $\Diff(\omega,E) \sim 0^{+} -i\omega \LocTensTwo(E)$ in the limit $\omega \rightarrow 0^{+}$ with $\LocTensTwo(E)$ a real positive definite tensor, the pole describes localization.
It leads to exponentially localized phase-space propagation kernel and probability of quantum transport at long distance.
In 2D,
\be
P(\vecR,t\rightarrow \infty|E) =
\frac{ K_0 \left( \sqrt{ \vect{R} \cdot \LocTens^{-2} (E) \cdot \vect{R}}\right)}
{2\pi \det \{ \LocTens(E) \} } \Theta(t)
\label{eq:prop-kernel-loc2d}
\ee
where $K_0$ is the modified Bessel function,
and in 3D,
\be
P(\vecR,t\rightarrow \infty|E) =
\frac{e^{-\sqrt{ \vect{R} \cdot \LocTens^{-2} (E) \cdot \vect{R}}}}
{4\pi \det \{ \LocTens(E) \} \sqrt{ \vect{R} \cdot \LocTens^{-2} (E) \cdot \vect{R}}} \Theta(t).
\label{eq:prop-kernel-loc}
\ee
In both 2D and 3D, the fonction $P(\vecR)$ decays exponentially\footnote{Note that $K_0(x)\sim e^{-x} \sqrt{\pi/2x}$ for $x \gg 1$.} over the characteristic length $\Lloc^u(E)$ along the eigenaxis $u$ of the localization tensor $\LocTens (E) \equiv \sqrt{\LocTensTwo (E)}$.

%% file: Diagrammes/diag-BSE.tex
\begin{fmffile}{diag-BSE-fmf}
\be
\parbox[height=10cm]{0.2\linewidth}{
	    \begin{fmfgraph*}(15,15)
		\fmfleft{v1,v4}
		\fmfright{v2,v3}
		\fmf{phantom}{v1,v2}
		\fmf{phantom}{v3,v4}
		\fmffreeze
		\fmfv{label=$\veck_-$,l.d=0.5}{v1}
		\fmfv{label=$\veck'_-$,l.d=0.5}{v2}
		\fmfv{label=$\veck'_+$,l.d=0.5}{v3}
		\fmfv{label=$\veck_+$,l.d=0.5}{v4}
		\fmf{plain}{v2,v1}
		\fmf{plain}{v4,v3}
		\fmf{plain,tension=0.}{v1,v4}
		\fmf{plain,tension=0.}{v2,v3}
		\fmffreeze
		\fmf{phantom,tension=1.}{v1,M,v3}
		\fmf{phantom,tension=1.}{v2,M,v4}
		\fmfv{label=$\Phi$,label.d=0}{M}
		\fmfkeep{diag-Phi}
	    \end{fmfgraph*}
	    }
 = \,
\parbox{0.15\linewidth}{
	    \begin{fmfgraph*}(15,15)
		\fmfleft{i1,i2}
		\fmfright{o1,o2}
		\fmf{plain_arrow,label=$\veck_-$,l.s=left,l.d=2,width=2}{o1,i1}
		\fmf{plain_arrow,label=$\veck_+$,l.s=left,l.d=2,width=2}{i2,o2}
		\fmfkeep{diag-free}
	    \end{fmfgraph*}
	    }
\, + \,
\parbox{0.3\linewidth}{
	    \begin{fmfgraph*}(30,15)
		\fmfleft{i1,i2}
		\fmfright{o1,o2}
		\fmf{phantom,tension=1.}{i1,v1}
		\fmf{phantom,tension=1.}{v1,v2}
		\fmf{phantom,tension=1.}{v2,o1}
		\fmf{phantom,tension=1.}{o2,v3}
		\fmf{phantom,tension=1.}{v3,v4}
		\fmf{phantom,tension=1.}{v4,i2}
		\fmffreeze
		\fmf{plain_arrow,label=$\veck_-$,l.s=left,l.d=2,width=2}{v1,i1}
		\fmf{plain}{v2,v1}
		\fmf{plain}{o1,v2}
		\fmf{plain}{v3,o2}
		\fmf{plain}{v4,v3}
		\fmf{plain_arrow,label=$\veck_+$,l.s=left,l.d=2,width=2}{i2,v4}
		\fmf{plain,tension=0.}{v1,v4}
		\fmf{plain,tension=0.}{v2,v3}
		\fmf{plain,tension=0.}{o1,o2}
		\fmffreeze
		\fmf{phantom,tension=1.}{v1,M,v3}
		\fmf{phantom,tension=1.}{v2,M,v4}
		\fmfv{label=$\mathrm{U}$,label.d=0}{M}
		\fmf{phantom,tension=1.}{v3,N,o2}
		\fmf{phantom,tension=1.}{v2,N,o1}
		\fmfv{label=$\Phi$,label.d=0}{N}
		\fmfv{label=$\veck'_+$,l.d=0.5}{o2}
		\fmfv{label=$\veck'_-$,l.d=0.5}{o1}
	    \end{fmfgraph*}
}
\label{Diag:BSE}
\ee
\end{fmffile}\\

%% file: Diagrammes/diag-U-vert.tex
\begin{fmffile}{diag-U-vert-fmf}
\be
\parbox{0.15\linewidth}{
	    \begin{fmfgraph*}(15,15)
		\fmfleft{v1,v4}
		\fmfright{v2,v3}
		\fmf{phantom}{v1,v2}
		\fmf{phantom}{v3,v4}
		\fmffreeze
		\fmf{plain}{v2,v1}
		\fmf{plain}{v4,v3}
		\fmf{plain,tension=0.}{v1,v4}
		\fmf{plain,tension=0.}{v2,v3}
		\fmffreeze
		\fmf{phantom,tension=1.}{v1,M,v3}
		\fmf{phantom,tension=1.}{v2,M,v4}
		\fmfv{label=$\mathrm{U}$,label.d=0}{M}
	    \end{fmfgraph*}
}
 = \,
\parbox{0.01\linewidth}{
	    \begin{fmfgraph*}(1,15)
		\fmfleft{i1,i2}
		\fmfright{o1,o2}
		\fmf{dashes,tension=0}{o1,o2}
		\fmfdot{o1,o2}
	    \end{fmfgraph*}
	    }
\, + \,
\parbox{0.15\linewidth}{
	    \begin{fmfgraph*}(15,15)
		\fmfleft{i1,i2}
		\fmfright{o1,o2}
		\fmf{plain_arrow,width=2}{i2,o2}
		\fmf{plain_arrow,width=2}{o1,i1}
		\fmf{dashes,tension=0}{i1,o2}
		\fmf{dashes,tension=0}{o1,i2}
		\fmfdot{o1,o2,i1,i2}
	    \end{fmfgraph*}
	    }
	    \, + \,
\parbox{0.3\linewidth}{
	    \begin{fmfgraph*}(30,15)
		\fmfleft{i1,i2}
		\fmfright{o1,o2}
		\fmftop{t1}
		\fmf{plain_arrow,width=2}{i2,v2,vv2,o2} 
		\fmf{plain_arrow,width=2}{vv1,v1} 
		\fmf{phantom}{o1,vv1}
		\fmf{phantom}{v1,i1}
		\fmf{dashes,tension=0}{v1,v2}
		\fmf{dashes,tension=0}{vv1,vv2}
		\fmf{dashes,left=0.5,tension=0}{i2,o2}
		\fmfdot{o2,i2,v1,vv1,v2,vv2}
	    \end{fmfgraph*}
	    }
\, + \, ...
\label{Diag:Uvert}
\ee
\end{fmffile}\\

%% file: TexFiles/Conductivity.06.tex
\subsection{Conductivity and Einstein's relation \label{part:Einst-rel}}

Finally, another quantity of interest for our problem -- in parallel of those studied in sections~\ref{par:spectral} and \ref{part:BSE} -- is the conductivity. 
In complete analogy to the usual conductivity of charge in condensed-matter systems~\cite{mahan2000}, we here define the conductivity tensor $\sigtens$ in our system as proportional to the current-current correlation function, via the Kubo formula\footnote{This corresponds to the more general definition $\sig^{i,j}(\omega,E) = \int_0^{\infty} \ud t \, e^{i\omega t} \textrm{Tr}\{ \delta(E-\Ham) j_{i}(x,t) j_{j}(x)\}$ ($j$ is the current operator) where the correlations between $\Gr$ and $\Ga$ have been dropped (see for example Ref.~\cite{akkermans2006}).}~\cite{akkermans2006}:
\be \label{eq:def-cond}
\sig^{i,j}(\omega,E) =
\int \frac{\ud \veck}{(2 \pi)^d} \frac{\ud \veck'}{(2 \pi)^d} \, \Re \left[ \, \av{{\upsilon}_i \langle \veck \vert \Gr(E_+) \vert \veck' \rangle {\upsilon}'_j \langle \veck' \vert \Ga(E_-) \vert \veck \rangle} \right],
\ee
where ${\upsilon}_i=\hbar k_i/m$ is the velocity along axis $i$.
As the structure of Eq.~(\ref{eq:def-cond}) is reminiscent of the definition of the four-point vertex $\Phi$ [see Eq.~(\ref{eq:phi-bb})], calculations of the conductivity tensor can also be represented diagrammatically.
The skeleton diagram, shown in Eq.~(\ref{diag:cond-skel}), consists of the in and out-going velocities ${\boldsymbol \upsilon}$ and ${\boldsymbol \upsilon}'$ and of a \emph{bubble} made of a retarded (top line) and an advanced (bottom line) Green function.
As for $\Phi$, the scattering events of the top and bottom lines can be correlated [see for example Eqs.~(\ref{Diag:BSE}) and (\ref{Diag:Uvert})].
\input{Diagrammes/diag-cond-skel}

Thanks to Einstein's classical argument, it was realized that, at thermal equilibrium, in a gas submitted to a force, the diffusion and drift currents have to be equal.
This relation holds in general for quantum systems in the linear response regime (see \eg\ Ref.~\cite{rammer1998}).
In particular, here we expect the DC conductivity and diffusion tensors to be proportional~: $\sigtens(\omega=0) \propto \Diff$.
Calculating $\sigtensB(\omega=0)$ in the Boltzmann and Born approximations for anisotropic disorder permits us to find the proportionality factor (see details in appendix~\ref{ap:Einstein}), which
in our system yields
\be
\sigtens=\frac{2 \pi N_0(E)}{\hbar} \Diff.
\label{eq:releinstein}
\ee

%% file: Diagrammes/diag-cond-skel.tex
\begin{fmffile}{diag-cond-skel-fmf}
\be
\parbox{0.3\linewidth}{
	    \begin{fmfgraph*}(30,15)
		\fmfleft{i1}
		\fmfright{o1}
		\fmfpoly{phantom,tension=0.}{v1,v2,v3,v4,v5,v6}
		\fmf{wiggly,tension=5.,label=${\boldsymbol \upsilon}$,l.s=right}{i1,v4}
		\fmf{plain_arrow,left=0.2,tension=0.,width=2}{v4,v3}
		\fmf{dashes,left=0.2,tension=0.,width=2}{v3,v2}
		\fmf{plain_arrow,left=0.2,tension=0.,width=2}{v2,v1}
		\fmf{plain_arrow,left=0.2,tension=0.,width=2}{v1,v6}
		\fmf{dashes,left=0.2,tension=0.,width=2}{v6,v5}
		\fmf{plain_arrow,left=0.2,tension=0.,width=2}{v5,v4}
		\fmf{wiggly,tension=5.,label=${\boldsymbol \upsilon}'$,l.s=right,l.d=3.5}{v1,o1}
		\fmf{phantom}{v3,v5}
		\fmf{phantom}{v2,v6}
	    \end{fmfgraph*}
}
\label{diag:cond-skel}
\ee
\end{fmffile}\\

%% file: TexFiles/CorrF.07.tex
\section{Disorder correlation function \label{part:correl}}

Having recalled the general theory of quantum transport in disordered media, we now specify the framework of our study.
We will consider ultracold matter waves in speckle potentials as realized in several experiments \cite{lye2005,clement2005,fort2005,schulte2005,clement2008,chen2008,billy2008,white2009,pasienski2010,mrsv2010,kondov2011,jendrzejewski2011}.

\begin{figure}[!t]
\begin{center}
\includegraphics[width=0.7\textwidth]{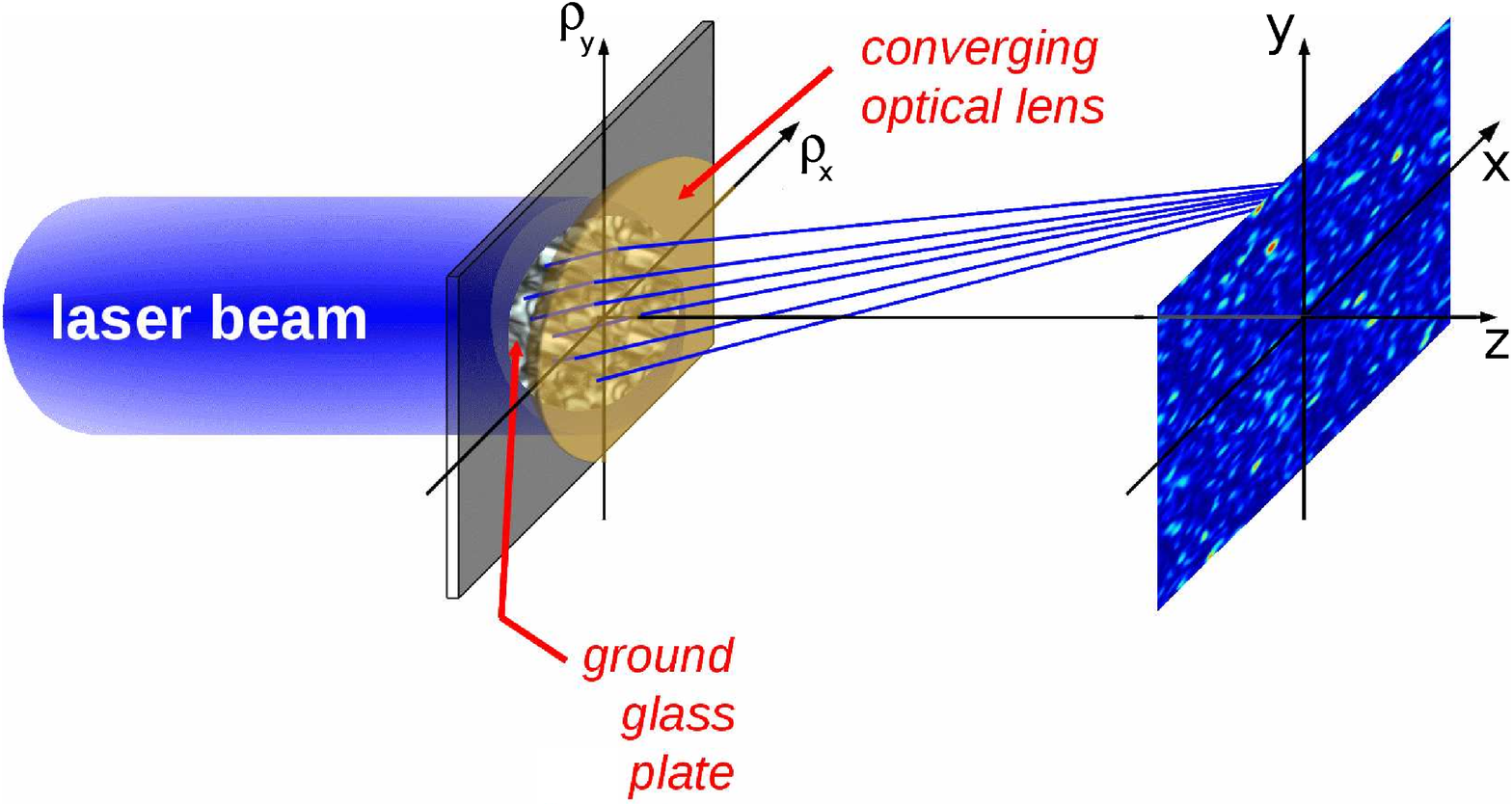}
\end{center}
\vspace{-0.3cm}
\caption{\small{(Color online)
Schematic of the apparatus used to create an optical speckle pattern.
A laser beam is diffracted by a ground-glass plate diffuser
of pupil function $\Id (\vect{\boldsymbol{\rho}})$, where $\vect{\boldsymbol{\rho}} \equiv (\rho_x,\rho_y)$ spans the diffuser,
which imprints a random phase on the various light paths.
The intensity field, $\Iscreen (\vect{r})$, observed
in the focal plane of
a converging lens, is a speckle pattern,
which creates a disordered potential $\Vopt(\vect{r})$ for the atoms.}
}
\label{fig:setup_speckle}
\end{figure}

In brief, a speckle pattern is created when a coherent light beam is shone through a diffusive plate and focused by an optical lens of focal distance $f$ (see Fig.~\ref{fig:setup_speckle} and Ref.~\cite{goodman2007}).
At each point of its surface, the diffusive plate imprints a random phase on the electric field.
The resulting electric field in the right-hand side of the lens is then the summation of many complex independent random components, and is therefore a Gaussian random variable according to the central limit theorem. 
The potential acting on the atoms is proportional to the intensity pattern (\ie\ the square modulus of the electric field).
It is thus a spatially (non Gaussian) random variable.
It is mainly characterized by the two-point correlation function $\Cor(\vect{r})=\av{\Vopt(\vect{r})\Vopt(\vect{0})}$.
For a fine-grain diffuser, the two-point correlation function $\Cor(\vect{r})$ is determined by the pupil function $\Id (\vect{\boldsymbol{\rho}})$ (\ie\ the intensity pattern just after the diffusive plate) \cite{goodman2007}.
For Gaussian laser beams of waists $w_{x,y}$ and plates with homogeneous transmission\footnote{\new{Here we assume that the diffuser covers the full area lit by the Gaussian beam. If it is not the case, a cut-off has to be introduced in the pupil function, which results into some oscillations in the wings of the correlation function. In experiments, if the diffusive plate is sufficiently large, this effect is small, and we disregard it in the following.}},
we have $\Id (\rho_x, \rho_y)=I_0 e^{-2(\rho_x^2/w_{x}^2+\rho_y^2/w_{y}^2)}$.
For the configuration of Fig.~\ref{fig:setup_speckle}, in the paraxial approximation, we find
\be
\Cor(\vect{r})=\Vr^2\cor_{\textrm{1sp}}(x,y,z)
\label{eq:corr1sp}
\ee
with
\be
\cor_{\textrm{1sp}}(x,y,z) = 
\frac{\exp{\left[ - \frac{x^2/\sigmaOrthx^2}{1+{4}z^2/\sigmaParax^2} \right]}}{\sqrt{1+{4}z^2/\sigmaParax^2}}\frac{ \exp{\left[ - \frac{y^2/\sigmaOrthy^2}{1+{4}z^2/\sigmaParay^2} \right]}}{\sqrt{1+{4}z^2/\sigmaParay^2}},
\label{eq:corr1sp_r}
\ee
${\sigmaPara}_{x,y}=4 \lambdaL f^2/\pi w_{x,y}^2$ and ${\sigmaOrth}_{x,y}=\lambdaL f/\pi w_{x,y}$ where $\lambdaL$ is the laser wavelength.
Here $x$ and $y$ are the coordinates orthogonal to the propagation axis $z$, and $z=0$ corresponds to the focal plane.
We chose $\Vr \equiv \sqrt{C(\vecr=0)}$ as definition of the amplitude of the disorder.

\subsection{Anisotropic Gaussian speckle (2D) \label{part:correl-2D}}

If the atoms are confined in a 2D geometry by a strong trapping potential along $z$ centered on $z=0$, they experience a disordered potential with correlation function 
$\Cor(x,y)=\Vr^2\cor_{\textrm{1sp}}(x,y,0)
= \Vr^2 \exp{\left[ - \frac{1}{\sigmaOrth^2}(x^2+\anifact^2 y^2) \right]}$,
with $\sigmaOrth=\sigmaOrthx$ and $\anifact=\sigmaOrthx/\sigmaOrthy$ the configuration anisotropy factor.
The Fourier transform gives the power spectrum
\be
\TFCor(\vect{k})=\Vr^2 \pi \frac{\sigmaOrth^2}{\anifact} \exp{\left[ - \frac{\sigmaOrth^2}{4} (k_x^2+\frac{k_y^2}{\anifact^2}) \right]}.
\label{2D-corrTF}
\ee
Without loss of generality, we assume that $\anifact \geq 1$.
When $|\veck| \ll \sigmaOrthx^{-1}, \sigmaOrthy^{-1}$, we get $\TFCor(\vect{k})\simeq \Vr^2 \pi \frac{\sigmaOrth^2}{\anifact}$ and we recover the power spectrum of white noise disorder, the only relevant parameter being $\Vr^2 \sigmaOrthx \sigmaOrthy$.
The power spectrum (\ref{2D-corrTF}) is obtained by shining an anisotropic Gaussian beam on the diffusive plate.
It also approximately holds in the case of Ref.~\cite{mrsv2010} where 
a quasi-2D Bose gas of width $l_z$ is subjected to a speckle created by an \emph{isotropic} Gaussian laser beam shone with an angle $\theta$ with respect to the plane of atoms, if $l_z \ll \sigmaOrth \ll \sigmaPara$.
In this case $\anifact\simeq 1/\sin\theta$ ($\theta \simeq \pi/6$ for the experiment of Ref.~\cite{mrsv2010}).

\subsection{Single speckle (3D) \label{part:correl-single}}

In the 3D case, the disorder correlation function $\Cor(\vecr)$ is given by Eq.~(\ref{eq:corr1sp}) with $w_{x}=w_{y}=w$.
The resulting speckle pattern has correlation lengths $\sigmaPara$ in the propagation axis ($z$) and $\sigmaOrth$ in the orthogonal plane ($x,y$).
In general $4f > w$, and $\Cor (\vecr)$ is elongated along $z$ .
The corresponding disorder power spectrum reads 
\be
\label{TFcorsingle}
\TFCor(\veck)=\Vr^2 \TFcor_{\textrm{1sp}}(\veck)
\ee
with
\be
\label{TFcorsingle2}
\TFcor_{\textrm{1sp}}(\veck)=\pi^{3/2} \frac{\sigmaOrth \sigmaPara}{\vert\kOrth\vert} e^{- \frac{\sigmaOrth^2}{4} \kOrth^2} e^{-\frac{1}{4} \left(\frac{\sigmaPara}{\sigmaOrth}\right)^2 \frac{k_z^2}{\kOrth^2}},
\ee
where $\veckOrth$ is the projection of $\veck$ in the ($k_x,k_y$) plane.
It is isotropic in the ($k_x,k_y$) plane but has a significantly different shape along the $k_z$ axis.
In particular, it shows a strong algebraic divergence when $k_z=0$ and $k_x^2+k_y^2 \rightarrow 0$.
It features absence of white-noise limit, which reflects the long-range correlations of the potential\footnote{\new{We assume that the disorder extends to infinity. In experiments the speckle pattern has a finite extension $L$, and the divergence is truncated at $\vert k \vert \sim 1/L$. If the parameters are such that those components play a role, the inhomogeneity of the disorder has to be taken into account.}}.
The consequences of this property, obtained in the paraxial approximation, will be further discussed in the following.

%% file: TexFiles/Scattering.06.tex
\subsection{Scattering mean-free time}

\be
\selfE(E)= \av{V \Gr_0(E) V}.
\ee
For homogeneous disorder,
$\langle \veck | \selfE(E) | \veck' \rangle = (2\pi)^d \delta(\veck-\veck') \selfE(E,\veck) \,  $ with
\be
\selfE(E,\veck) = \int \frac{\ud \veck''}{(2 \pi)^d} \,
\TFCor(\veck - \veck'') \, \Gr_0(E,\veck''),
\label{eq:selfE-Born}
\ee
where $\TFCor(\veck)$ is the disorder power spectrum.
Using Eq.~(\ref{eq:tau_s-def}) and the disorder-free Green function, we thus have
\be
\smft(E,\veck) = \frac{\hbar}{2\pi}
\frac{1}{\big\langle \TFCor(\veck - \veck') \big\rangle_{\veck'\vert E}},
\label{eq:tau_s}
\ee
where 
\be
\big\langle ... \big\rangle_{\veck'\vert E} = \int \frac{\ud \veck'}{(2 \pi)^d} \,
... \, \delta\left[E - \eps{\veck'} \right]
\label{eq:def-moy}
\ee
represents the integration over the $\veck$-space shell defined by $\eps{\veck}=E$.
In the following we discuss anisotropic properties of the scattering time for the 2D case (the 3D cases are presented in Sec.~\ref{part:scatt-3D}).

In the case of isotropic disorder [\ie\ $\TFCor(\veck - \veck')= \TFCor(\vert \veck - \veck'\vert)$] the scattering time does not depend on the direction of the incoming wave vector $\veck$.
In general, the scattering is however anisotropic, \ie\ the probability that the particle acquires a direction $\veck'$ depends on the direction of $\veck'$ relative to $\veck$.
Isotropic scattering is found only for $\delta$-correlated disorder
In the case of anisotropic disorder we are interested in, not only the scattering depends on the relative direction of $\veck'$ and $\veck$, but it also depends on the direction of the incoming wave $\veck$.

%% file: TexFiles/ScatteringExample2DGauss.06.tex
\subsection{Anisotropic Gaussian speckle (2D) \label{part:2Dex-scat}}

\begin{figure}[!t] 
\begin{center}
\includegraphics[width=0.7\textwidth]{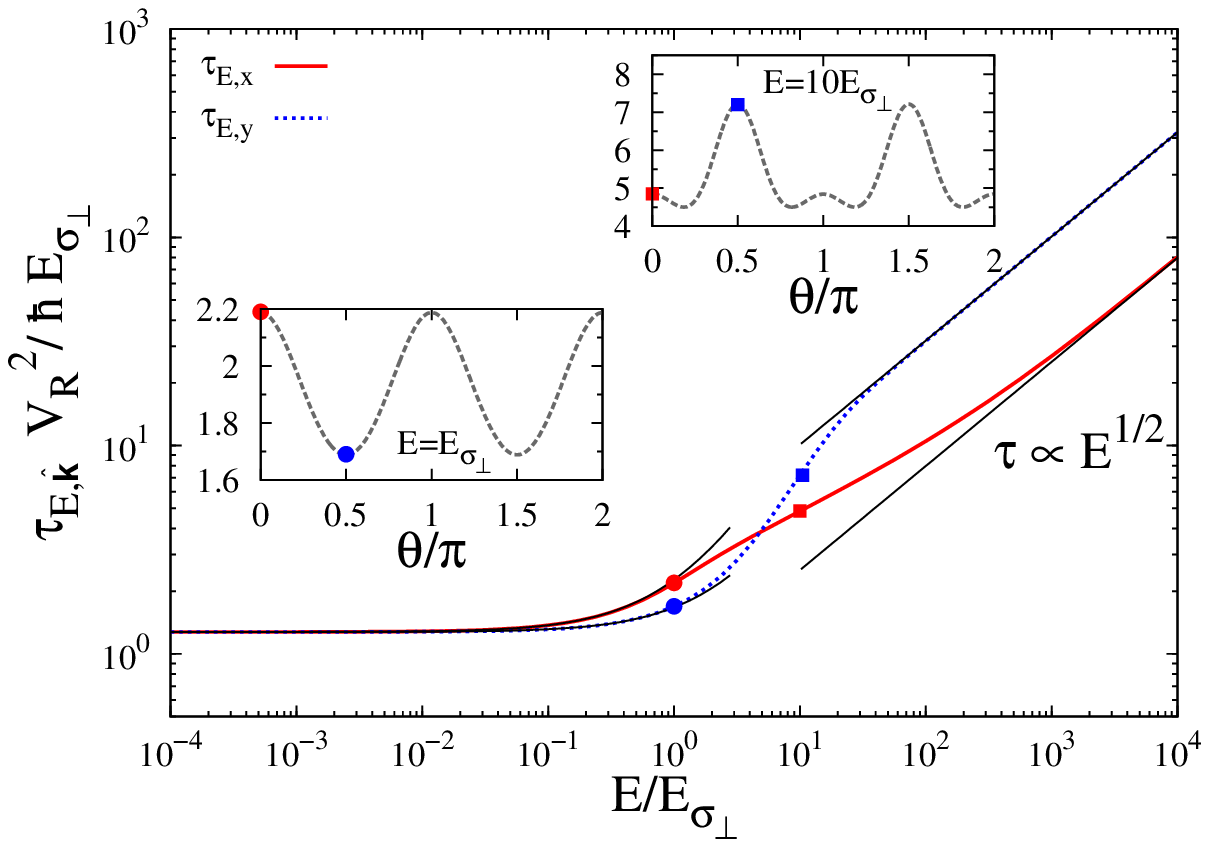}
\end{center} 
\caption{\small{(Color online) On-shell scattering mean free time $\tau_{E,\uveck}\equiv\smft(E,\kE\uveck)$ [Eq.~(\ref{eq:taus2Dgauss}) for $|\veck|=\kE$] along the $\uveck_x$ (solid red line) and $\uveck_y$ directions (dotted blue line) for the 2D speckle potential defined in Sec.~\ref{part:correl-2D} with $\anifact=4$.
The solid black lines are the isotropic low-energy limits obtained for $\kE\sigmaOrth \ll 1$ [Eq.~(\ref{ls2D_smallk})] and the high-energy limit obtained for $\kE\sigmaOrth \gg \anifact$ [Eq.~(\ref{ls2D_largek})].
The insets show the angular dependance of $\tau_{E,\uveck}$ at two different energies [with the parametrization $\uveck\equiv(\cos\theta,\sin\theta)$].
The points on the lines are color- and shape-coded to match those in the insets.}} \label{Figure:ls_gauss2D}
\end{figure}
Let us consider the 2D anisotropic speckle potential of geometrical anisotropy factor $\anifact$ introduced in Sec.~\ref{part:correl-2D}.
Replacing $\TFCor(\veck)$ by Eq.~(\ref{2D-corrTF}) in Eq.~(\ref{eq:tau_s}) and using the disorder-free dispersion relation of the vacuum
in Eq.~(\ref{eq:def-moy}), we obtain the scattering mean free time
\be \label{eq:taus2Dgauss}
\smft(E,\veck)= \frac{\hbar E_{\sigmaOrth}}{V_R^2}
\frac{2\anifact}{\int \ud\Omega_{\uveck'} \, 
e^{-\frac{\sigmaOrth^2}{4} (\kE \hat{k}'_x- k_x )^2}
e^{-\frac{\sigmaOrth^2}{4\anifact^2} (\kE \hat{k}'_y - k_y )^2}},
\ee
where $\uveck\equiv \veck/|\veck|$ is the unit vector pointing in the direction of $\veck$,
$\Omega_{\uveck}$ is the $\veck$-space solid angle, $\kE \equiv \sqrt{2mE}/\hbar$ is the momentum associated to energy $E$ in free space and $E_{\sigmaOrth} \equiv \hbar^2/m \sigmaOrth^2$ is the correlation energy of the disorder.
The scattering time (\ref{eq:taus2Dgauss}) is plotted in Fig.~\ref{Figure:ls_gauss2D} as a function of energy along the two main axes, for $|\veck|=\kE$ and for a fixed geometrical anisotropy $\anifact=4$.
Let us discuss some limiting cases and use the notation $\tau_{E,\uveck}\equiv\smft(E,\kE\uveck)$.

In the low-energy limit, $\kE \sigmaOrth \ll 1$, we have
\be
\tau_{E,\uveck} = \frac{\hbar E_{\sigmaOrth}}{V_R^2} \frac{\anifact}{\pi} + \frac{\hbar E}{4\pi V_R^2}  \bigg[ \anifact+\frac{2}{\anifact}+2\left( \anifact \hat{k}_x^2+\frac{\hat{k}_y^2}{\anifact} \right)
+ \Oo \left(\frac{E^2}{\anifact^4 E_{\sigmaOrth}^2} \right) \bigg],
\label{ls2D_smallk}
\ee
which is displayed in Fig.~\ref{Figure:ls_gauss2D} (left-hand-side black lines).
In this limit the de Broglie wavelength of the particle ($2\pi/\kE$) exceeds the correlation lengths of the disorder ($\sigmaOrthx$ and $\sigmaOrthy$) and the speckle can be approximated by a white-noise (uncorrelated) disordered potential. Equation~(\ref{2D-corrTF}) becomes $\TFCor(\vect{k})\simeq \Vr^2 \pi \frac{\sigmaOrth^2}{\anifact}$ (see Sec.~\ref{part:correl-2D}) and $\tau_{E,\uveck}$ is isotropic, constant, and it only depends on the product $\Vr^2 \sigmaOrthx \sigmaOrthy$ (up to corrections of relative order $E/E_{\sigmaOrth}$).

In the opposite, high-energy limit, $\kE\sigmaOrth \gg \anifact$, the de Broglie wavelength of the particle is much smaller than the 
smallest correlation length of the disorder 
and the particle behaves `classically'.
Since $\TFCor(\vect{k})$ has a wider extension in the $\uveck_y$ direction than in the $\uveck_x$ direction (for $\anifact>1$), there are more scattering channels for particles travelling along $x$ so that $\tau_{E,\uveck_x}<\tau_{E,\uveck_y}$.
More precisely, we find
\be \label{ls2D_largek}
\tau_{E,\uveck} \simeq \frac{\hbar E_{\sigmaOrth}}{V_R^2} \,
\frac{\kE \sigmaOrth}{\sqrt{\pi}} \sqrt{\hat{k}_x^2+\anifact^2 \hat{k}_y^2},
\ee
which is shown in Fig.~\ref{Figure:ls_gauss2D} (right-hand-side black lines).
In particular, we find that in the high-energy limit $\tau_{E,\uveck} \propto \sqrt{E}$.

\begin{figure}[!t] 
\begin{center}
\includegraphics[width=0.7\textwidth]{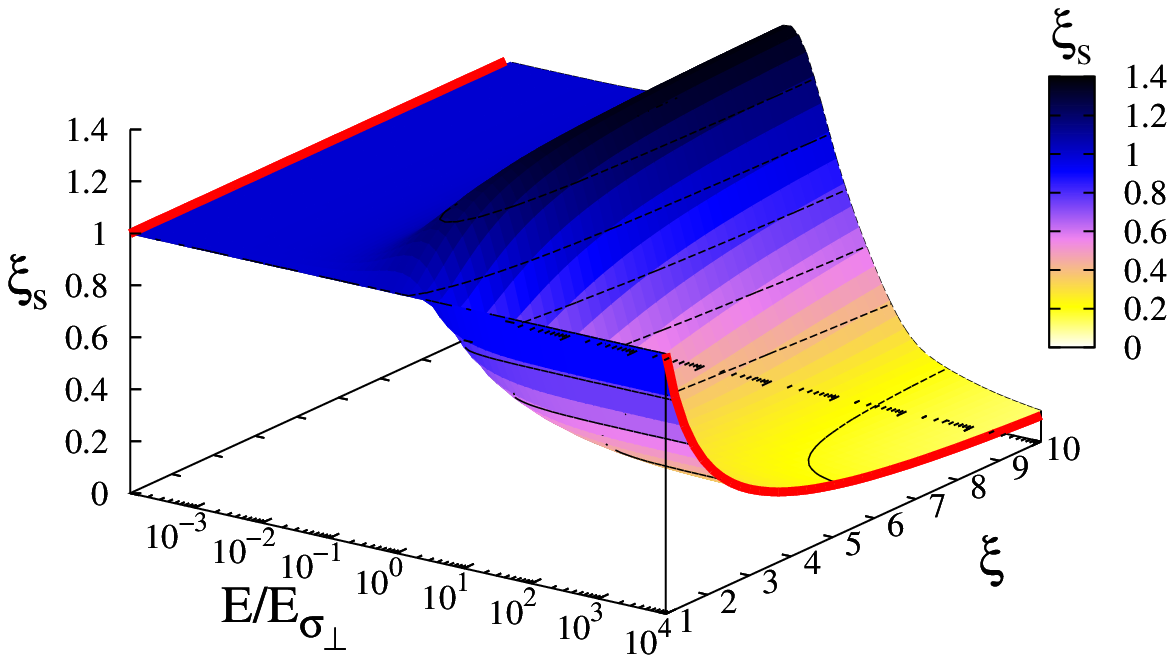}
\end{center} 
\caption{\small{(Color online) Anisotropy factor of the scattering time, $\anifact_s=\tau_{E,\uveck_x}/\tau_{E,\uveck_y}$, as a function of $E/E_{\sigmaOrth}$ and $\anifact$, for the 2D speckle potential of Sec.~\ref{part:correl-2D}.
The red lines are the low ($\anifact_s \rightarrow 1$) and high energy limits ($\anifact_s \rightarrow \frac{1}{\anifact}$) [see Eqs.~(\ref{ls2D_smallk}) and (\ref{largek_gauss})].}}
\label{Figure:lambdas2D}
\end{figure}
It is also interesting to study the anisotropy factor of the scattering time
\be \label{lambdas2D}
\anifact_s \equiv \frac{\tau_{E,\uveck_x}}{\tau_{E,\uveck_y}},
\ee 
which is shown in Fig.~\ref{Figure:lambdas2D} as a function of $E/E_{\sigmaOrth}$ and $\anifact$.
As already mentioned $\tau_{E,\uveck}$ is isotropic in the white-noise limit, so that $\anifact_s \simeq 1$ for $\kE \sigmaOrth \ll 1$ (left-hand-side red line in Fig.~\ref{Figure:lambdas2D}).
When increasing the energy, the scattering time first increases along the direction with the largest correlation length, \ie\ the direction in which $\TFCor(\vect{k})$ is narrower ($x$ for $\anifact>1$). 
Therefore, $\anifact_s$ increases with $E$, for sufficiently small values of $E/E_{\sigmaOrth}$, and we have $\anifact_s>1$. 
Using Eq.~(\ref{ls2D_smallk}), an explicit calculation yields
\be \label{smallk_gauss}
\anifact_s \simeq 1 + \frac{E}{E_{\sigmaOrth}} \frac{\anifact^2 - 1}{2\anifact^2} + \Oo \left(\frac{E^2}{\anifact^4 E_{\sigmaOrth}^2} \right).
\ee
For $\kE \sigmaOrth \gg \anifact$, using Eq.~(\ref{ls2D_largek}), we obtain
\be \label{largek_gauss}
\anifact_s \simeq \frac{1}{\anifact},
\ee 
which shows that the anisotropy factor of scattering is proportional to the inverse of the geometrical anisotropy (right-hand-side red line in Fig.~\ref{Figure:lambdas2D}).
Note that the classical limit relation~(\ref{largek_gauss}) is universal provided that the configuration anisotropy factor is well defined,
\ie\ that the disorder correlation function can be obtained by the anisotropic homothety  of an isotropic one,
$\Cor(x,y)=\Cor_{\textrm{iso}}(x,\anifact y)$.
In this high-energy limit, $\anifact_s < 1$ (contrary to the low-energy limit case).
Therefore, for any value of $\anifact$, $\tau_{E,\uveck}$ exibits an inversion of anisotropy when the energy increases, typically at $E\sim E_{\sigmaOrth}$.

\begin{figure}[!t] 
\begin{center}
\includegraphics[width=0.8\textwidth]{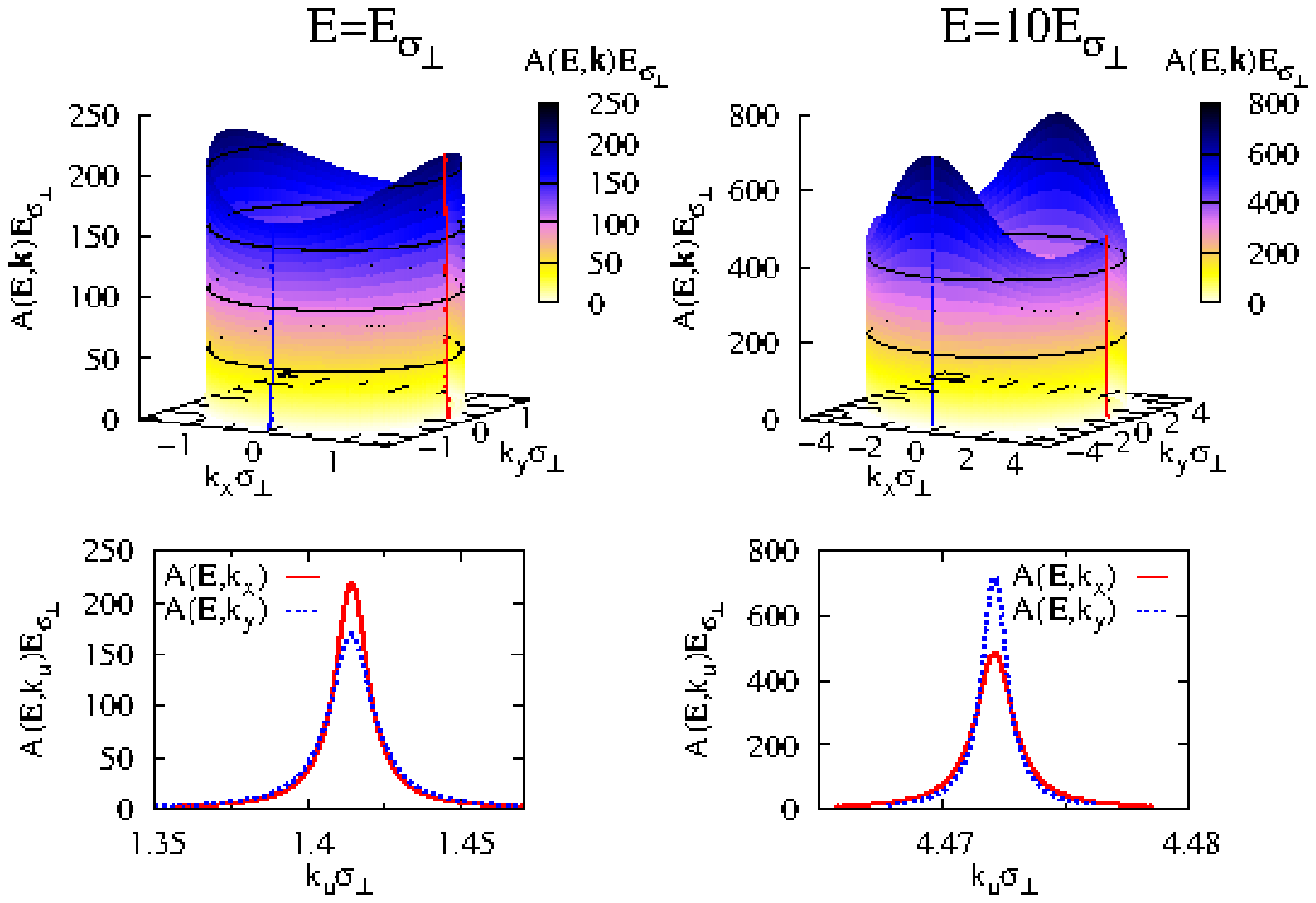}
\end{center} 
\caption{\small{(Color online) On-shell spectral function as a function of $\veck$
for the 2D speckle potential of Sec.~\ref{part:correl-2D}, with $\Vr=0.2E_{\sigmaOrth}$ and $\anifact=4$.
The top row shows the full spectral function.
The bottom row shows cuts along the $k_x$ (solid red lines) and $k_y$ axis (dotted blue lines).
The two columns refer to different energies:
$E=E_{\sigmaOrth}$ (left) and $E=10E_{\sigmaOrth}$ (right),
which correspond to the dots and the squares in Fig~\ref{Figure:ls_gauss2D}, respectively.}} \label{Figure:Spectral2D}
\end{figure}
As described in section~\ref{par:spectral} the scattering time is the width of the spectral function.
It can be measured in a 2D experiment such as that of Ref.~\cite{mrsv2010} by monitoring the momentum distribution of an almost energy-resolved wavepacket \cite{cherroret2012}.
To illustrate this, a plot of the spectral function as a function of momentum and at fixed energy is shown in Fig.~\ref{Figure:Spectral2D}.
In each direction $\uveck$ the spectral function peaks at $4\tau_{E,\uveck}/\hbar$ and has a width proportional to $1/\tau_{E,\uveck}$.
The anisotropy of the scattering time is revealed in the angle-dependence of both these quantities.
It is more apparent in the angular dependence of the amplitude, which shows marked peaks.
At low energy, the maxima are located on the $k_x$ axis, while at high energy, they are located on the $k_y$ axis, which signals inversion of the scattering anisotropy.

%% file: TexFiles/Diffusion.06.tex
\subsection{Solution of the Bethe-Salpeter equation \label{part:diff}}
In the independent scattering (Boltzmann) and weak disorder (Born) approximation,
only the first term in Eq.~(\ref{Diag:Uvert}) is retained and the irreducible vertex function $\mathrm{U}$ equals the disorder structure factor~\cite{rammer1998}: $\mathrm{U} \simeq \mathrm{U}_{\mathrm{B}} = \av{V \otimes V}$ and
\be
U_{\veck,\veck'} (\vecq, \omega, E) \simeq {\UB}_{\veck,\veck'} = \TFCor (\veck - \veck'),
\label{eq:U-diff}
\ee
or equivalently
\input{Diagrammes/diag-U-diff}
Then, incorporating Eq.~(\ref{eq:U-diff})-(\ref{Diag:U-diff}) into the BSE~(\ref{eq:BSE})-(\ref{Diag:BSE}) and expanding it in series of $\mathrm{U}$, one finds
\vspace{0.25cm}
\input{Diagrammes/diag-diff}
where the diffuson $\Gamma$ reduces to ladder diagrams:
\input{Diagrammes/diag-gamma}
It describes an infinite series of independent scattering events, which leads to Drude-like diffusion.

In appendix~\ref{part:ap-BSE}, explicit calculations are detailed.
In brief, in the long-time ($\omega \to 0$) and large-distance ($|\vecq| \to 0$) limit the vertex $\Phi$ is the sum of a regular term and a singular term~\cite{woelfe1984,bhatt1985}:
\be \label{BSEsolution}
\Phi_{\veck,\veck'}(\vecq,\omega,E) = 
\Phi^{\mathrm{sing}}_{\veck,\veck'}(\vecq,\omega,E) + \Phi^{\mathrm{reg}}_{\veck,\veck'}(0,0,E).
\ee
The regular part is given by 
\be \label{BSEregular}
\Phi^{\mathrm{reg}}_{\veck,\veck'}(0,0,E) = \sum_{\lambda_E^n \neq 1} \frac{1}{1-\lambda_E^n} f_{E,\veck} \phi_{E,\veck}^n \phi_{E,\veck'}^n f_{E,\veck'},
\ee
where $f_{E,\veck} \equiv f_{\veck}(\vecq=0,\omega=0,E)$ [see Eq.~(\ref{eq:fkE})]
and $\phi_{E,\veck}^n$ ($\lambda_E^n$) are the eigenvectors (eigenvalues) of an integral operator involving the
disorder correlation function and\footnote{This operator is in fact $1-\Oplamb$, taken in the Born and Boltzmann approximations, where $\Oplamb$ has been introduced in paragraph~\ref{part:BSE}.} $f_{E,\veck}$:
\be \label{eigeq}
\int \frac{\ud \veck'}{(2 \pi)^d} \, \TFCor(\veck-\veck') \, f_{E,\veck'} \,
\phi_{E,\veck'}^n = \lambda_E^n \phi_{E,\veck}^n.
\ee
The regular part contributes to the finite time and finite distance propagation of the density, which we disregard here.
The existence of the singular part is a direct consequence of the Ward identity~\cite{vollhardt1980b} which expresses the conservation of particle number, and which guarantees that one of the eigenvalues of Eq.~(\ref{eigeq}) is equal to one, $\lambda^{n=1}_E=1$ (see appendix~\ref{part:ap-BSE}).
In the framework of the on-shell approximation, such that $\eps{\veck}=\eps{\veck'}=E$,
in the long time and large distance limit $(|\vecq|,\omega) \to 0$,
the vertex $\Phi$ is given by
\be \label{BSEsingular}
\Phi^{\mathrm{sing}}_{\veck,\veck'}(\vecq,\omega,E) 
= \frac{2\pi}{\hbar N_0(E)} \frac{\gamma_{\veck}(\vecq,E) \, \gamma_{\veck'}(\vecq,E)}
{- i \omega + \vecq \! \cdot \! \DiffTensB(E) \! \cdot \! \vecq}
\ee
with $N_0(E)$ the disorder-free density of states, and
\beq 
&& \gamma_{\veck}(\vecq,E) =
\frac{A_0(E,\veck)}{2\pi}
\bigg\{ 1-\frac{2 \pi i}{\hbar}
\label{gamma} \\
&& ~~~~~ \times \sum_{\lambda^n_E \neq 1} \frac{\lambda^n_E}{1-\lambda^n_E} \, \tau_{E,\uveck} \phi_{E,\uveck}^n \,  
\langle \vecq \cdot {\boldsymbol \upsilon'} \tau_{E,\uveck'} \phi_{E,\uveck'}^n \rangle_{\veck'\vert E} \bigg\},
\nonumber
\eeq
where $A_0(E,\veck)=2\pi \delta[E-\eps{\veck}]$ is the disorder-free spectral function.
Equation~(\ref{BSEsingular}) shows that the vertex $\Phi$ is dominated by the diffusion pole $(i \hbar \omega - \hbar \vecq \! \cdot \! \DiffTensB(E) \! \cdot \! \vecq)^{-1}$.
The Boltzmann diffusion tensor $\DiffTensB(E)$ has components~\cite{woelfe1984}
\beq 
&& \DB^{i,j}(E) = 
\frac{1}{N_0(E)} 
\bigg\{  
\Big\langle \tau_{E,\uveck} \, \upsilon_i \, \upsilon_j \Big\rangle_{\veck\vert E} 
\label{DBE} \\
&& ~~~~~ + \frac{2 \pi}{\hbar } 
\sum_{\lambda^n_E \neq 1} \frac{\lambda^n_E}{1 \! - \! \lambda^n_E} 
\Big\langle \tau_{E,\uveck} {\upsilon}_i \phi_{E,\uveck}^n \Big\rangle_{\veck\vert E}
\Big\langle \tau_{E,\uveck} {\upsilon}_j \phi_{E,\uveck}^n \Big\rangle_{\veck\vert E}
\bigg\},
\nonumber
\eeq
where ${\upsilon}_i = \hbar k_i/m$,
$\tau_{E,\uveck} \equiv\smft(E,\kE\uveck) = \hbar/2\pi \langle \tilde{C}(\kE\uveck-\veck') \rangle_{\veck'\vert E}$
is the on-shell scattering mean free time [see Eq.~(\ref{eq:tau_s})], and
$\langle ... \rangle_{\veck\vert E}$ represents integration over the $\veck$-space shell defined by $\eps{\veck}=E$ [see Eq.~(\ref{eq:def-moy})].
The functions $\phi_{E,\uveck}^n$ and the real-valued positive numbers $\lambda^n_E$ are the solutions of the integral eigenproblem (\ref{eigeq}), which becomes, in the on-shell approximation (see appendix~\ref{part:ap-BSE}),
\be
\label{eigeq-os}
\frac{2\pi}{\hbar} \Big\langle \tau_{E,\uveck'} \tilde{C}(\kE\uveck-\veck') \phi_{E,\uveck'}^n 
\Big\rangle_{\veck'\vert E} = \lambda^n_E \, \phi_{E,\uveck}^n \,,
\ee 
normalized by
$\frac{2\pi}{\hbar} \big\langle \tau_{E,\uveck} \phi_{E,\uveck}^n\phi_{E,\uveck}^m \big\rangle_{\veck \vert E} = \delta_{n,m}$~\cite{woelfe1984}.
It follows from Eq.~(\ref{DBE}) that the incoherent (Boltzmann) diffusion tensor $\DiffTensB(E)$ is obtained from the two-point disorder correlation function $\Cor(\vecr)$, which determines $\tau_{E,\uveck}$ [see Eq.~(\ref{eq:tau_s})] as well as $\phi_{E,\uveck}^n$ and $\lambda^n_E$ [see Eq.~(\ref{eigeq-os})].

In the isotropic case (for details see appendix~\ref{part:limit-iso}), Eq.~(\ref{eigeq-os}) is solved by the cylindrical, $Z_l^{\pm 1}$, (2D; see appendix~\ref{part:limit-iso}) or spherical, $Y_l^m$, (3D) harmonics, the same level harmonics [\ie~with the same $l$] being degenerate in $\lambda^n_E$.
Then, it follows from the symmetries of the cylindrical/spherical harmonics that only the first term  in Eq.~(\ref{DBE}) plus the $p$-level harmonics ($Z_1^{\pm 1}$ in 2D and $Y_1^m$ with $m=-1,0,1$ in 3D) couple to $\upsilon$ and contribute to  $\DB(E)$.
Incorporating the explicit formulas for $\phi_{E,\uveck}^n$ and $\lambda^n_E$ [see Eqs.~(\ref{eq:iso-eigen}) to (\ref{eq:iso-diff-tens})], we then recover
well-known expressions for isotropic disorder~\cite{kuhn2005,kuhn2007,miniatura2009,beilin2010}.

In the anisotropic case, harmonics couple, and the $\phi_{E,\uveck}^n$ are no longer cylindrical/spherical harmonics.

%% file: Diagrammes/diag-U-diff.tex
\begin{fmffile}{diag-U-diff-fmf}
\be
\parbox{0.15\linewidth}{
	    \begin{fmfgraph*}(15,15)
		\fmfleft{v1,v4}
		\fmfright{v2,v3}
		\fmf{phantom}{v1,v2}
		\fmf{phantom}{v3,v4}
		\fmffreeze
		\fmf{plain}{v2,v1}
		\fmf{plain}{v4,v3}
		\fmf{plain,tension=0.}{v1,v4}
		\fmf{plain,tension=0.}{v2,v3}
		\fmffreeze
		\fmf{phantom,tension=1.}{v1,M,v3}
		\fmf{phantom,tension=1.}{v2,M,v4}
		\fmfv{label=$\mathrm{U}_{\mathrm{B}}$,label.d=0}{M}
	    \end{fmfgraph*}
}
\,  = \,
\parbox{0.01\linewidth}{
	    \begin{fmfgraph*}(1,15)
		\fmfleft{i1,i2}
		\fmfright{o1,o2}
		\fmf{dashes,tension=0}{o1,o2}
		\fmfdot{o1,o2}
	    \end{fmfgraph*}
	    }
\;\;\; .
\label{Diag:U-diff}
\ee
\end{fmffile}\\

%% file: Diagrammes/diag-diff.tex
\begin{fmffile}{diag-diff-fmf}
\be
\parbox{0.2\linewidth}{
	    \fmfreuse{diag-Phi}
	    }
=
\parbox{0.15\linewidth}{
	    \fmfreuse{diag-free}
	    }
\, + \,
\parbox{0.3\linewidth}{
	    \begin{fmfgraph*}(30,15)
		\fmfleft{i1,i2}
		\fmfright{o1,o2}
		\fmf{phantom,tension=1.}{i1,v1,v2,o1}
		\fmf{phantom,tension=1.}{o2,v3,v4,i2}
		\fmffreeze
		\fmf{plain_arrow,label=$\veck_-$,l.s=left,l.d=2,width=2}{v1,i1}
		\fmf{plain}{v2,v1}
		\fmf{plain_arrow,label=$\veck'_-$,l.s=left,l.d=2,width=2}{o1,v2}
		\fmf{plain_arrow,label=$\veck'_+$,l.s=left,l.d=2,width=2}{v3,o2}
		\fmf{plain}{v4,v3}
		\fmf{plain_arrow,label=$\veck_+$,l.s=left,l.d=2,width=2}{i2,v4}
		\fmf{plain,tension=0.}{v1,v4}
		\fmf{plain,tension=0.}{v2,v3}
		\fmffreeze
		\fmf{phantom,tension=1.}{v1,M,v3}
		\fmf{phantom,tension=1.}{v2,M,v4}
		\fmfv{label=$\Gamma$,label.d=0}{M}
		\fmfkeep{diag-gamma}
	    \end{fmfgraph*}
}
\label{Diag:BSE-diff}
\ee
\end{fmffile}\\

%% file: Diagrammes/diag-gamma.tex
\begin{fmffile}{diag-gamma-fmf}
\be
\parbox{0.15\linewidth}{
	    \begin{fmfgraph*}(15,15)
		\fmfleft{v1,v4}
		\fmfright{v2,v3}
		\fmf{phantom}{v1,v2}
		\fmf{phantom}{v3,v4}
		\fmffreeze
		\fmf{plain}{v2,v1}
		\fmf{plain}{v4,v3}
		\fmf{plain,tension=0.}{v1,v4}
		\fmf{plain,tension=0.}{v2,v3}
		\fmffreeze
		\fmf{phantom,tension=1.}{v1,M,v3}
		\fmf{phantom,tension=1.}{v2,M,v4}
		\fmfv{label=$\Gamma$,label.d=0}{M}
	    \end{fmfgraph*}
}
\,  = \,
\parbox{0.01\linewidth}{
	    \begin{fmfgraph*}(1,15)
		\fmfleft{i1,i2}
		\fmfright{o1,o2}
		\fmf{dashes,tension=0}{o1,o2}
		\fmfdot{o1,o2}
	    \end{fmfgraph*}
	    }
\, + \,
\parbox{0.15\linewidth}{
	    \begin{fmfgraph*}(15,15)
		\fmfleft{i1,i2}
		\fmfright{o1,o2}
		\fmf{plain_arrow,width=2}{i2,o2}
		\fmf{plain_arrow,width=2}{o1,i1}
		\fmf{dashes,tension=0}{i1,i2}
		\fmf{dashes,tension=0}{o1,o2}
		\fmfdot{o1,o2,i1,i2}
	    \end{fmfgraph*}
	    }
\, + \,
\parbox{0.22\linewidth}{
	    \begin{fmfgraph*}(22,15)
		\fmfleft{i1,i2}
		\fmfright{o1,o2}
		\fmf{plain_arrow,width=2}{i2,v2,o2}
		\fmf{plain_arrow,width=2}{o1,v1,i1}
		\fmf{dashes,tension=0}{i1,i2}
		\fmf{dashes,tension=0}{v1,v2}
		\fmf{dashes,tension=0}{o1,o2}
		\fmfdot{o1,o2,i1,i2}
	    \end{fmfgraph*}
	    }
\, + \, ...
\label{Diag:gamma}
\ee
\end{fmffile}\\

%% file: TexFiles/DiffusionExample2DGauss.06.tex
\subsection{Anisotropic Gaussian speckle (2D) \label{part:diff-ex}}

Consider again the 2D anisotropic speckle potential of Sec.~\ref{part:correl-2D}.
The first step in the calculation of $\DiffTensB$ is to determine the eigenfunctions $\phi_{E,\uveck}^n$ and the associated eigenvalues $\lambda^n_E$ of Eq.~(\ref{eigeq-os}).
We solve Eq.~(\ref{eigeq-os}) numerically, by a standard algorithm of diagonalization, with $2^{9}=512$ points, regularly spaced on the $\veck$-space shell $\vert\veck\vert=\kE$.
The diffusion tensor is diagonal in the basis made by the symmetry axes of the correlation function~(\ref{2D-corrTF}): $\{ \uvecu_x,\uvecu_y \}$.

\begin{figure}[!t] 
\begin{center}
\includegraphics[width=0.8\textwidth]{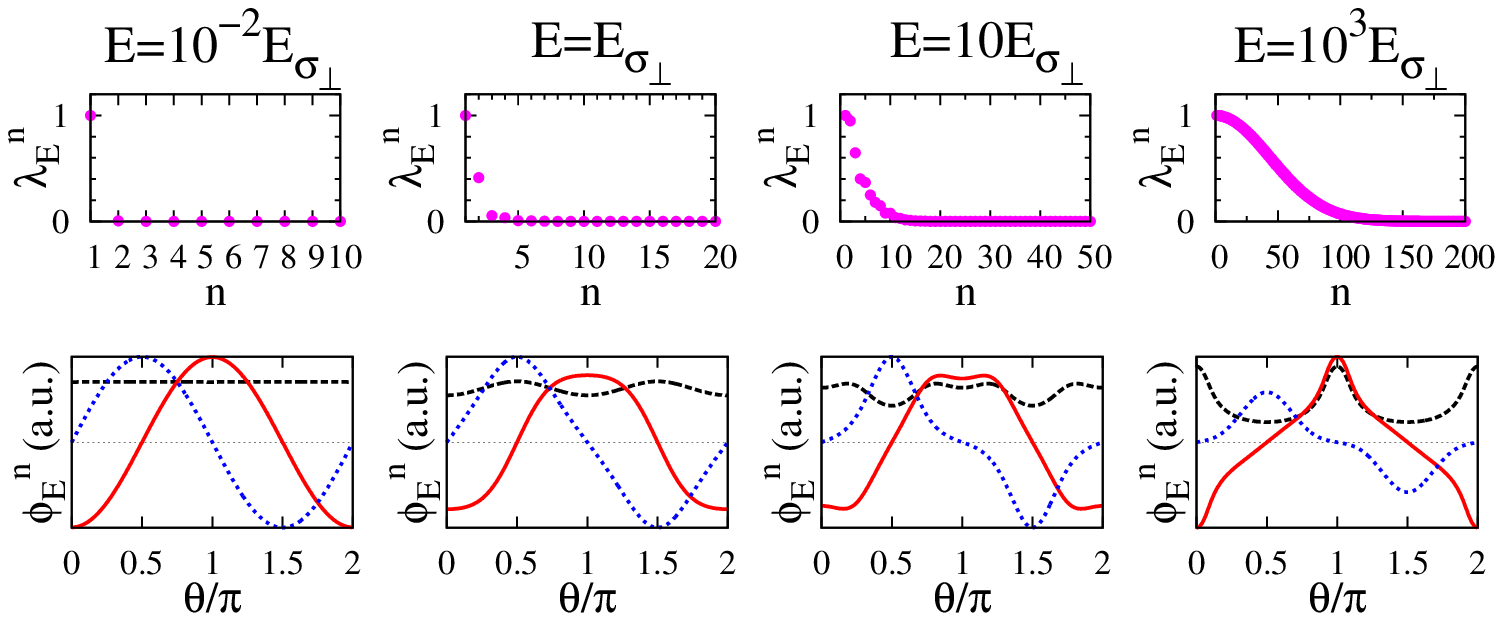}
\end{center} 
\caption{\small{(Color online) Top row: Eigenvalues of Eq.~(\ref{eigeq-os}) for the 2D speckle potential of Sec.~\ref{part:correl-2D} with $\anifact=4$.
Bottom row: Angular dependence of 
the eigenfunctions $\phi_{E,\uveck}^n$ for $n=1$ (dashed black line), $2$ (solid red line) and $3$ (dotted blue line).
We use the parametrization $\uveck\equiv(\cos\theta,\sin\theta)$.
The different columns refer to different energies (indicated on top of the figure).}}
\label{Figure:spectr2DGauss}
\end{figure}
The eigenvalues and some eigenfunctions obtained numerically are shown in Fig.~\ref{Figure:spectr2DGauss} for various values of $E/E_{\sigmaOrth}$.
As discussed above, we find $\lambda^{n=1}_E=1$.
For $E \ll E_{\sigmaOrth}$, only the first term in the right-hand side of Eq.~(\ref{DBE}) contributes to the diffusion tensor since all $\lambda_E^{n>1}$ are vanishingly small.
When the energy increases, the values of the coefficients $\lambda_E^{n>1}$ increase.
It corresponds to an increase of the weight of the terms associated to the orbitals with $n>1$ in Eq.~(\ref{DBE}), and \emph{a priori} all the orbitals with $n>1$ might have an increasing contribution.
However, the symmetry properties of the functions $\phi_{E,\uveck}^n$ cancel the contributions of most of them, and only the orbitals with $n=2$ and $3$ do contribute (see below).

In the low energy limit, one can develop Eq.~(\ref{2D-corrTF}) in powers of $|\veck|$.
Up to order $\Oo(E^2/\anifact^4 E_{\sigmaOrth}^2)$, the first three eigenfunctions are given by: 
\be
\phi^1_{E,\uveck}
= 1 - \frac{E}{2\anifact^2E_{\sigmaOrth}} 
\left[ 1+ (\anifact^2-1) {\hat{k}_x}^2 \right]+\Oo \left(\frac{E^2}{\anifact^4 E_{\sigmaOrth}^2}\right),
\ee
with eigenvalue $\lambda_E^1=1$;
\be
\phi^2_{E,\uveck} = \hat{k}_x \left[ \sqrt{2} + B_2\frac{E}{\anifact^2 E_{\sigmaOrth}} \right]+\Oo \left(\frac{E^2}{\anifact^4 E_{\sigmaOrth}^2}\right)
\label{eq:2D-eigenf2}
\ee
with eigenvalue $\lambda_E^2=E/2 E_{\sigmaOrth}$; and 
\be
\phi^3_{E,\uveck} = \hat{k}_y \left[ \sqrt{2} +B_3 \frac{E}{\anifact^2 E_{\sigmaOrth}} \right]+\Oo\left(\frac{E^2}{\anifact^4 E_{\sigmaOrth}^2}\right)
\label{eq:2D-eigenf3}
\ee
with eigenvalue $\lambda_E^3=E/2 \anifact^2 E_{\sigmaOrth}$,
where $B_2$ and $B_3$ are constant values that do not intervene in the following.
In this limit the numerical results agree very well with the analytical findings (which for clarity are not shown on Fig.~\ref{Figure:spectr2DGauss}).
In the very low energy limit, the disorder power spectrum becomes isotropic and constant, $\TFCor(\vect{k})\simeq \Vr^2 \pi \sigmaOrth^2/\anifact$, [see Sec.~\ref{part:correl-2D} and Eq.~(\ref{2D-corrTF})].
The orbitals $\phi_{E,\uveck}^n$ are thus proportional to the cylindrical harmonics,
which are exact solutions of Eq.~(\ref{eigeq-os}) in the isotropic case (see appendix~\ref{part:limit-iso}, and use the parametrization $\hat{k}_x=\cos\theta$ and $\hat{k}_y=\sin\theta$).
In contrast to the isotropic case where the values of $\lambda_E^n$ are degenerated in a given $l$-level, here we find that the degeneracy inside a $l$ level is lifted for any anisotropy $\anifact \neq 1$ [see the values of $\lambda_E^{2,3}$ below Eqs.~(\ref{eq:2D-eigenf2}) and (\ref{eq:2D-eigenf3})].
When the energy further increases, the anisotropy plays a more important role and the harmonics are more and more distorted (see Fig.~\ref{Figure:spectr2DGauss}).
However their topology remains the same, and in particular the number of nodal points and their positions are unchanged.
In the following, we thus refer to $Z_l^{\pm 1}$-like orbitals.

\begin{figure}[!t] 
\begin{center}
\includegraphics[width=0.7\textwidth]{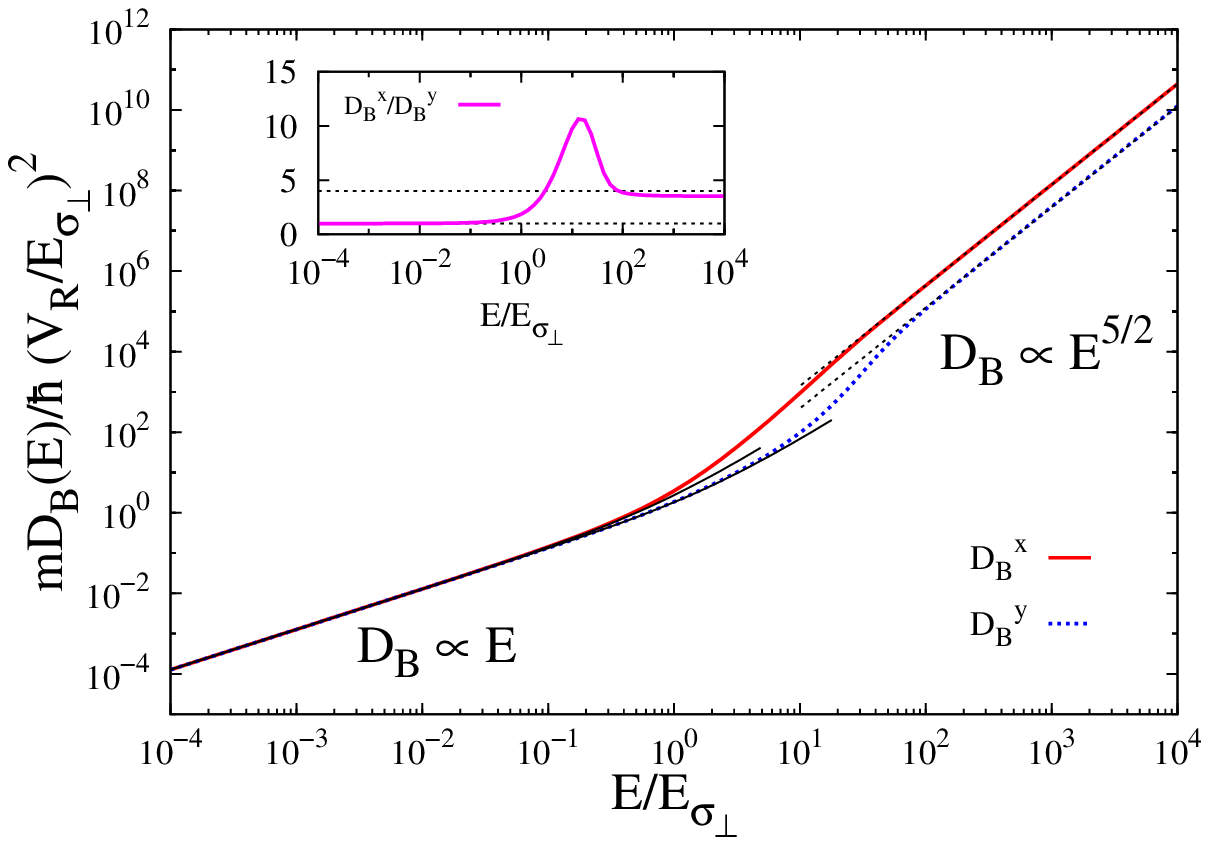}
\end{center} 
\caption{\small{(Color online) Components of the diffusion tensor: $\DB^x$ (soild red line) and $\DB^y$ (dotted blue line) for the 2D speckle potential of Sec.~\ref{part:correl-2D} with $\anifact=4$.
Solid black lines are limit values at small $E/E_{\sigmaOrth}$ [Eqs.~(\ref{DxLowK}) and (\ref{DyLowK})], with the 
isotropic white-noise limit $\DB^x(E) = \DB^y(E) \sim \hbar \anifact E E_{\sigmaOrth}/m \pi \Vr^2$.
For large $E/E_{\sigmaOrth}$ we find $\DB(E)\sim E^{5/2}$ (see text); a fit of the numerical data gives the prefactors $\DB^x=4.43 \, E^{5/2}/\Vr^2 E_{\sigmaOrth}^{1/2}$ and $\DB^y=1.24\,E^{5/2}/\Vr^2 E_{\sigmaOrth}^{1/2}$ (see dotted black lines).
The inset shows the transport anisotropy factor $\anifact_{\textrm{B}}=\DB^{x}/\DB^{y}$.}}
 \label{Figure:Dx_Dy_gauss}
\end{figure}
Incorporating the values of $\lambda_E^n$, $\phi^n_{E,\uveck}$ and $\tau_{E,\uveck}$ in Eq.~(\ref{DBE}), we can determine the Boltzmann diffusion tensor.
Figure~\ref{Figure:Dx_Dy_gauss} shows the resulting eigencomponents of the diffusion tensor.
In the low energy limit ($E \ll E_{\sigmaOrth}$), using Eqs.~(\ref{ls2D_smallk}), (\ref{eq:2D-eigenf2}) and (\ref{eq:2D-eigenf3}), we find that the first term in the right-hand side of Eq.~(\ref{DBE}) gives the leading contribution to $\DiffTensB(E)$ (of order $E/E_{\sigmaOrth}$).
This contribution is isotropic owing to the isotropy of $\tau_{E,\uveck}$ at low energy and of the underlying medium.
At very low energy, in the white-noise limit, we recover an isotropic diffusion tensor
$\DB^x(E)= \DB^y(E) \sim \hbar \anifact E E_{\sigmaOrth}/m \pi \Vr^2$.
The scaling $\DB^u(E)\propto E$ is universal for 2D disorder in the white-noise limit (when it exists).
The $Z_1^{+1}$-like orbital $\phi^2_{E,\uveck}$ contributes to the next order of $\DB^x$ and the $Z_1^{-1}$-like orbital $\phi^3_{E,\uveck}$ to $\DB^y$.
Up to order $\Oo(E^3/\anifact^6 E_{\sigmaOrth}^3)$, we obtain
\be
\DB^x(E) = \frac{\hbar}{m} \frac{E_{\sigmaOrth}^2}{\Vr^2}
\left[ \frac{\anifact E}{\pi E_{\sigmaOrth}} + \frac{E^2}{\pi E_{\sigmaOrth}^2} \, \frac{9\anifact^2+3}{8\anifact}+\Oo \left(\frac{E^3}{\anifact^6 E_{\sigmaOrth}^3}\right) \right]
\label{DxLowK},
\ee
and
\be
\DB^y(E) = \frac{\hbar}{m} \frac{E_{\sigmaOrth}^2}{\Vr^2}
\left[ \frac{\anifact E}{\pi E_{\sigmaOrth}} + \frac{E^2}{\pi E_{\sigmaOrth}^2} \, \frac{3\anifact^2+9}{8\anifact}+\Oo \left(\frac{E^3}{\anifact^6 E_{\sigmaOrth}^3} \right) \right]
\label{DyLowK},
\ee
which are displayed on Fig.~\ref{Figure:Dx_Dy_gauss} (left-hand-side solid lines).
When the energy increases, the anisotropy first comes from the anisotropic contribution of the scattering time $\tau_{E,\uveck}$, and from the lift of the degeneracy between $\lambda_E^2$ and $\lambda_E^3$.
When the energy further increases, the harmonics are distorted,
-- but their symmetries (\ie\ periodicity and parity) are preserved (see Fig.~\ref{Figure:spectr2DGauss}).
Hence, for the same reasons as in the isotropic case (see appendix~\ref{part:limit-iso}) only the $Z_1^{\pm 1}$-like orbitals couple to ${\boldsymbol \upsilon}$ in Eq.~(\ref{DBE}) and contribute to $\DB$ while the others don't.
The associated $\lambda_E^n$ increase (see Fig.~\ref{Figure:spectr2DGauss}),
the weight of the second term in Eq.~(\ref{DBE}) increases, and the components of the diffusion tensor show a very different behavior in the large-$E$ limit.
For $\kE \sigmaOrth \gg \anifact$, we found $\tau_{E,\uveck} \propto \kE$ (see Sec.~\ref{part:2Dex-scat}).
In addition, we find numerically a weak topological change of the orbitals with energy for $E/E_{\sigmaOrth} \gtrsim 10^2$.
Therefore the evaluation of $\DiffTensB$ with $E$ is mainly determined by the normalization condition [see formula below Eq.~(\ref{DBE})], which yields $\phi^n_{E,\uveck} \propto 1/\sqrt{\kE}$.
Then, assuming the scaling $1-\lambda_E^n \propto 1/E$, also verified numerically, we obtain $\DB^u(E) \propto E^{5/2}$, which matches the numerical results (see dotted black lines in Fig.~\ref{Figure:Dx_Dy_gauss}).
This scaling is similar to that found for isotropic disorder \cite{kuhn2007}.
As shown in Fig.~\ref{Figure:Dx_Dy_gauss}, the change of slope between the low- and high-energy regimes is different in the two directions.
For this reason, the anisotropy factor of the diffusion tensor, $\anifact_{\textrm{B}}=\DB^{x}/\DB^{y}$ shows a nonmonotonous behaviour versus $E$, with a marked peak (see inset of Fig.~\ref{Figure:Dx_Dy_gauss}).

\begin{figure}[!t] 
\begin{center}
\includegraphics[width=0.7\textwidth]{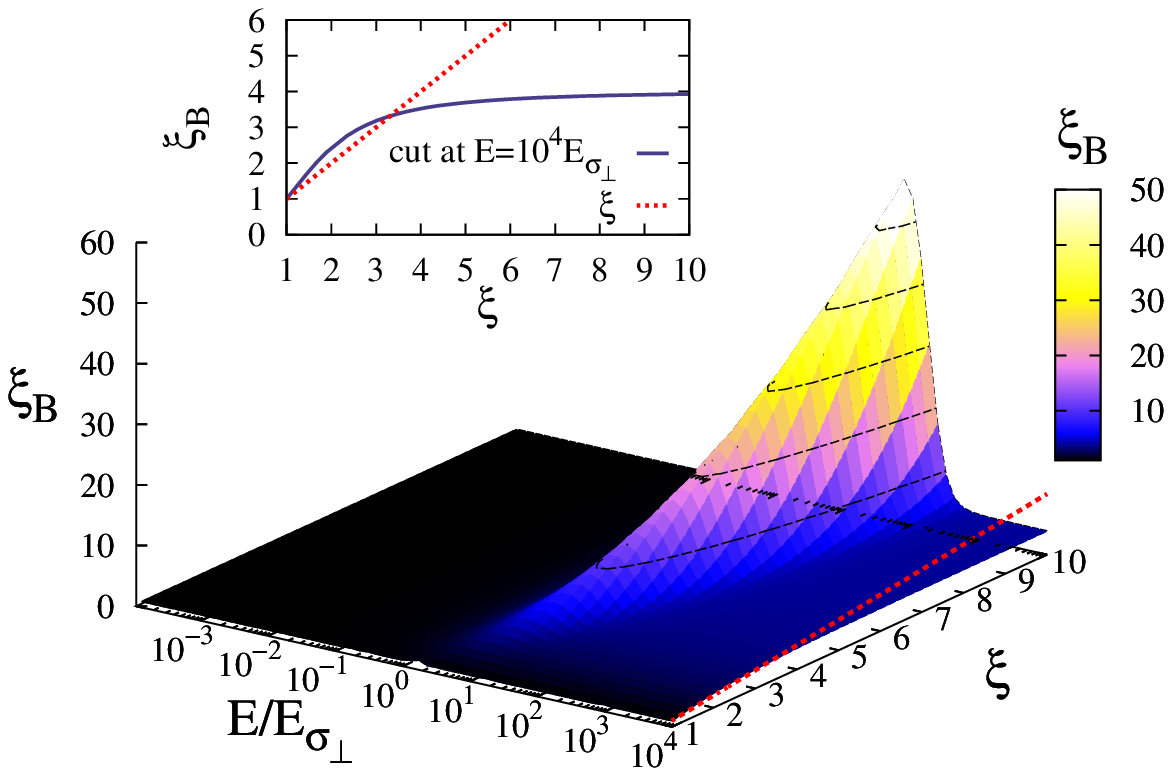}
\end{center} 
\caption{\small{(Color online) Boltzmann transport anisotropy factor $\anifact_{\textrm{B}}=\DB^{x}/\DB^{y}$ as a function of $E/E_{\sigmaOrth}$ and $\anifact$ for the 2D speckle potential of Sec.~\ref{part:correl-2D}.
The inset shows the high energy asymptotic value (cut at $E=10^4E_{\sigmaOrth}$).
The dotted red line in both the figure and the inset is $\anifact$.
}}
\label{fig:DyyDxx_gauss}
\end{figure}
The Boltzmann transport anisotropy factor $\anifact_{\textrm{B}}$ is shown in Fig.~\ref{fig:DyyDxx_gauss} for various configuration anisotropies $\anifact$.
As it is well-known, the scattering and transport mean free times are different quantities in correlated disorder, due to angle-dependent scattering~\cite{abrikosov1975,chapman1991,rammer1998}.
In particular, in the 2D speckle we consider, we do not find any inversion of the anisotropy of the diffusion, contrary to the scattering time, \ie\ the component $\DB^x(E)$ of the diffusion tensor is always larger than the component $\DB^y(E)$. 
For large values of $E/E_{\sigmaOrth}$, the Boltzmann transport anisotropy $\anifact_{\textrm{B}}$ reaches a constant value (see the inset of Fig.~\ref{Figure:Dx_Dy_gauss} for a cut at $\anifact=4$), which increases with the geometrical anisotropy $\anifact$ (see inset of Fig.~\ref{fig:DyyDxx_gauss}).
This asymptotic value is larger than $\anifact$ for small $\anifact$ and smaller for larger $\anifact$.
Therefore the anisotropy of the diffusion in the classical regime is not simply related to the configuration anisotropy.

The two distinct regimes found in the behaviour of $\DiffTensB$ and the non-trivial anisotropy effects make the Boltzmann diffusion regime in anisotropic 2D potentials very interesting for future experiments.
Those properties could be probed by imaging directly the atoms in the 2D speckle (as in Ref.~\cite{mrsv2010}) and controlling the width of the atomic energy distribution.

%% file: TexFiles/Localization.05.tex
subsection{Weak localization correction \label{part:weakloc}}

We calculate corrections to Boltzmann diffusion by taking into account quantum interference terms between the multiple-scattering paths.
Those interferences appear when the correlated scattering events do not occur in the same order in the propagation of the field and its conjuguate.
This is diagrammatically translated into crossing correlation lines as in the second term of Eq.~(\ref{Diag:Uvert}) for example.
In the weak scattering regime only the two-point correlations are retained in the scattering diagrams and the leading scale-dependent corrections to the classical conductivity are given by the maximally crossed diagrams~\cite{hikami1981,woelfe1984,bhatt1985,akkermans2006}: the cooperon [Eq.~(\ref{Diag:cooperon})] and the first two Hikami boxes [Eqs.~(\ref{Diag:hikami}) and (\ref{Diag:hikami2})].
\input{Diagrammes/diag-coop-hikami}
where the cooperon $X$ is the sum of maximally crossed diagrams
\input{Diagrammes/diag-X-vert}
and
\input{Diagrammes/diag-renorm-vertex-single}\\
is the renormalized vertex function (see appendix~\ref{ap:Vrenorm}).

Using time-reversal invariance~\cite{abrahams1979,vollhardt1980a,vollhardt1980b,rammer1998}, the cooperon $X$ can be expressed in terms of the diffuson $\Gamma$ [defined in Eq.~(\ref{Diag:gamma})]
\be
X_{\veck,\veck'}(\vecq,\omega,E)=\Gamma_{\frac{\veck-\veck'}{2}+\frac{\vecq}{2},\frac{\veck'-\veck}{2}+\frac{\vecq}{2}}(\veck+\veck',\omega,E).
\label{eq:coop}
\ee
The diffusion pole carried by $\Gamma$ in the limit $(\omega,\vecq) \rightarrow 0$ leads to a divergence of $X$ when $\omega, \veck+\veck' \rightarrow 0$.
In appendix~\ref{ap:CorrCond} we translate those diagrams into equations, and show that 
\be
\Delta \sigtens(\omega,E) = - \frac{\sigtensB(E)}{\pi N_0(E)}
\int \frac{\ud \vecQ}{(2 \pi)^d} \, \frac{1}{-i\hbar \omega + \hbar \vecQ \cdot \DiffTensB(E) \cdot \vecQ}.
\label{eq:corr-cond}
\ee
Using Einstein's relation~(\ref{eq:releinstein}) we then obtain the dynamic diffusion tensor $\DiffTens(\omega,E)=\DiffTensB(E)+\DiffTensCor(\omega,E)$, with~\cite{woelfe1984}
\be
\frac{\DiffTensCor(\omega,E)}{\DiffTensB(E)}=
- \frac{1}{\pi N_0(E)}
\int \frac{\ud \vecQ}{(2 \pi)^d} \, \frac{1}{-i\hbar \omega + \hbar \vecQ \cdot \DiffTensB(E) \cdot \vecQ}.
\label{eq:weak-loc}
\ee
Note that the quantum corrections $\DiffTensCor (\omega,E)$ do not explicitly depend on the disorder [\ie\ on $\TFCor(\veck)$], but only on the Boltzmann diffusion tensor $\DiffTensB(E)$~\cite{woelfe1984}.
In other words, in this approach, Boltzmann incoherent diffusion sets a diffusing medium, which contains all necessary information to compute coherent terms\footnote{This property is a consequence of the on-shell approximation.}.
In particular, it follows from Eq.~(\ref{eq:weak-loc}) that the weak localization quantum correction tensor $\DiffTensCor (\omega,E)$ has the same eigenaxes and anisotropies as the Boltzmann diffusion tensor $\DiffTensB(E)$.
Thus the anisotropy can be removed by rescaling distances along the transport eigenaxes $u$ by $\sqrt{\DB^{u}/\geomav{\DB}}$ (\ie\ momenta are rescaled by $\sqrt{\geomav{\DB}/\DB^{u}}$) with $\geomav{\DB} \equiv \det\{\DiffTensB\}^{1/d}$ the geometric average of the Boltzmann diffusion constants.
Since $\DiffTensCor$ is always negative in the limit $\omega \rightarrow 0^+$, the weak localization correction features slower diffusion than the one obtained from incoherent diffusion.
Equivalently, as long as the correction~(\ref{eq:weak-loc}) is small, one can write
\be
\frac{\DiffTensB(E)}{\DiffTens(\omega,E)}=1+ \frac{1}{\pi N_0(E)}
\int \frac{\ud \vecQ}{(2 \pi)^d} \, \frac{1}{-i\hbar \omega + \hbar \vecQ \cdot \DiffTensB(E) \cdot \vecQ},
\label{eq:weak-loc2}
\ee
which is the lowest-order term of a perturbative expansion of $1/\DiffTens(\omega,E)$.

\subsection{Strong localization \label{part:strongloc}}

The quantum interference correction~(\ref{eq:weak-loc}) has been derived perturbatively and is therefore valid as long as the correction itself is small.
In order to extend this approach and eventually describe the localization regime where $\DiffTens$ vanishes, Vollhardt and W\"olfle~\cite{vollhardt1980a,vollhardt1980b} proposed to self-consistently replace $\DiffTensB(E)$ by the dynamic diffusion tensor $\DiffTens(\omega,E)$ in the right-hand side of Eq.~(\ref{eq:weak-loc2}).
For isotropic scattering this procedure amounts to resumming
more divergent diagrams than the cooperon (which contain a square of a diffusion pole), thus contributing to localization~\cite{vollhardt1980b,vollhardt1992}.
Generalizing this standard approach to anisotropic disorder yields
\be
\frac{\DiffTensB}{\DiffTens(\omega)}=1 + \frac{1}{\pi N_0(E)}
\int \frac{\ud \vecQ}{(2 \pi)^d} \, \frac{1}{-i\hbar \omega + \hbar \vecQ \cdot \DiffTens(\omega) \cdot \vecQ}.
\label{self-const-cor}
\ee
In dimension $d\geq2$ the integral in the right-hand side of Eq.~(\ref{self-const-cor}) features ultraviolet divergence.
Since the diffusive dynamics is relevant only on length scales larger than the Boltzmann mean free path $l_\textrm{B}^u (E) \equiv d \sqrt{m/2E}\,D_\textrm{B}^u(E)$ along each transport eigenaxis, 
we regularize this divergence
by setting an upper ellipsoidal cut-off of radii $1/l_\textrm{B}^u$ in the integral domain\footnote{\new{Although somewhat arbitrary the factor unity between the cut-off radius and $1/l_\textrm{B}^u(E)$ is justified by the agreement we find with another approach in the isotropic case, provided that the real part of the self energy is included, see Sec.~\ref{part:comp-iso}.}}.
It corresponds to an isotropic cut-off in the space rescaled according to the anisotropy factors of $\DiffTensB$ as described above.

%% file: Diagrammes/diag-coop-hikami.tex
\begin{fmffile}{diag-coop-hikami-fmf}
\be
\Delta \sigtens_{(X)}=
\parbox{0.45\linewidth}{
	     \begin{fmfgraph*}(45,20)
		\fmfleft{i1}
		\fmfright{o1}
		\fmfpoly{phantom,label=$X$,l.d=1,tension=0.}{v1,x1,v2,v3,x3,v4,x4,v5,v6,x6}
		\fmf{wiggly,tension=3.,label=$\vecJ_{\veck}/\hbar$,l.s=right}{i1,v4}
		\fmf{plain,left=0.2,tension=0.,width=2}{v4,x3}
		\fmf{plain_arrow,left=0.2,tension=0.,width=2}{x3,v3}
		\fmf{plain,left=0.2,tension=0.}{v3,v2}
		\fmf{plain_arrow,left=0.2,tension=0.,width=2}{v2,x1}
		\fmf{plain,left=0.2,tension=0.,width=2}{x1,v1}
		\fmf{plain,left=0.2,tension=0.,width=2}{v1,x6}
		\fmf{plain_arrow,left=0.2,tension=0.,width=2}{x6,v6}
		\fmf{plain,left=0.2,tension=0.}{v6,v5}
		\fmf{plain_arrow,left=0.2,tension=0.,width=2}{v5,x4}
		\fmf{plain,left=0.2,tension=0.,width=2}{x4,v4}
		\fmf{wiggly,tension=3.,label=$\vecJ_{\veck'}/\hbar$,l.s=right}{v1,o1}
		\fmf{phantom}{v3,v5}
		\fmf{phantom}{v2,v6}
		\fmf{plain,tension=0.}{v3,v6}
		\fmf{plain,tension=0.}{v2,v5}
		\fmf{dots,tension=0.}{x1,x6}
		\fmf{dots,tension=0.}{x3,x4}
	    \end{fmfgraph*}

}
\label{Diag:cooperon}
\ee

\be
\Delta \sigtens_{(H_1)}=
\parbox{0.45\linewidth}{
	     \begin{fmfgraph*}(45,20)
		\fmfleft{i1}
		\fmfright{o1}
		\fmfpoly{phantom,label=$X$,l.d=1,tension=0.}{v1,x1,v2,v3,x3,v4,x4,v5,v6,x6}
		\fmf{wiggly,tension=3.,label=$\vecJ_{\veck}/\hbar$,l.s=right}{i1,v4}
		\fmf{plain,left=0.2,tension=0.,width=2}{v4,x3}
		\fmf{plain_arrow,left=0.2,tension=0.,width=2}{x3,v3}
		\fmf{plain,left=0.2,tension=0.}{v3,v2}
		\fmf{plain_arrow,left=0.2,tension=0.,width=2}{v2,x1}
		\fmf{plain,left=0.2,tension=0.,width=2}{x1,v1}
		\fmf{plain,left=0.2,tension=0.,width=2}{v1,x6}
		\fmf{plain_arrow,left=0.2,tension=0.,width=2}{x6,v6}
		\fmf{plain,left=0.2,tension=0.}{v6,v5}
		\fmf{plain_arrow,left=0.2,tension=0.,width=2}{v5,x4}
		\fmf{plain,left=0.2,tension=0.,width=2}{x4,v4}
		\fmf{wiggly,tension=3.,label=$\vecJ_{\veck'}/\hbar$,l.s=right}{v1,o1}
		\fmf{phantom}{v3,v5}
		\fmf{phantom}{v2,v6}
		\fmf{plain,tension=0.}{v3,v6}
		\fmf{plain,tension=0.}{v2,v5}
		\fmf{dots,tension=0.}{x1,x6}
		\fmf{dots,tension=0.}{x3,x4}
		\fmf{dashes,right=0.8,tension=0.}{v2,v3}
	    \end{fmfgraph*}

}
\label{Diag:hikami}
\ee

\be
\Delta \sigtens_{(H_2)}=
\parbox{0.45\linewidth}{
	     \begin{fmfgraph*}(45,20)
		\fmfleft{i1}
		\fmfright{o1}
		\fmfpoly{phantom,label=$X$,l.d=1,tension=0.}{v1,x1,v2,v3,x3,v4,x4,v5,v6,x6}
		\fmf{wiggly,tension=3.,label=$\vecJ_{\veck}/\hbar$,l.s=right}{i1,v4}
		\fmf{plain,left=0.2,tension=0.,width=2}{v4,x3}
		\fmf{plain_arrow,left=0.2,tension=0.,width=2}{x3,v3}
		\fmf{plain,left=0.2,tension=0.}{v3,v2}
		\fmf{plain_arrow,left=0.2,tension=0.,width=2}{v2,x1}
		\fmf{plain,left=0.2,tension=0.,width=2}{x1,v1}
		\fmf{plain,left=0.2,tension=0.,width=2}{v1,x6}
		\fmf{plain_arrow,left=0.2,tension=0.,width=2}{x6,v6}
		\fmf{plain,left=0.2,tension=0.}{v6,v5}
		\fmf{plain_arrow,left=0.2,tension=0.,width=2}{v5,x4}
		\fmf{plain,left=0.2,tension=0.,width=2}{x4,v4}
		\fmf{wiggly,tension=3.,label=$\vecJ_{\veck'}/\hbar$,l.s=right}{v1,o1}
		\fmf{phantom}{v3,v5}
		\fmf{phantom}{v2,v6}
		\fmf{plain,tension=0.}{v3,v6}
		\fmf{plain,tension=0.}{v2,v5}
		\fmf{dots,tension=0.}{x1,x6}
		\fmf{dots,tension=0.}{x3,x4}
		\fmf{dashes,right=0.8,tension=0.}{v5,v6}
	    \end{fmfgraph*}

}
\label{Diag:hikami2}
\ee
\end{fmffile}\\

%% file: Diagrammes/diag-X-vert.tex
\begin{fmffile}{diag-X-vert-fmf}
\be
\parbox{0.15\linewidth}{
	    \begin{fmfgraph*}(15,15)
		\fmfleft{v1,v4}
		\fmfright{v2,v3}
		\fmf{phantom}{v1,v2}
		\fmf{phantom}{v3,v4}
		\fmffreeze
		\fmf{plain}{v2,v1}
		\fmf{plain}{v4,v3}
		\fmf{plain,tension=0.}{v1,v3}
		\fmf{plain,tension=0.}{v2,v4}
		\fmffreeze
		\fmf{phantom,tension=1.}{v1,M,v3}
		\fmf{phantom,tension=1.}{v2,M,v4}
		\fmfv{label=$X$,label.d=3}{M}
	    \end{fmfgraph*}
}
 = \,
\parbox{0.125\linewidth}{
	    \begin{fmfgraph*}(12.5,15)
		\fmfleft{i1,i2}
		\fmfright{o1,o2}
		\fmf{plain_arrow,width=2}{i2,o2}
		\fmf{plain_arrow,width=2}{o1,i1}
		\fmf{dashes,tension=0}{i1,o2}
		\fmf{dashes,tension=0}{o1,i2}
		\fmfdot{o1,o2,i1,i2}
	    \end{fmfgraph*}
	    }
	    \, + \,
\parbox{0.2\linewidth}{
	    \begin{fmfgraph*}(20,15)
		\fmfleft{i1,i2}
		\fmfright{o1,o2}
		\fmf{plain_arrow,width=2}{i2,v2,o2}
		\fmf{plain_arrow,width=2}{o1,v1,i1}
		\fmf{dashes,tension=0}{i1,o2}
		\fmf{dashes,tension=0}{o1,i2}
		\fmf{dashes,tension=0}{v1,v2}
		\fmfdot{o1,o2,i1,i2,v1,v2}
	    \end{fmfgraph*}
	    }
	    \, + \,
\parbox{0.24\linewidth}{
	    \begin{fmfgraph*}(24,15)
		\fmfleft{i1,i2}
		\fmfright{o1,o2}
		\fmf{plain_arrow,width=2}{i2,v2,vv2,o2}
		\fmf{plain_arrow,width=2}{o1,vv1,v1,i1}
		\fmf{dashes,tension=0}{v1,vv2}
		\fmf{dashes,tension=0}{vv1,v2}
		\fmf{dashes,tension=0}{i1,o2}
		\fmf{dashes,tension=0}{o1,i2}
		\fmfdot{o1,i1,o2,i2,v1,vv1,v2,vv2}
	    \end{fmfgraph*}
	    }
\, + \, ...
\label{Diag:Xvert}
\ee
\end{fmffile}\\

%% file: Diagrammes/diag-renorm-vertex-single.tex
\begin{fmffile}{diag-renorm-vertex-single-fmf}
\be
\parbox{0.2\linewidth}{
	     \begin{fmfgraph*}(20,10)
		\fmfleft{i1}
		\fmfright{o1}
		\fmfpoly{phantom,tension=0.}{v1,x1,v2,x3,v4,x4,v5,x6}
		\fmf{wiggly,tension=3.,label=$\vecJ_{\veck}/\hbar$,l.s=right}{i1,v4}
		\fmf{plain,left=0.2,tension=0.,width=2}{v4,x3}
		\fmf{plain_arrow,left=0.2,tension=0.,width=2}{x3,v2}
		\fmf{plain_arrow,left=0.2,tension=0.,width=2}{v5,x4}
		\fmf{plain,left=0.2,tension=0.,width=2}{x4,v4}
		\fmf{phantom,tension=100.}{v1,o1}
		\fmf{phantom}{v2,v5}
		\fmf{phantom}{v2,v5}
		\fmf{dots,tension=0.}{x3,x4}
		\fmfkeep{diag-Vrenorm}
	    \end{fmfgraph*}
	    }
\ee
\end{fmffile}

%% file: TexFiles/LocalizationExample2DGauss.05.tex
\subsection{Anisotropic Gaussian speckle (2D) \label{part:locex2D}}

We now solve the self-consistent equation~(\ref{self-const-cor}) for the inverse dynamic diffusion tensor in the 2D case.
In the long time limit $\omega \rightarrow 0^+$, the unique solution of Eq.~(\ref{self-const-cor}) is of the form $\DiffTens(\omega,E) \sim 0^{+} -i\omega \LocTens^2(E)$, where $\LocTens(E)$ is a real positive definite tensor.
As described in Sec.~\ref{part:BSE}, it leads to the exponentially decreasing propagation kernel~(\ref{eq:prop-kernel-loc2d}).
Solving Eq.~(\ref{self-const-cor}) then yields the anisotropic localization tensor, 
\be
\LocTens(E)= \geomav{l_\textrm{B}}(E) \sqrt{\frac{\DiffTensB(E)}{\geomav{D_\textrm{B}}(E)} } \left( e^{\pi \kE \geomav{l_\textrm{B}}(E)}-1 \right)^{1/2}
\label{eq:locTens-2D}
\ee
where $\geomav{l_\textrm{B}}(E) \equiv d \sqrt{m/2E}\,\geomav{\DB}(E)$.
The eigenaxes of the localization tensor are thus the same as those of the Boltzmann diffusion tensor and its anisotropy factor is the square root of that of $\DiffTensB(E)$, \ie\ $\anifact_{\textrm{loc}}\equiv\Lloc^x/\Lloc^y=\sqrt{\anifact_{\textrm{B}}}$.

\begin{figure}[!t] 
\begin{center}
\includegraphics[width=0.7\textwidth]{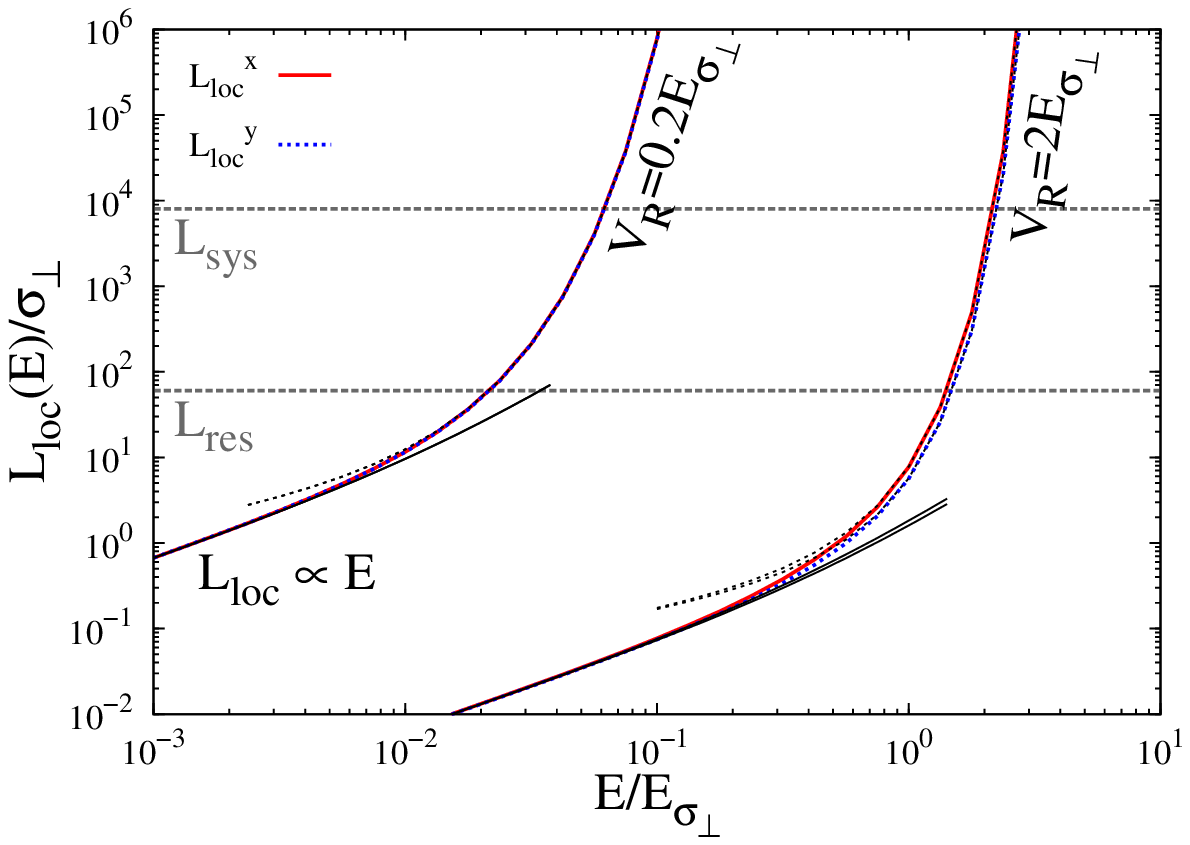}
\end{center} 
\caption{\small{(Color online) Components of the localization tensor $\Lloc^x$ (solid red line) and $\Lloc^y$ (dotted blue line) for the 2D speckle potential of Sec.~\ref{part:correl-2D}, with $\anifact=4$ and $\Vr=0.2E_{\sigmaOrth}$ and $2E_{\sigmaOrth}$.
The solid black lines are the limiting behaviour for small values of $E/E_{\sigmaOrth}$ [Eq.~(\ref{Lloc-lowE})] and the dotted ones for high values of $E/E_{\sigmaOrth}$ [Eq.~(\ref{Lloc-highE})].
The dashed grey lines indicate typical values of the imaging resolution ($L_{\textrm{res}}$) and the system size ($L_{\textrm{sys}}$) in ultracold-atom experiments, see text at the end of Sec.~\ref{part:locex2D}.}}
\label{Figure:loc2D}
\end{figure}
We now apply the self-consistent theory to our running example: the 2D anisotropic speckle potential with correlation function~(\ref{2D-corrTF}).
Including the results for the Boltzmann diffusion tensor $\DiffTensB(E)$ obtained in Sec.~\ref{part:diff-ex} into Eq.~(\ref{eq:locTens-2D}), we find the localization tensor $\LocTens(E)$.
Figure~\ref{Figure:loc2D} presents the eigencomponents of $\LocTens$ in its eigenbasis $\{ \uvecu_x,\uvecu_y \}$ as a function of energy, for a configuration anisotropy of $\anifact=4$ and two different amplitudes of the disorder, $\Vr/E_{\sigmaOrth}=0.2$ and $2$.
At low energy ($E \ll E_{\sigmaOrth}, \Vr, \Vr^2/E_{\sigmaOrth}$), using Eqs.~(\ref{DxLowK}) and (\ref{DyLowK}), we find
\beq
 \Lloc^{x,y}(E)&=&\sigmaOrth\frac{E_{\sigmaOrth}^3}{\Vr^3} \frac{\anifact^{3/2}}{\pi} \frac{2E}{E_{\sigmaOrth}}
\bigg[1+ \frac{\anifact E E_{\sigmaOrth}}{2\Vr^2}  \nonumber \\
&+& \frac{E}{E_{\sigmaOrth}} \frac{(18 \pm 3)\anifact^2+(18 \mp 3)}{16 \anifact^2} \nonumber \\
&+& \Oo \left(\frac{E^2}{\anifact^4 E_{\sigmaOrth}^2}, \frac{E^2}{\anifact^2 \Vr^2}, \frac{E^2 E_{\sigmaOrth}^2}{\Vr^4}\right) \bigg],
\label{Lloc-lowE}
\eeq
where the upper sign holds for direction $x$, and the lower sign for direction $y$.
Equation~(\ref{Lloc-lowE}) corresponds to the solid black lines in Fig.~\ref{Figure:loc2D}.
As $\DiffTensB$ is almost isotropic for $E/E_{\sigmaOrth} \lesssim 1$ (see Fig.~\ref{Figure:Dx_Dy_gauss}), $\LocTens$ is also almost isotropic in the whole range presented in Fig.~\ref{Figure:loc2D}.
Equation~(\ref{Lloc-lowE}) describes an isotropic localization tensor with an anisotropic correction which is significant only if $\Vr/E_{\sigmaOrth} \gtrsim \anifact^{3/2}/\sqrt{\anifact^2-1}$ ($\simeq 2$ for $\anifact=4$).
At higher energy, when $\kE \geomav{l_\textrm{B}}(E)= 2 m \geomav{\DB}(E)/\hbar \gtrsim 1$, we expect
\be
\Lloc^{u}(E) \simeq \frac{2m\sqrt{\geomav{\DB}(E) \DB^u(E)}}{\kE \hbar} \, e^{\pi m \geomav{\DB}(E)/\hbar},
\label{Lloc-highE}
\ee
which is plotted as dotted black lines in Fig.~\ref{Figure:Dx_Dy_gauss}.
According to Eqs.~(\ref{DxLowK}) and (\ref{DyLowK}) (retaining only the lowest-energy term), this regime appears for $E/E_{\sigmaOrth} \gtrsim (\pi/2\anifact) (\Vr/E_{\sigmaOrth})^2$.
When $\anifact=4$ (as in Fig.~\ref{Figure:loc2D}), it gives $E/E_{\sigmaOrth} \gtrsim 0.015$ for $\Vr/E_{\sigmaOrth}=0.2$ and $E/E_{\sigmaOrth} \gtrsim 1.5$ for $\Vr/E_{\sigmaOrth}=2$.
As predicted by the scaling theory of Anderson Localization~\cite{abrahams1979} and explicitely seen in Eq.~(\ref{Lloc-highE}), the 2D localization length increases exponentially at large energy (hence the limited energy range in Fig.~\ref{Figure:loc2D}).
Therefore measuring it experimentally with ultracold atoms~\cite{pezze2011b,plisson2011,allard2012} is very challenging and can be done in a rather narrow energy window, in which $\Lloc$ is larger than the resolution of the imaging system ($L_{\textrm{res}}$) but smaller than the size of the sample ($L_{\textrm{sys}}$).
This is illustrated for $\sigmaOrth=0.25\, \mu \textrm{m}$ on Fig.~\ref{Figure:loc2D} by the grey dashed lines corresponding to $L_{\textrm{res}} \simeq 15\, \mu \textrm{m}$ and $L_{\textrm{sys}} \simeq 2\, \textrm{mm}$, which are typical values extracted from Refs.~\cite{mrsv2010,jendrzejewski2011}.

\new{One can finally note that 2D speckle potentials bear a classical percolation threshold at energy $E_\textrm{p} \simeq -\Vr/2$~\cite{pezze2011b}. In the classical regime ($1/k < \sigmaOrthx, \sigmaOrthy$), genuine Anderson localization has to be distinguished from classical trapping, which happens for $E < E_\textrm{p}$. However, classical percolation is not relevant for the parameters used in Fig.~\ref{Figure:loc2D}. Indeed, for $|\Vr| \leq 2E_{\sigmaOrth}$ (as in the figure) and for $E<E_\textrm{p}$, we have $E \lesssim |\Vr|/2 \leq E_{\sigma_{\perp}}$, so that $k\sigmaOrthy \leq k\sigmaOrthx = k\sigmaOrth \lesssim 1$, which is not in the classical regime. }

%% file: TexFiles/ScatteringExample3D.04.tex
\subsection{Single-scattering \label{part:scatt-3D}}

\begin{figure}[!t] 
\begin{center}
\includegraphics[width=0.7\textwidth]{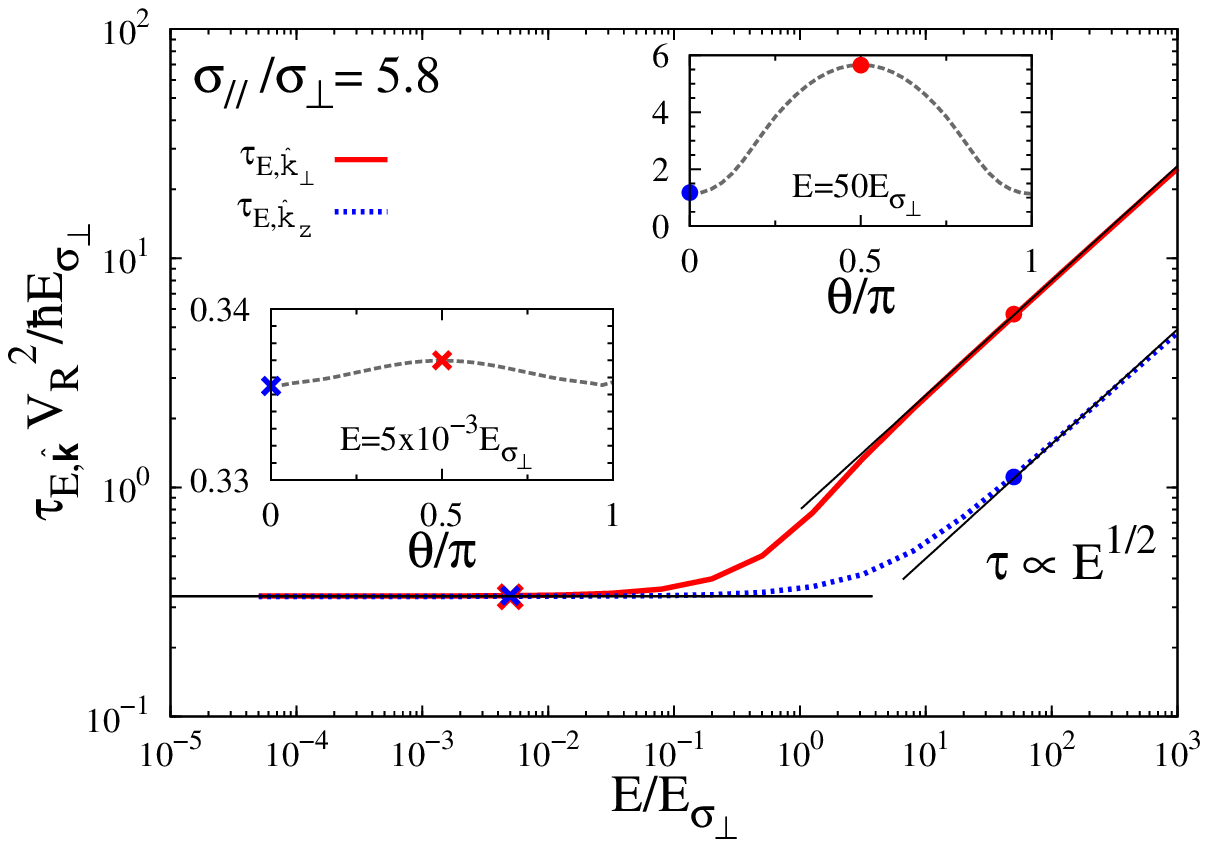}
\end{center} 
\caption{\small{(Color online) Scattering mean free time $\tau_{E,\uveck}$ in the 3D case [Eq.~(\ref{eq:tau-single})] with $\sigmaPara/\sigmaOrth = 5.8$ with $|\veck|=\kE$ , in the $(\uveck_x,\uveck_y)$ plane (solid red line) and along the $\uveck_z$ direction (dotted blue line).
The black lines are the low-energy [$\kE \sigmaOrth \ll 1$, see Eq.~(\ref{eq:tau-single-lowE})] and the high energy [$\kE \sigmaOrth \gg 1$, see Eq.~(\ref{eq:tau-single-highE})] limits.
Note that in both limits $\tau_{E,\uveck}$ is anisotropic, although for $\kE \sigmaOrth \ll 1$, the anisotropy is very small, $\anifact_s\simeq 1.002$.
The insets show the angular dependence of $\tau_{E,\uveck}$ at different energies [with $\theta=(\uveck;\uveck_z)$].
The points on the lines are color- and shape-coded to match those in the insets.
}}
\label{Figure:ls_3Dsingle}
\end{figure}

Inserting Eqs.~(\ref{TFcorsingle}) and (\ref{TFcorsingle2}) into Eq.~(\ref{eq:tau_s}), we find the scattering mean free time
\be
\smft(E,\veck) = \frac{\hbar E_{\sigmaOrth}}{V_R^2}
\frac{(2\pi)^2/\kE \sigmaOrth}{\int \ud\Omega_{\uveck'} \, \TFcor_{\textrm{1sp}}(\kE \uveck'-\veck)/\sigmaOrth^3},
\label{eq:tau-single}
\ee
which is shown in Fig.~\ref{Figure:ls_3Dsingle} for $|\veck|=\kE$ [we use the definition $\tau_{E,\uveck}\equiv\smft(E,\kE\uveck)$].

Since $\TFCor(\veck)$ is isotropic in the $(k_x,k_y)$ plane, $\tau_{E,\uveck}$ only depends on the polar angle $\theta$ between $\veck$ and $\uveck_z$ and not on the azimutal angle $\phi$.
We find that the scattering time is an increasing function of energy.
It is also shorter for particles travelling along the $z$ direction ($\tau_{E,\uveck_z} < \tau_{E,\uveck_{\perp}}$) for all values of $E$.
As for the 2D case analyzed in Sec.~\ref{part:2Dex-scat}, this is due to the wider extension of $\TFCor(\veck)$ in the plane $(k_x,k_y)$, which offers more scattering channels to particles travelling along $z$.
In contrast to the 2D speckle case however, $\tau_{E,\uveck}$ shows no inversion of anisotropy.

In the low energy limit ($\kE \sigmaOrth \ll 1$), $\tau_{E,\uveck}$ converges to a constant value.
In contrast to the 2D case, it signals the absence of a 3D white-noise limit\footnote{In the case of a white-noise limit in 3D, the scattering time is isotropic with the scaling $\tau_{E,\uveck} \propto 1/\sqrt{E}$ (\ie\ ${\ls}_{E,\uveck}$ is constant).
This can be found by inserting a constant $\TFCor(\veck)$ in Eq.~(\ref{eq:tau_s}).}.
This can be attributed to the strong anisotropic divergence of $\TFCor(\veck)$ when $|\veck|\rightarrow 0$, which reflects the long-range correlations of the disorder (see Sec.~\ref{part:correl-single}).
More precisely, for $|\veck|\sigmaOrth \ll 1$, we have
\be
\TFcor_{\textrm{1sp}}(\veck) \simeq \pi^{3/2} \frac{\sigmaOrth \sigmaPara}{|\veck|} \TFcor(\uveck) = \pi^{3/2} \frac{\sigmaOrth \sigmaPara}{|\veck|} \frac{e^{-\frac{1}{4} \left(\frac{\sigmaPara}{\sigmaOrth}\right)^2 \frac{\hat{k}_z^2}{\hat{k}_\perp^2}}}{\vert\hat{k}_\perp\vert}.
\label{eq:scaling-corr-lowE}
\ee
Replacing this expression into Eq.~(\ref{eq:tau-single}) we then find
\be
\tau_{E,\uveck} = \frac{\hbar E_{\sigmaOrth}}{V_R^2}
\frac{4 \sqrt{\pi}}{\int \ud\Omega_{\uveck'} \, \TFcor(\uveck'-\uveck)},
\label{eq:tau-single-lowE}
\ee
which is independent of $E$.
Equation~(\ref{eq:tau-single-lowE}) is plotted as solid black lines on the left-hand side of Fig.~\ref{Figure:ls_3Dsingle}.
Note that $\tau_{E,\uveck}$ does not become strictly isotropic in this limit.
However, the residual anisotropy of the scattering time, found from Eq.~(\ref{eq:tau-single-lowE}) and from the anisotropy of $\TFcor(\uveck)$ in Eq.~(\ref{eq:scaling-corr-lowE}), is very small, and practically unobservable ($\tau_{E,\uveck_{\perp}}/\tau_{E,\uveck_z}\simeq1.002$).
When the energy increases, 
the scattering time in the $(x,y)$ plane is the first to deviate significantly from the low-energy behaviour at $E \sim E_{\sigmaPara}$($=3\times10^{-2}E_{\sigmaOrth}$ for the parameters of Fig.~\ref{Figure:ls_3Dsingle}), while the scattering time in the $z$ direction increases  only at $E \sim E_{\sigmaOrth}$.
This can be understood again by the narrower width of the power spectrum $\TFCor(\veck)$ in the $k_z$ direction.

In the high-energy limit ($\kE\sigmaOrth \gg 1$)
the $\veck$-space shell integral of Eq.~(\ref{eq:tau-single}), which is done on a sphere of radius $\kE$ containing the origin, can be reduced to integrating $\TFcor_{\textrm{1sp}}$ on the plane which is tangent to the sphere at the origin.
We then find
\be
\tau_{E,\uveck} \simeq \frac{\hbar E_{\sigmaOrth}}{V_R^2} \frac{\sigmaOrth}{\sigmaPara}
\frac{4 \sqrt{\pi} \kE \sigmaOrth}{\int \ud \kappa \ud \kappa' \,
\frac{e^{-\frac{\kappa^2 \hat{k}_z^2 +{\kappa'}^2}{4}} e^{-\frac{1}{4} \left(\frac{\sigmaPara}{\sigmaOrth}\right)^2 \frac{\kappa^2 \hat{k}_\perp^2}{\kappa^2 \hat{k}_z^2 +{\kappa'}^2}}}{\sqrt{\kappa^2 \hat{k}_z^2 +{\kappa'}^2}} } .
\label{eq:tau-single-highE}
\ee
In particular, we find $\tau_{E,\uveck_{\perp}} =\hbar E_{\sigmaOrth} \kE \sigmaOrth /2\Vr^2 \sqrt{\pi}$, $\tau_{E,\uveck_{z}}=\hbar E_{\sigmaOrth} \kE \sigmaOrth^2 /\Vr^2 \pi \sigmaPara$
(both shown as the right-hand-side solid black lines in Fig.~\ref{Figure:ls_3Dsingle}).
The anisotropy of the scattering then becomes significant for the parameters of Fig.~\ref{Figure:ls_3Dsingle}, $\tau_{E,\uveck_{\perp}}/\tau_{E,\uveck_z}=\sqrt{\pi} \sigmaPara/2 \sigmaOrth$ in this limit.
The high-energy scaling $\tau_{E,\uveck} \propto \kE$, which was also found in our 2D speckle, is quite universal: as long as the power spectrum is of finite integral in all the planes (lines in 2D) crossing the origin, the procedure described above can be applied to Eq.~(\ref{eq:tau_s}).
Then $\tau_{E,\uveck}$ only depends on the dispersion relation $\eps{\veck}$ and, in particular, it is independent of the space dimension.

%% file: TexFiles/DiffusionExample3D.04.tex
\subsection{Boltzmann diffusion \label{part:diff-ex3D}}

The Boltzmann diffusion is obtained, as in the 2D case analyzed previously, by solving Eq.~(\ref{eigeq-os}) numerically and incorporating the results into Eq.~(\ref{DBE}).
For the diagonalization of the integral operator~(\ref{eigeq-os}) we use $2^7 \times 2^7=128 \times 128$ points regularly spaced on the $\veck$-space shell $\vert\veck\vert=\kE$.
We have studied several values of the configuration anisotropy $\sigmaPara/\sigmaOrth$, which all show the same behaviour discussed below.

\begin{figure}[t] 
\begin{center}
\includegraphics[width=0.7\textwidth]{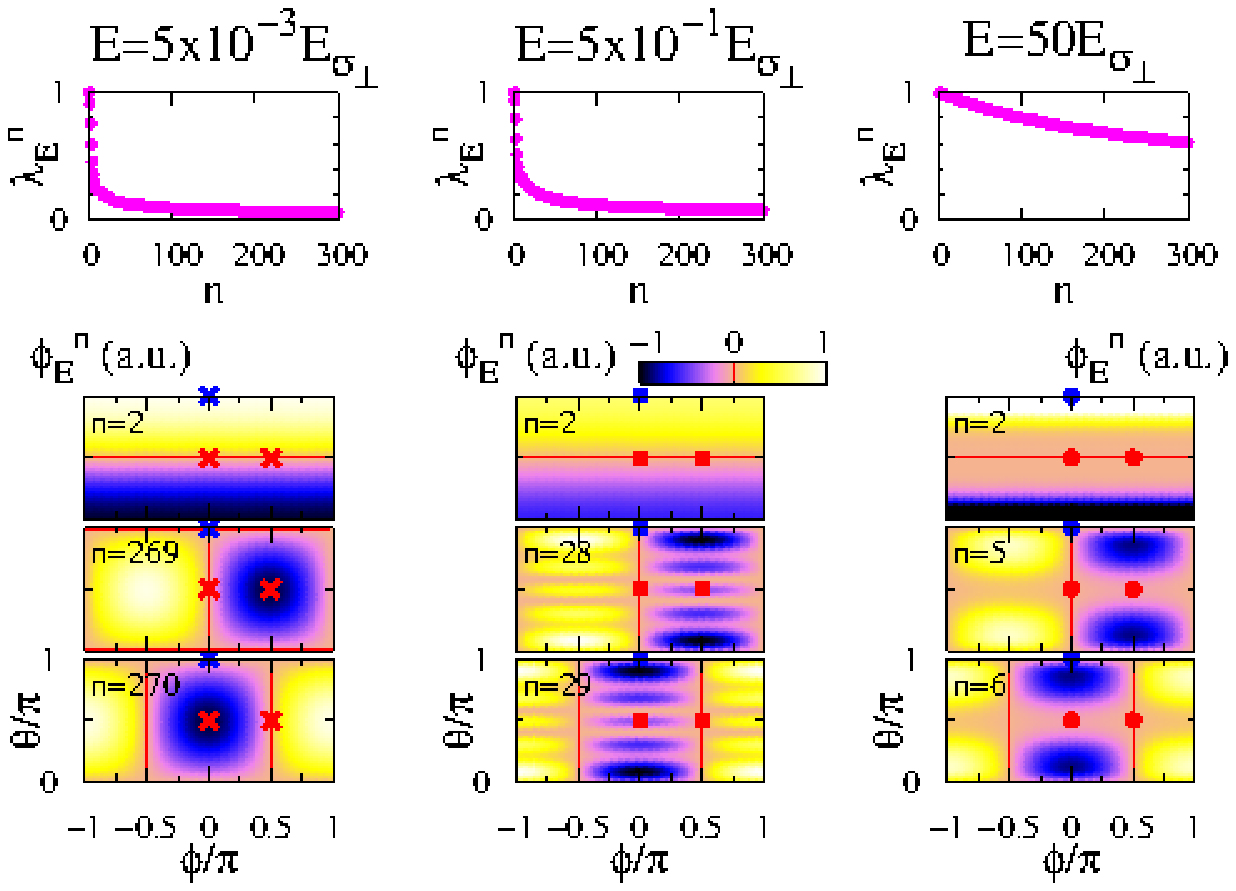}
\end{center} 
\caption{\small{(Color online) Three dimensional case with $\sigmaPara/\sigmaOrth= 5.8$.
Eigenvalues of Eq.~(\ref{eigeq-os}) at various energies indicated on the figure (top row).
Topography of the eigenvectors $\phi_{E,\uveck}^n$, at the same energies, which mainly contribute to $\DB^x$ (bottom row), $\DB^y$ (2$^{\textrm{nd}}$ row) and $\DB^z$ (3$^{\textrm{rd}}$ row) respectively [with the parametrization $\uveck=(\uveck_x,\uveck_y,\uveck_z) \equiv (\sin\theta \cos\phi, \sin\theta \sin\phi, \cos\theta)$].
The values of $n$ are indicated on the figure, the red lines locate the nodal lines.
The points are color- and shape-coded to match those of Fig.~\ref{Figure:ls_3Dsingle}.
}}
\label{Figure:spectr3DSingle}
\end{figure}

The eigenvalues $\lambda_E^n$ of Eq.~(\ref{eigeq-os}) for different energies, as well as the topography of the eigenvectors of Eq.~(\ref{eigeq-os}) that dominate $\DB^x$ (bottom row), $\DB^y$ ($2^\textrm{nd}$ row), and $\DB^z$ ($3^\textrm{rd}$ row) are shown in Fig.~\ref{Figure:spectr3DSingle} for $\sigmaPara/\sigmaOrth=5.8$.
Similarly as for the 2D case, we fond that $\lambda^{n}_E$ decays from 1 to 0 when $n$ increases, more sharply for low energy. The $\phi_{E,\uveck}^n$ are topologically similar to the spherical harmonics at all energies, \ie\ they show similar nodal surfaces, but the associated $\lambda^{n}_E$ are not degenerated in a given $l$-like level.
More precisely, due to the cylindrical symmetry of the power spectrum, the value of $\lambda^{n}_E$ associated to the $Y_l^{+m}$-like and $Y_l^{-m}$-like orbitals are the same for a given $m$, but the degeneracy between the different values of $|m|$ is lifted.

\begin{figure}[t] 
\begin{center}
\includegraphics[width=1.2\textwidth]{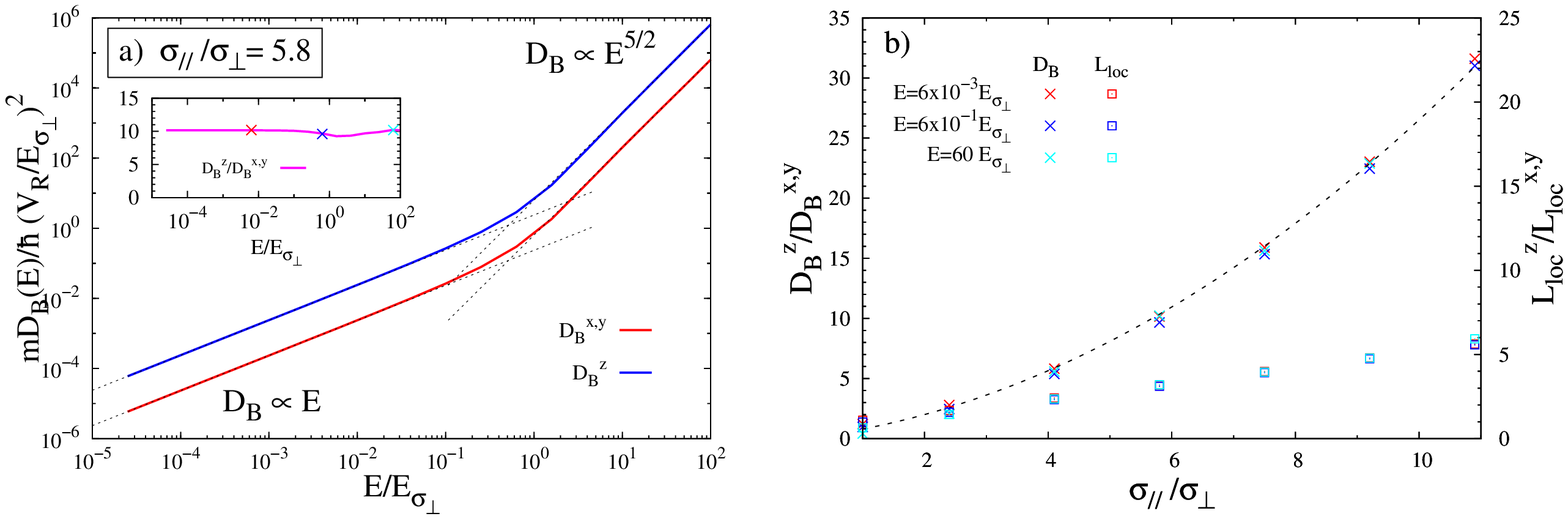}
\end{center}
\vspace{-0.4cm}
\caption{\small{(Color online) (a) Boltzmann diffusion coefficients along the transport eigenaxes (eigencomponents of $\DiffTensB$) for the 3D configuration with $\sigmaPara/\sigmaOrth = 5.8$.
The dotted lines are power-law fits ($\DB^u \propto E^{\gamma_u}$) to the data in the low and high energy limits.
The inset show the transport anisotropy factor $\DB^z/\DB^{x,y}$, and the crosses match those of the right panel.
(b) Anisotropy factors $\anifact_{\textrm{B}}=\DB^z/\DB^{x,y}$ and $\anifact_\textrm{\tiny loc}=\Lloc^z/\Lloc^{x,y}=\sqrt{\DB^z/\DB^{x,y}}$ as a function of the configuration anisotropy $\anifact=\sigmaPara/\sigmaOrth$, at $E/E_{\sigmaOrth}=6\times10^{-3}, 6\times10^{-1}$ and $60$. The dotted line is a fit of all the data which gives $\anifact_{\textrm{B}}=0.59 \anifact+0.21 \anifact^2$.
}}
\label{Figure:Dx_Dy_single}
\end{figure}
Figure~\ref{Figure:Dx_Dy_single}(a) shows the resulting eigencomponents of the diffusion tensor in the 3D case for $\sigmaPara/\sigmaOrth=5.8$.
It is isotropic in the ($x,y$) plane, because of the cylindrical-invariance of the correlation function $\TFCor(\veck)$ around the axis $\uveck_z$.
For the same symmetry reasons as in the isotropic case (see appendix~\ref{part:limit-iso}) and as in the 2D case, only the $p$-level-like orbitals couple to $\boldsymbol{ \upsilon}$.
For $\kE \sigmaOrth \ll 1$, we find that $\DB^{x,y}$ is dominated by the first term in Eq.~(\ref{DBE}) and $\DB^z$ by the $Y_1^0$-like orbital ($n=2$ at all energies).
For $\kE \sigmaOrth \gg 1$, the situation changes:
while $\DB^z$ is still dominated by the $Y_1^0$-like orbital, $\DB^{x}$ is now dominated by the $Y_1^{+1}$-like orbitals and $\DB^{y}$ by the $Y_1^{-1}$-like orbitals (respectively $n=6$ and $5$ at $E=50 E_{\sigmaOrth}$ in Fig.~\ref{Figure:spectr3DSingle})
with a contribution of the $Y_3^{\pm 1}$-like orbitals increasing with $E$. At high energy, we find that the nodal lines of the $Y_3^{\pm 1}$-like orbitals calculated numerically are displaced compared to the associated spherical harmonics.
Therefore their contribution in Eq.~(\ref{DBE}) does not cancel out for symmetry reasons.
Those properties explain the main features of $\DiffTensB$.

Firstly, we find that the diffusion tensor is larger along axis $z$ ($\DB^z > \DB^{x,y}$) for all values of $E$ [see Fig.~\ref{Figure:Dx_Dy_single}(a)], and the anisotropy of $\DiffTensB$ is thus reversed with respect to that of $\tau_{E,\uveck}$ (we recall that we found $\tau_{E,\uveck_{z}}<\tau_{E,\uveck_{\perp}}$ for any $E$, see Sec.~\ref{part:scatt-3D}).
This is due to the fact that
the ($Y_1^0$-like) orbitals contributing to $\DB^z$ are associated to values of $\lambda_E^n$ larger than those contributing to $\DB^{x,y}$ (in Fig.~\ref{Figure:spectr3DSingle}, the $\phi_{E,\uveck}^n$ are numbered by decreasing eigenvalues).

Secondly, $\TFCor (\veck)$ shows a strong anisotropic, infrared divergence in the paraxial approximation (see Secs.~\ref{part:correl-single} and \ref{part:scatt-3D}).
Following-up with the scaling of $\TFcor_{\textrm{1sp}}(\veck)$, Eq.~(\ref{eq:scaling-corr-lowE}), used to show that $\tau_{\uveck,E}$ is independent of energy for $\kE\sigmaOrth \ll 1$, and inserting it into Eq.~(\ref{eigeq-os}) and the associated normalization, we find that $\lambda^n_E$ does not depend on $E$, and $\phi_{E,\uveck}^n$ is of the form $\varphi^n (\uveck)/\sqrt{\kE}$.
Then, all terms in Eq.~(\ref{DBE}) are topologically unchanged and scale as $E$ at low energy.
The anisotropy of $\DiffTensB$ thus persists down to arbitrary low values of $E$ and $\DB^u \propto E$, as observed in the left-hand side of Fig.~\ref{Figure:Dx_Dy_single}(a)
for $\kE \sigmaOrth \ll 1$ (\ie\ $E \ll E_{\sigmaOrth}$).
This is another manifestation of the absence of white-noise limit\footnote{A 3D white-noise limit would lead to the scaling $\DB^u(E)\propto \sqrt{E}$ and an isotropic limit at low energy.}.

Thirdly, for $\kE \sigmaOrth \gg 1$, we found $\tau_{E,\uveck} \propto \sqrt{E}$. Then, assuming weak topological change of the orbitals and the scaling $1-\lambda_E^n \propto 1/E$ (confirmed numerically),
we get $\phi_{E,\uveck}^n \propto 1/\kE$ and $\DB^u(E) \propto E^{5/2}$.
This scaling is confirmed in Fig.~\ref{Figure:Dx_Dy_single}(a) by fits to the data for $E \gg E_{\sigma_\perp}$ (right-hand-side dotted lines).
Remarkably, in spite of the different contributing terms in Eq.~(\ref{DBE}) at low and high values of $E$, the transport anisotropy is nearly independent of $E$ with $\DB^z/\DB^{x,y}\simeq 10$ [see inset of Fig.~\ref{Figure:Dx_Dy_single}(a)].

We have repeated the same study for various values of the configuration anisotropy, $\anifact=\sigmaPara/\sigmaOrth$.
They all show a similar behaviour as a function of energy as reported in Fig.~\ref{Figure:Dx_Dy_single}(a) for $\anifact=5.8$.
In particular, we found the same scalings with energy and a diffusion anisotropy $\anifact_{\textrm{B}}=\DB^z/\DB^{x,y}$ that is nearly independent of energy.
In Fig.~\ref{Figure:Dx_Dy_single}(b), we plot $\anifact_{\textrm{B}}$ versus $\anifact$ for three values of the energy.
We find that the diffusion anisotropy monotonously increases with the configuration anisotropy, as could be intuited.
In order to guess a fitting function for $\anifact_{\textrm{B}}$, one may rely on a simplified model of random walk in an anisotropic lattice of anisotropy factor $\anifact$.
If the transition time is governed by the travel duration between two wells, one expect $\anifact_{\textrm{B}}\propto \anifact$.
If it is governed by the trapping time, one expects $\anifact_{\textrm{B}}\propto \anifact^2$.
In our continuous model of disorder, the situation may be expected to be somehow intermediate.
For our considered range of $\anifact$, we find that the fit $\anifact_{\textrm{B}}=0.59 \anifact + 0.21 \anifact^2$ reproduces well our results as shown in Fig.~\ref{Figure:Dx_Dy_single}(b).

%% file: TexFiles/LocalizationExample3D.05.tex
\subsection{Localization \label{part:loc-ex3D}}

In order to analyze strong localization effects, we now solve the self-consistent equation~(\ref{self-const-cor}) for the 3D case in the long time limit ($\omega \rightarrow 0$).
A threshold energy $\Emob$ (mobility edge) appears, solution of $\geomav{D_\textrm{B}}(\Emob) \equiv \det\{\DiffTensB(\Emob)\}^{1/3}=\hbar/\sqrt{3\pi}m$.
For $E<\Emob$, one finds $\DiffTens(\omega,E) \sim 0^+ -i\omega \LocTens^2(E)$ for $\omega \rightarrow 0$, where $\LocTens(E)$ is a real positive definite tensor.
It characterizes exponential localization within the propagation kernel~(\ref{eq:prop-kernel-loc}) with the anisotropic localization tensor $\LocTens(E)$.
The localization tensor is diagonal in the same basis as the Boltzmann diffusion tensor $\DiffTensB$.
Explicitely, we have
\be
\Lloc^u = \geomav{\Lloc} \sqrt{ \frac{\DB^u}{\geomav{\DB}} },
\label{eq:Lloc-3D-0}
\ee
where $\geomav{\Lloc}=\det \{ \LocTens(E) \}^{1/3}$ is the unique solution of
\be
\frac{\geomav{\Lloc}}{\geomav{l_\textrm{B}}} \left[ 1-\frac{\pi}{3} (\kE \geomav{l_\textrm{B}})^2 \right] = \arctan \left(\frac{\geomav{\Lloc}}{\geomav{l_\textrm{B}}}\right).
\label{eq:Lloc-3D}
\ee
For $E>\Emob$, $\DiffTens(\omega,E)$ converges to a real definite positive tensor when $\omega \rightarrow 0$. It describes anisotropic normal diffusive dynamics, characterized by the propagation kernel~(\ref{eq:prop-kernel-diff}) where $\Diff(E)$ is replaced by the quantum-corrected diffusion tensor
\beq
\label{eq:Diff-3D}
\DiffTens (E) &\equiv& \lim_{\omega \rightarrow 0}\DiffTens(\omega,E)\\
&=&\left[ 1-\frac{\hbar^2}{3 \pi m^2 \left\{ \geomav{\DB}(E)\right\}^2 } \right]\DiffTensB(E). \nonumber
\eeq

As already mentionned in Sec.~\ref{part:weakloc} the behavior of $\LocTens$ and $\DiffTens$ is completely determined by that of $\DiffTensB$ in our approach.
The anisotropies of $\LocTens (E)$ are the square roots of those of $\DiffTensB (E)$ [see Eq.~(\ref{eq:Lloc-3D-0})] and the anisotropies of $\DiffTens(E)$ are the same as those of $\DiffTensB(E)$ [see Eq.~(\ref{eq:Diff-3D})].
Therefore, as for $\DiffTensB$,
for the 3D configuration, the anisotropy factors of $\LocTens$ and $\DiffTens$ are nearly independent of $E$.
The localization anisotropy $\anifact_\textrm{\tiny loc}=\Lloc^z/\Lloc^{x,y}$ is plotted versus the configuration anisotropy on Fig.~\ref{Figure:Dx_Dy_single}(b).
At low energy, using the scaling of $\DB^u(E)$ obtained previously we predict $\Lloc^u(E) \propto \big(\DB^u/\geomav{\DB}\big)^{1/2} E^{3/2}$.
When $E$ increases, $\Lloc^u(E)$ grows and finally diverges at $\Emob$.
In the diffusive regime the quantum corrections are significant only close to $\Emob$, while for higher values of $E$, $\DiffTens (E) \simeq \DiffTensB (E)$.
Therefore, in the high $E$ limit we have $D_*^u(E) \propto (D_\textrm{B}^u/\geomav{D_\textrm{B}}) E^{5/2}$ as found previously (see Sec.~\ref{part:diff-ex3D}).

%% file: TexFiles/MobilityEdge.07.tex
The self-consistent approach used above
is expected to fairly describe the quantum transport properties~\cite{woelfe1984,vollhardt1992,kuhn2007}.
It gives some quantitative estimates consistent with numerical calculations~\cite{kroha1990} and experimental data~\cite{hu2008,jendrzejewski2012}.
However it has two main flaws.

On the one hand, it predicts that, just below the mobility edge, the localization length diverges as $\Lloc^u(E) \propto (\Emob-E)^{-\nu}$ with $\nu=1$ and,
just above the mobility edge $\Emob$, the corrected diffusion tensor increases as $D_*^u (E) \propto (E-\Emob)^s$ with $s=1$.
Those values of the critical exponents $\nu$ and $s$ are consistent with the prediction $s=\nu (d-2)$ of the scaling theory~\cite{abrahams1979,vollhardt1982} and they are independent of the choice of cut-off that we made.
However, it is known, from  advanced numerical calculations on the Anderson model~\cite{kramer1993,slevin1999} and from experiments~\cite{lopez2012}, that they are not correct.
The correct value of the critical exponents in 3D is $\nu=s=1.58\pm 0.01$~\cite{kramer1993,slevin1999}.
In order to reproduce this value, it seems necessary to take into account the fractal nature of the wave functions at the critical point~\cite{evers2008}, which is beyond the self-consistent theory of AL.

On the other hand, in contrast to critical exponents, the mobility edge, $\Emob$ is a non-universal quantity and should be determined from microscopic theory.
In this respect, the on-shell approximation is questionnable because it neglects the strong modification of the spectral function induced by the disorder.
This renormalizes energies and may thus strongly affect the value of $\Emob$.

\subsubsection{Energy renormalization\label{part:e-renorm}}
In order to improve the self-consistent method, one could in principle use the more sophisticated approach of Ref.~\cite{kroha1990}, which does incorporate the spectral function,
and provides values of $\Emob$ in agreement with numerical calculations in the Anderson model.
For continuous disorder, one may rely on the approach of Refs.~\cite{skipetrov2008,yedjour2010}, which has been applied to several standard models of disorder.
However, since we are interested in continuous disordered potentials with fine and anisotropic structures, these methods are hardly practicable.
From a numerical point of view, estimates of necessary ressources seem out of present-day possibilities.
In order to overcome this issue, we have proposed in Ref.~\cite{piraud2012a}
an alternative method based on the assumption that the leading term missing in the on-shell approximation is the real part of the self energy,
\be
\selfE^\prime(E,\veck) \equiv \textrm{P} \int \frac{\ud \veck'}{(2\pi)^d} \,  \frac{\tilde{\Cor}(\veck - \veck')}{E - \epsilon_{\veck^\prime}},
\ee
where $\textrm{P}$ is the Cauchy principal value, see Eq.~(\ref{eq:selfE-Born}).
A quasi-particle of momentum $\veck$ has an energy $E$, solution of $E-\eps{\veck}-\selfE^\prime(E,\veck)=0$.
Here, we incorporate $\selfE^\prime(E,\veck)$ into the theory self-consistently and by averaging,
in first approximation, its $\veck$-angle dependence.
It amounts to replace the on-shell prescription by $\eps{\veck} = E^\prime \equiv E - \Delta (E)$ with
\be \label{eq:shift}
\Delta (E) \equiv \frac{1}{4\pi} \int_{\eps{\veck} = E -\Delta (E)} \ud\Omega_{\uveck}\ \selfE^\prime(E,\veck).
\ee
Within this approach, all previous quantities [$\smft(\veck),\DiffTensB,\LocTens,\DiffTens$] are now regarded as functions of $E^\prime$ instead of $E$.
It does not change the overall energy dependence of the quantities discussed above, but may be important for direct comparison to energy-resolved experimental measurements.
In the following we concentrate on the 3D mobility edge $\Emob$.
It is the solution of $\Emob - \Delta (\Emob) = \EmobNoShift$,
where $\EmobNoShift$ is determined using the on-shell approach and $\Delta$ can be regarded as an energy shift, which renormalizes the energies.

\subsubsection{Isotropic disorder \label{part:comp-iso}}
\begin{figure}[!t]
\begin{center}
\includegraphics[width=0.7\textwidth]{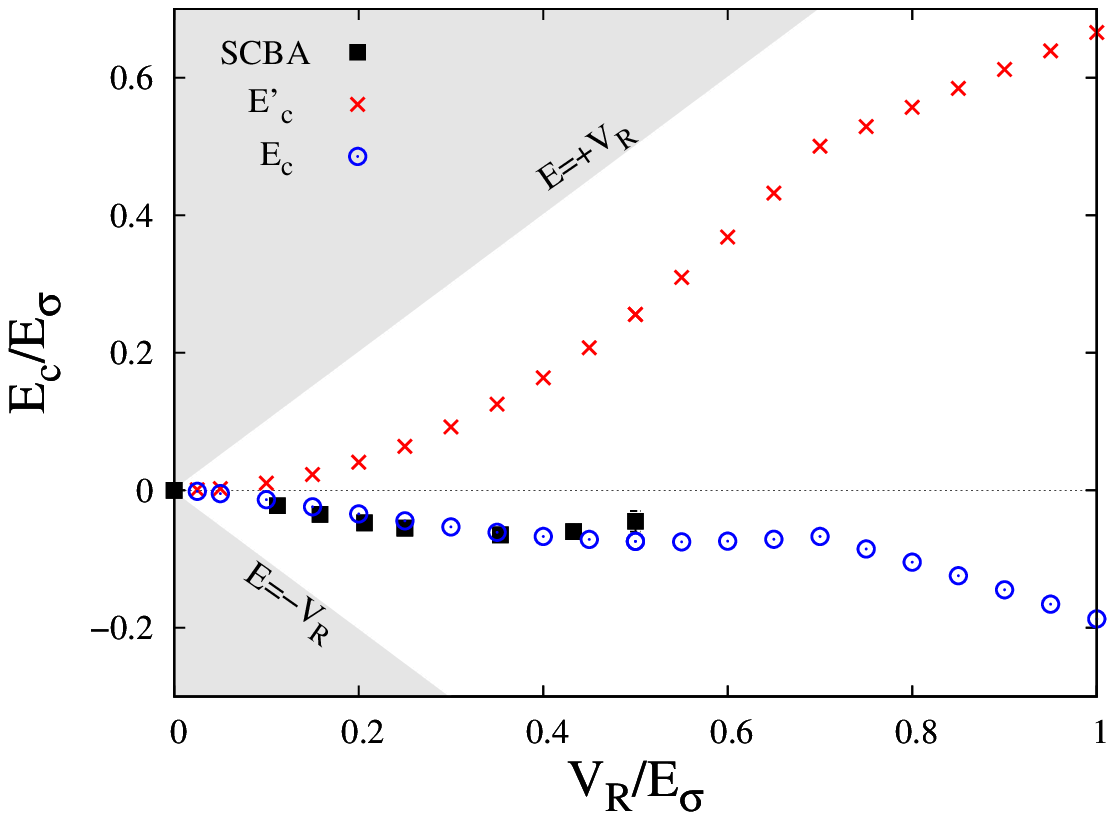}
\end{center}
\vspace{-0.3cm}
\caption{\label{fig:comp_yedj}
\small{(Color online)
Comparison of the mobility edges as calculated with the SCBA method (the full black squares are the results obtained by A. Yedjour and B. van Tiggelen in Ref.~\cite{yedjour2010}, that we reproduce here), with the on-shell method ($\EmobNoShift$, red crosses) and with the renormalized self-consistent approach (corrected $\Emob$, thick blue circles), for an isotropic 3D speckle potential.
When comparing to Fig.~8 of Ref.~\cite{yedjour2010}, note that in Ref.~\cite{yedjour2010} the reference of energy is the minimum value of the disorder and that we have the correspondences $E_\xi=E_\sigma/2$ and $U=\Vr^2$.
}}
\end{figure}
Here, we validate the above approach by a direct comparison to an alternative method applicable to \emph{isotropic} disorder.
Consider a speckle disorder obtained inside an integrating sphere lit with a laser beam, the real-space correlation function of which reads~\cite{kuhn2007,yedjour2010}
\be
\Cor(\vect{r})=\Vr^2\frac{\sin\left(|\vecr|/\sigmar\right)^2}{\left( |\vecr|/\sigmar \right)^2},
\label{eq:corr3Diso}
\ee
with $\sigma$ the correlation length.
The associated power spectrum (see appendix~\ref{part:limit-iso}) is isotropic and bears the same infrared divergence as the anisotropic model of 3D disorder considered in this work as well as other configurations~\cite{piraud2012a}: $\TFCor(\veck)\propto 1/|\veck|$ when $|\veck| \rightarrow 0$.
Figure~\ref{fig:comp_yedj} shows the on-shell mobility edge $\EmobNoShift$ calculated as in Sec.~\ref{part:loc-ex3D} (see also Ref.~\cite{kuhn2007}), the renormalized mobility edge $\Emob$ calculated by our method (see Sec.~\ref{part:e-renorm}), and the mobility edge found using the self-consistent Born approximation (SCBA) in Ref.~\cite{yedjour2010}.
As it is clearly seen in Fig.~\ref{fig:comp_yedj}, the disorder-induced modification of the spectral function plays a major role for the prediction of the mobility edge.
While the on-shell mobility edge, $\EmobNoShift$, is \new{above the statistical average of the potential ($\av{V}=0$ for our choice of energy reference)}, 
the corrected mobility edge, $\Emob$, as calculated either by the method of Ref.~\cite{yedjour2010} or by our self-consistent renormalized approach, is \new{below the statistical average of the potential}.
In addition, we find that the renormalized self-consistent approach predicts values of $\Emob$ in very good agreement (within $5-7\%$) with those of Ref.~\cite{yedjour2010}.
These results support our method to estimate $\Emob$.

\subsubsection{Anisotropic disorder}
\begin{figure}[!t]
\begin{center}
\includegraphics[width=0.7\textwidth]{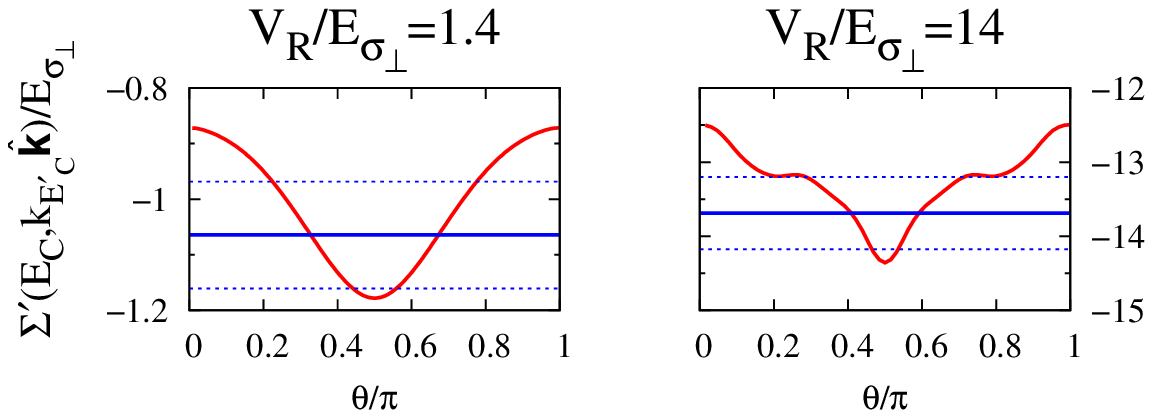}
\end{center}
\vspace{-0.6cm}
\caption{\small{(Color online)
Angular dependence of $\selfE^\prime(\Emob, k_{\EmobNoShift}\uveck)$ (thin solid lines) for the
3D single-speckle for $\sigmaPara/\sigmaOrth= 5.8$ [with $\theta=(\uveck,\uveck_z)$], and for different values of $\Vr$ (indicated on the figure).
The horizontal solid blue line is the mean value and the dashed blue lines represent the standard deviation around the mean, both calculated over the $\uveck$-solid angle.
}}
\label{Figure:varangle3D}
\end{figure}
We now apply our method to anisotropic disorder in the 3D single-speckle configuration.
The mobility edge is found by searching the root of the self-consistent equation~(\ref{eq:shift}).
Note that the averaging of the angular dependence of $\selfE^\prime$ in Eq.~(\ref{eq:shift}) is justified {\it a posteriori} by the weak $\uveck$-angle variations of $\selfE^\prime$ found around its mean value at $\Emob$ (with standard deviations less than $10-15\%$).
This is illustrated in Fig.~\ref{Figure:varangle3D},
which presents the angular variations obtained numerically in the calculation of $\Delta (\Emob)$, 
for typical values of $\Vr$ and for an anisotropy of $\sigmaPara/\sigmaOrth=5.8$.

\begin{figure}[!t] 
\begin{center}
\includegraphics[width=0.7\textwidth]{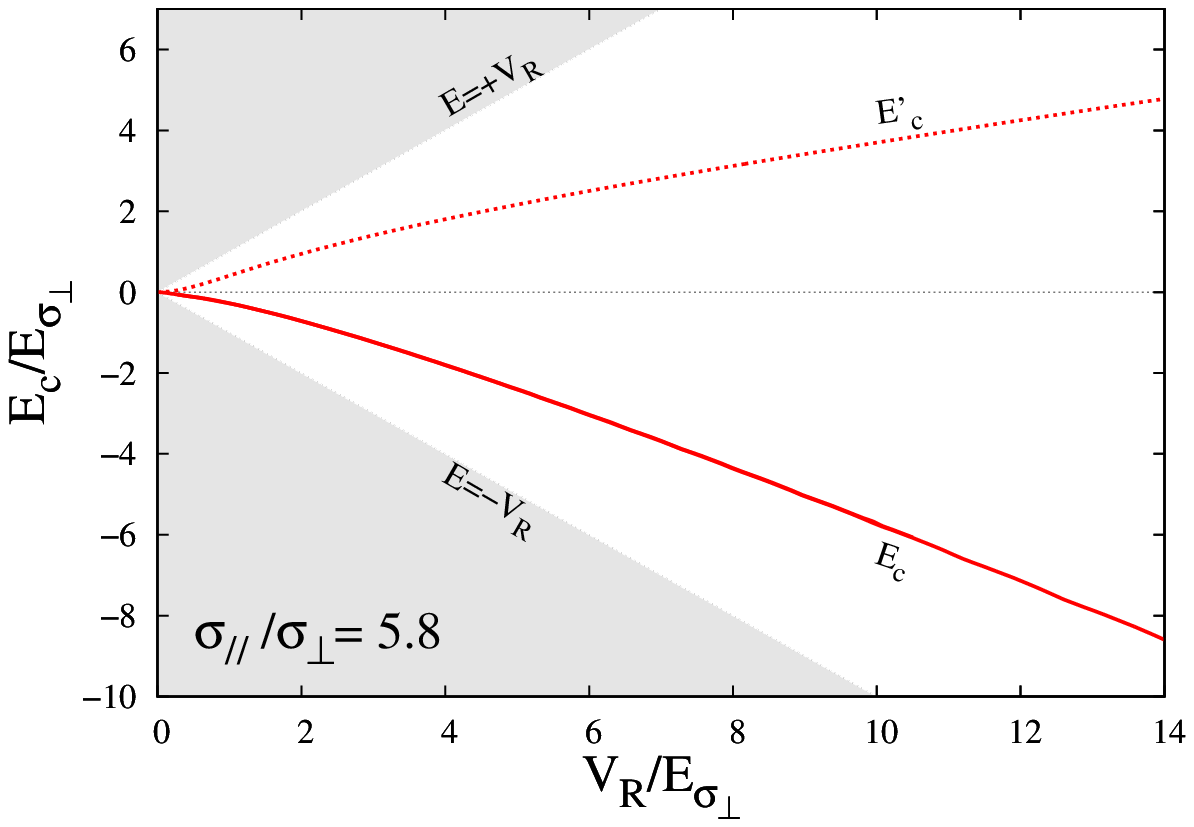}
\end{center} 
\caption{\small{(color online)
On-shell ($\EmobNoShift$) and renormalized ($\Emob$) mobility edges
versus the disorder amplitude $\Vr$ 
for the 3D (single-speckle) case with $\sigmaPara/\sigmaOrth= 5.8$.}
}
\label{Figure:Emob}
\end{figure}
The on-shell ($\EmobNoShift$) and renormalized ($\Emob$) mobility edges
are shown in Fig.~\ref{Figure:Emob}.
As for isotropic disorder, it is eye-catching that the shift of the energy states completely changes the behavior of the mobility edge.
While the on-shell mobility edge, $\EmobNoShift$, is \new{above the statistical average of the potential},  the renormalized mobility edge, $\Emob$, is \new{below}.
This behaviour seems very robust for 3D speckle disorder. It was found for isotropic 3D speckles (see Ref.~\cite{yedjour2010} and Sec.~\ref{part:comp-iso}), as well as other models of speckle potentials with structured correlations~\cite{piraud2012a}.

%% file: TexFiles/Conclusion.04.tex
Disordered potentials with finite-range correlations are often characterized by a counter-intuitive and interesting behaviour~\cite{billy2008,lugan2009,plodzien2011,piraud2011a,piraud2011b,piraud2012a}.
These are directly related to the microscopic statistical properties of the potential, hallmarked by the disorder correlation function.
In this paper we have focused on anisotropy effects in 2D and 3D correlated disorder.  We have quantitatively studied the transport and localization of matter waves by using mesoscopic transport theory~\cite{rammer1998} and a standard on-shell self-consistent perturbative approach~\cite{woelfe1984}.
The latter, first pioneered by  Vollhardt and W\"olfe~\cite{vollhardt1980a,vollhardt1980b}, remains the most powerful, quantitative, microscopic approach to Anderson localization in dimension higher than one ($d \geqslant 2$), in spite of the unavoidable problem of describing the physics inside the critical region in $d>2$.
Within this approach, we have characterized incoherent diffusion, quantum corrected diffusion and localization tensors versus the particle energy and found rich diffusion and localization properties.
We have supported the general theory with application to speckle potentials in 2D and 3D.

In the 2D case, we have considered an anisotropic Gaussian correlation function as used in Refs.~\cite{mrsv2010,pezze2011b}.
The energy-dependences of relevant quantities are studied:
For $E\ll E_{\sigmaOrth}$, in the white-noise limit, we find $\tau_{E,\uveck} \propto 1$ for the scattering time and $\DiffTensB \propto E$ for the Boltzmann diffusion tensor, which are both isotropic. 
For $E\gg E_{\sigmaOrth}$, we find $\tau_{E,\uveck} \propto \sqrt{E}$ and $\DiffTensB \propto E^{5/2}$.
As a general rule, the anisotropy of the disorder ($\anifact$), of the scattering time ($\anifact_s$) and of Boltzmann diffusion ($\anifact_{\textrm{B}}$) are all different.
The scattering time shows an inversion of anisotropy from $\anifact_s>1$ (for $\anifact>1$) at low energy to $\anifact_s=1/\anifact$ ($<1$) at high energy.
In contrast, the transport anisotropy is always $\anifact_{\textrm{B}}>1$ (for $\anifact>1$) but shows a strongly nonmonotonic behaviour as a function of energy with a marked maximum at $E\sim E_{\sigmaOrth}$.
The anisotropy of localization is simply the square root of that of transport.
For typical experimental parameters, we found that it is very small in observable regimes, except for very strongly anisotropic disorder.
So far, experiments have only studied the classical regime~\cite{mrsv2010,pezze2011b} and our study offers scope for future studies of quantum transport and localization in 2D speckle potentials.

In the 3D case we have considered the strongly anisotropic correlation function of speckle potentials obtained with a single laser.
Here, the energy dependence of relevant quantities are the following:
For $E\ll E_{\sigmaOrth}$, we find $\tau_{E,\uveck} \propto 1$ and is slightly anisotropic, while $\DiffTensB \propto E$ and is significantly anisotropic, which is due to anisotropic suppression of the white-noise limit in the model we used.
For $E\gg E_{\sigmaOrth}$, we find $\tau_{E,\uveck} \propto \sqrt{E}$ and $\DiffTensB \propto E^{5/2}$, both being anisotropic.
We have also analyzed the anisotropy of transport as a function of the configuration anisotropy. We found that it is almost independent of the energy, and has a the behaviour $\anifact_{\textrm{B}}=0.59 \anifact+0.21 \anifact^2$.
In our approach, the anisotropy of the localization tensor is the square root of that of the Boltzmann diffusion tensor.
We have also studied the behaviour of the 3D mobility edge.
To do so, we have extended the on-shell approach and proposed a way to renormalize energies.
We have found a striking agreement of our method with the more involved method based on SCBA developed in Ref~\cite{yedjour2010} for isotropic disorder.
The effect of renormalizing energies does not alter the overall energy dependence of the quantities discussed above, but may be important for direct comparison to energy-resolved experimental measurements.
As regards the mobility edge, we have found that the renormalization of energies has both a quantitative and qualitative impact.
In particular, we find that, as for isotropic disorder, the renormalized mobility edge is below the average value of the disorder.

Finally, our results and method may provide a guide line to future experiments investigating the so-far unexplored effect of anisotropy in quantum transport of matter waves.
In the case of ultracold atoms, to which our study directly applies, the transport properties can be probed by direct imaging of the atoms and control of the energy.
First experimental studies of Anderson localization of 3D matter waves in anisotropic speckle potentials have been reported~\cite{kondov2011,jendrzejewski2011}. Our study is directly relevant to these experiments.
For a detailed comparison of theoretical predictions and experimental observations, see Ref~\cite{piraud2012a}.
In addition,
the effects discussed in this manuscript  can be expected for other kinds of waves and/or other models of disorder, and are particularly relevant to new systems where the disorder correlations can be controlled~\cite{lsp2010,kondov2011,jendrzejewski2011,clement2006,kuhl2000,kuhl2008,barthelemy2008}.

%% file: TexFiles/Appendix-BSE.03.tex
\section{Intensity kernel \label{part:ap-BSE}}

In this section we show the step-by-step calculation of the long-time and large-distance limit of the intensity kernel given by Eqs.~(\ref{BSEsolution}), (\ref{BSEregular}) and (\ref{BSEsingular}) and the diffusion tensor Eq.~(\ref{DBE}).

As explained in Sec.~\ref{part:BSE}, the solution of the Bethe-Salpeter equation~(\ref{eq:BSE})-(\ref{Diag:BSE}) can be obtained by inverting the operator $\Oplamb \equiv 1 - \av{\Gr} \otimes \av{\Ga} \,\U$ [see Eq.~(\ref{eq:sol-BSE})].
To this aim, we diagonalize the operator $\av{\Gr} \otimes \av{\Ga} \,\U$ in the $(\vecq,\omega)=(0,0)$ limit.
We thus solve
\be
\int \frac{\ud \veck'}{(2 \pi)^d} \, U^E_{\veck,\veck'} \, f_{E,\veck'} \, \phi_{E,\veck'}^n = \lambda_E^n \phi_{E,\veck}^n
\label{eq:eigeq-ap}
\ee
where $U^E_{\veck,\veck'}=U_{\veck,\veck'}(\vecq=0,\omega=0,E)$ and $f_{E,\veck} = \av{\Gr}(E,\veck) \av{\Ga}(E,\veck)$ [see Eq.~(\ref{eq:fkE}) for $\vecq=0$ and $\omega=0$].

\subsection{Preliminary remark}
First, let us notice that we have
\be \label{eq:fk}
f_{E,\veck} = \frac{\smft(E,\veck)}{\hbar} \, A(E,\veck),
\ee
where $A(E,\veck)$ is the spectral function defined in Eq.~(\ref{eq:AkE0}) and $\smft(E,\veck)$ is the scattering mean free time defined in Eq.~(\ref{eq:tau_s-def}).

\subsection{Properties of Eq.~(\ref{eq:eigeq-ap})}
The main properties of Eq.~(\ref{eq:eigeq-ap}) and of its eigenfuctions are listed below:
\begin{enumerate}
\item The eigenvalues $\lambda_E^n$ and the eigenvectors $\phi_{E,\veck}^n$ of Eq.~(\ref{eq:eigeq-ap}) are real.

\begin{proof}
By multiplying Eq.~(\ref{eq:eigeq-ap}) by $\av{\Ga}(E,\veck)$, we obtain 
\be \label{real}
\int \frac{\ud \veck'}{(2 \pi)^d} \, 
M^E_{\veck,\veck'} \,
\av{\Ga}(E,\veck') \, \phi_{E,\veck'}^n = \lambda_E^n
\, \av{\Ga}(E,\veck) \, \phi_{E,\veck}^n,
\ee
where $M^E_{\veck,\veck'} \equiv \av{\Ga}(E,\veck) \, U^E_{\veck,\veck'} \, \av{\Gr}(E,\veck')$.
The latter is Hermitian since $\av{\Ga}(E,\veck)^* = \av{\Gr}(E,\veck)$ and $U^E_{\veck,\veck'}$ is real and symmetric.
Therefore all the eigenvalues $\lambda_E^n$ are real. 
By taking the complex conjugate of Eq.~(\ref{real}), dividing by $\av{\Gr}(E,\veck)$ and comparing it to Eq.~(\ref{eq:eigeq-ap}), we obtain that the functions 
$\phi_{E,\veck}^n$ are real.


\medskip

If $U_{\veck,\veck'}^E$ is positive-definite, the eigenvalues $\lambda_E^n$ are positive.
In particular, this is always true in the Born approximation\footnote{In this case, $U^E_{\veck,\veck'} = \TFCor(\veck - \veck')$ is symmetric and positive-definite.
This latter property is assured for any disordered potential by the fact that the power spectrum $\TFCor(\veck)$, being the Fourier Transform of the autoconvolution product of the potential, is positive for any $\veck$.}.
When $U_{\veck,\veck'}^E$ is symmetric and positive-definite, we can write it as 
$U_{\veck,\veck'}^E  = 
\int \frac{\ud \veck''}{(2 \pi)^d} 
Q_{\veck,\veck''} 
d_{\veck''} 
Q_{\veck'',\veck'}^T$, where $d_{\veck''}>0$
and $Q$ is an orthogonal operator.
For any vector of components $x_{\veck}$, we have 
$\int \frac{\ud \veck}{(2 \pi)^d}
\frac{\ud \veck'}{(2 \pi)^d}
x_{\veck} M_{\veck, \veck'}^E x_{\veck'} =
\int \frac{\ud \veck}{(2 \pi)^d}
d_{\veck} | y_{\veck}|^2 >0$, where 
$y_{\veck} \equiv \int \frac{\ud \veck'}{(2 \pi)^d}
\av{\Ga}(E,\vect{k}')
x_{\veck'} Q_{\veck',\veck}$.
It shows that $M_{\veck, \veck'}^E$ is positive definite.
Its eigenvalues $\lambda_E^n$ are therefore positive.
\end{proof}

\item The eigenvectors $\phi_{E,\veck}^n$ can be chosen to satisfy the 
orthonormalization condition 
\be \label{comp1}
\int \frac{\ud \veck}{(2 \pi)^d} \, f_{E,\veck} \,
\phi_{E,\veck}^n \phi_{E,\veck}^m
= \delta_{n,m}.
\ee

\begin{proof}
This is an immediate consequence of the fact that, according to 
Eq.~(\ref{real}), the functions $\av{\Ga}(E,\veck)  \, \phi_{E,\veck}^n$ are 
eigenfunctions of the Hermitian operator $M^E_{\veck,\veck'}$. 
\end{proof}

\item The eigenvectors $\phi_{E,\veck}^n$ satisfy the completeness relation 
\be \label{comp2}
f_{E,\veck} \, \sum_n \phi_{E,\veck}^n \, \phi_{E,\veck'}^n = 
(2 \pi)^d \, \delta(\veck-\veck').
\ee

\begin{proof}
This follows from the fact that the eigenfuntions 
$\av{\Ga}(E,\veck) \, \phi_{E,\veck}^n$ of the matrix $M^E_{\veck,\veck'}$, Eq.~(\ref{real}),
form a complete basis.
\end{proof}

\item The irreducible vertex function $U^E_{\veck,\veck'}$ 
can be expressed as
\be \label{stfact}
U^E_{\veck,\veck'} = \sum_n \lambda_E^n \, \phi_{E,\veck}^n \, \phi_{E,\veck'}^n.
\ee

\begin{proof}
We multiply both terms of Eq.~(\ref{eq:eigeq-ap}) by 
$\phi_{E,\veck'}^n$ and sum over $n$. Equation (\ref{stfact}) is recovered by 
using the completeness relation Eq.~(\ref{comp2}).
\end{proof}

\item The most important property of Eq.~(\ref{eq:eigeq-ap})
is that one of the eigenvalues is 
\be
\lambda_E^{n=1} = 1 \label{lambda0},
\ee
and the corresponding eigenvector is 
proportional to the inverse scattering mean free time:
\beq \label{phi0}
\phi_{E,\veck}^{n=1} = \sqrt{\hbar} 
\frac{ [\smft(E,\veck)]^{-1} }{\sqrt{ \frac{\ud \veck'}{(2 \pi)^d} \, A(E,\veck) \, [\smft(E,\veck)]^{-1} }}.
\eeq

\begin{proof}
This is a direct consequence of the Ward identity \cite{vollhardt1980b}:
\be \label{WardId00}
\Delta \selfE_{\veck} (\vecq,\omega,E) = \int \frac{\ud \veck'}{(2 \pi)^d} \, 
U_{\veck,\veck'} (\vecq, \omega, E) \, \Delta \Gr_\veck(\vecq,\omega,E),
\ee
where $\Delta \selfE_{\veck}(\vecq,\omega,E)= \selfE(E_+,\veck_+) - \selfEa(E_-,\veck_-)$ and $\Delta \Gr_{\veck}(\vecq,\omega,E)= \av{\Gr}(E_+,\veck_+) - \av{\Ga}(E_-,\veck_-)$.
For $(\vecq,\omega) = (0,0)$ it becomes
\be \label{WardId01}
\Delta \selfE_{\veck} (0,0,E) = \int \frac{\ud \veck'}{(2 \pi)^d} \, 
U^E_{\veck,\veck'}\, f_{E,\veck} \, \Delta \selfE_{\veck} (0,0,E).
\ee
When comparing Eq.~(\ref{WardId01}) to Eq.~(\ref{eq:eigeq-ap}), 
we obtain that $\Delta \selfE_{\veck}(0,0,E)= - i \hbar /\smft(E,\veck)$ is a 
solution of Eq.~(\ref{eq:eigeq-ap}) with unit eigenvalue.
Using Eq.~(\ref{eq:fk}) and the orthonormalization condition~(\ref{comp1}) one then easily finds Eq.~(\ref{phi0}).
\end{proof}

\item The eigenfunctions $\phi_{E,\veck}^n$ have the parity properties: 
\beq
\phi_{E,-\veck}^{n=1} &=&\phi_{E,\veck}^{n=1} \label{parity0}\\
\phi_{E,-\veck}^n & =& -\phi_{E,\veck}^n \qquad \mathrm{for} \quad n>1. \label{parityn}
\eeq

\begin{proof}
This is a consequence of the parity of the vertex $U^E_{\veck,\veck'}$,
in particular, $U^E_{-\veck,-\veck'} = U^E_{\veck,\veck'}$.
Using Eq.~(\ref{stfact}) we have 
$\sum_n \lambda_E^n \, \phi_{E,\veck}^n \, \phi_{E,\veck'}^n = \sum_n \lambda_E^n \, \phi_{E,-\veck}^n \, \phi_{E,-\veck'}^n$, 
which can only be satisfied if the eigenfunctions $\phi_{E,\veck}^n$ have a well defined parity.
The eigenfunction $\phi_{E,\veck}^{n=1}$ is given by Eq.~(\ref{phi0}) and it is even. 
In addition, using Eqs.~(\ref{eq:fk}) and (\ref{phi0}) in the 
orthonormalization condition~(\ref{comp1}), we have $\int \frac{\ud \veck}{(2 \pi)^d} \, A(E,\veck) \, \phi_{E,\veck}^n =0$ for $n>1$. Which shows that $\phi_{E,\veck}^n$ are odd functions of $\veck$.
\end{proof}
\end{enumerate}


\subsection{Solution of the BSE}
Note first that, if Eq.~(\ref{eq:eigeq-ap}) could be diagonalized with all eigenvalues 
different from one ($\lambda_E^n \neq 1$ for all $n$), it is straightforward to show, using Eq.~(\ref{comp2}), that we would have
$\Oplamb^{-1}_{\veck,\veck'}(0,0,E)=\sum_n [1/(1 - \lambda_E^n)] f_{\veck} \phi_\veck^n \phi_{\veck'}^n $.
In this case no diffusion would be observed. 
As noticed above, however, the conservation of particle number, through the Ward identity, 
imposes that there is one eigenvalue equal to one.
As there is no other conserved quantity in the system we are considering, we can assume that the eigenvalue $\lambda=1$ is not degenerated and that there is a finite gap between this eigenvalue and the rest of the spectrum when $(\vecq,\omega) \to 0$~\cite{barabanenkov1991,barabanenkov1995}. 
This suggests the following ansatz for the solution of the BSE~(\ref{eq:BSE})-(\ref{Diag:BSE}) [see Eq.~(\ref{eq:sol-BSE})],
in the small (but non-zero) $\vecq$ and $\omega$ limit:
\beq \label{ansatzGamma}
\Phi_{\veck,\veck'}(\vecq,\omega,E) = & & 
f_{E,\veck} \frac{\phi^1_{\veck}(\vecq,\omega,E)\phi^1_{\veck'}(\vecq,\omega,E)}{\lambda(\vecq,\omega,E)} f_{E,\veck'} \nonumber \\
&& + \sum_{\lambda_E^n \neq 1} \frac{1}{1-\lambda_E^n} 
f_{E,\veck} \phi_{E,\veck}^n \phi_{E,\veck'}^n f_{E,\veck'}, \nonumber\\
\eeq 
where $\phi^1_{\veck}(\vecq,\omega,E)$ and $1+\lambda(\vecq,\omega,E)$ are solutions of the eigenequation
\beq \label{eigeq01}
\int \frac{\ud \veck'}{(2 \pi)^d} & U^E_{\veck,\veck'} \,  f_\veck(\vecq,\omega,E) \,
\phi^1_{\veck'}(\vecq,\omega,E) \nonumber \\
\qquad &=\big[ 1+\lambda(\vecq,\omega,E) \big] \phi^1_{\veck}(\vecq,\omega,E).
\eeq
The latter are the first eigenvalue and eigenvector at small $(\vecq,\omega)$, and reduce to Eqs.~(\ref{lambda0}) and (\ref{phi0}) when $(\vecq,\omega)=(0,0)$, respectively.
We then write $f_{\veck}(\vecq,\omega,E)=f_{E,\veck}+F_{\veck}(\vecq,\omega,E)$ the
expansion of $f_{\veck}(\vecq,\omega,E)$.
Making the ansatz $\phi^1_{\veck}(\vecq,\omega,E) = \sum_n a_n(\vecq,\omega,E) \phi_{E,\veck}^n$, 
we find 
\be \label{lambdaqw}
\lambda(\vecq,\omega,E) = 
\sum_{n} \frac{a_n(\vecq,\omega,E)}{a_1(\vecq,\omega,E)} \, 
\int \frac{\ud \veck}{(2 \pi)^d} \, \phi_{E,\veck}^0 \, 
F_{\veck}(\vecq,\omega,E) \, \phi_{E,\veck}^n.
\ee
Finally, the coefficients $a_n(\vecq,\omega,E)$ are 
found by imposing that Eq.~(\ref{ansatzGamma}) solves the BSE. 
After some algebra one finds $a_1(\vecq,\omega,E) = 1$ and
$a_n(\vecq,\omega,E) = \frac{\lambda_E^n}{1-\lambda_E^n} 
\int \frac{\ud \veck}{(2 \pi)^d} \, \phi_{E,\veck}^0 \, F_{\veck}(\vecq,\omega,E) \, \phi_{E,\veck}^n$, for $n>1$.

\subsection{On-shell approximation\label{app:o-s}}
We now proceed to the on-shell (weak disorder) approximation, and we neglect the effect of disorder on the spectral function.
Equation~(\ref{eq:fk}) becomes
\be \label{eq:fk-os}
f_{E,\veck} \approx \frac{\tau_{E,\uveck}}{\hbar} \, A_0(\veck,E),
\ee
where $\tau_{E,\uveck}$ is the on-shell scattering mean free time [$\tau_{E,\uveck}\equiv\smft(E,\kE\uveck)$], $A_0(\veck, E) = 2\pi \, \delta[E-\eps{\veck}]$ and $\eps{\veck}$  are, respectively, the disorder-free particle spectral function and dispersion relation.
An explicit calculation of the small $(\vecq, \omega)$ expansion of $f_{\veck}(\vecq,\omega,E)$, gives\footnote{The small $(\vecq,\omega)$ expansion of $f_{\veck}(\vecq,\omega,E)$ requires special attention in the on-shell approximation.
Let us consider for instance the first order term in $\omega$.
We find
$F_{\veck}(\vecq,\omega,E) \approx \frac{\hbar \omega}{2} [f_{E,\veck} \av{\Ga}(E,\veck)- f_{E,\veck} \av{\Gr}(E,\veck)]$.
In the on-shell approximation this equation appears to go as the square of a $\delta$-function, and one has to handle this divergence correctly \cite{mahan2000}:
we assume that $f_{E,\veck} \av{\Gr}(E,\veck) \sim 2 \pi \, c \, \delta(E-\eps{\veck})$,
where the factor $c$ is calculated by imposing that the integral over energy of $f_{E,\veck} \av{\Gr}(E,\veck)$ remains invariant, \ie\ $c=\int \frac{\ud E}{2\pi} \, f_{E,\veck} \av{\Gr}(E,\veck)$.
With this method, we find $f_{E,\veck} \av{\Gr}(E,\veck) =  i (\tau_{E,\uveck}^2/\hbar^2) A_0(\veck,E)$
and therefore
$F_{\veck}(\vecq,\omega,E) \approx \hbar \omega \, i (\tau_{E,\uveck}^2/\hbar^2) A_0(\veck,E)$, as in Eq.~(\ref{Fw}).
Following the same method, we can calculate the other terms in Eq.~(\ref{Fw}).
Finally note that Eq.~(\ref{Fw}) also assumes that $\smft(E,\veck)$ is a smooth 
function of $\veck$, such that $\nabla_\veck \smft(E,\veck) \approx 0$.}
\beq \label{Fw}
F_{\veck}(\vecq,\omega,E) &=& \Bigg\{ \frac{ i \tau_{E,\uveck}^2 }{\hbar^2} 
\big[ \hbar\omega - \vecq \cdot \nabla_\veck \eps{\veck}  \big] \nonumber\\
 &+& \frac{2 \tau_{E,\uveck}^3}{\hbar^3} \, \hbar\omega \, \big[\vecq \cdot \nabla_\veck \eps{\veck} \big] - \frac{\tau_{E,\uveck}^3}{\hbar^3} \, \big[ \vecq \cdot \nabla_\veck \eps{\veck} \big]^2 \Bigg\} \nonumber\\
&\times& A_0(\veck,E) + O(\omega^2,q^3,q^2\omega).
\eeq
Then, making use 
of the parity properties of the functions $\phi_{E,\uveck}^n$ 
[Eqs.~(\ref{parity0}) and (\ref{parityn})], 
$\tau_{E,\uveck}$ (even function of $\uveck$) and $\nabla_\veck \eps{\veck}$ (odd function of $\veck$), 
we finally obtain $\phi^1_{\veck}(\vecq,\omega,E) f_{E,\veck}=2\pi \gamma_{\veck}(\vecq,E)/\sqrt{\hbar \langle \tau_{E,\uveck}^{-1} \rangle} $ where $\gamma_{\veck}$ is given by Eq.~(\ref{gamma}) and 
$\lambda(\vecq,\omega,E)=2 N_0(E) \left[i \hbar \omega - \hbar \vecq \! \cdot \! \Diff(E) \! \cdot \! \vecq \right]/\hbar \langle \tau_{E,\uveck}^{-1} \rangle$
with the diffusion tensor of Eq.~(\ref{DBE}).
The solution of the BSE is thus given by Eq.~(\ref{BSEsolution}) with Eqs.~(\ref{BSEregular}) and (\ref{BSEsingular}).
Note that this expression for the diffusion constant is quite general (only the on-shell approximation has been made), provided that the full irreducible vertex function $\mathrm{U}$ is considered in the eigenequation~(\ref{eq:eigeq-ap}).
In Sec.~\ref{part:diff} the Born and Boltzmann approximations are made $\mathrm{U}=\mathrm{U}_{\mathrm{B}}$ [see Eq.~(\ref{eigeq-os})].

%% file: TexFiles/Appendix-iso.02.tex
\section{Isotropic disorder \label{part:limit-iso}}
For disorder with isotropic correlation function, we define, as in Ref.~\cite{kuhn2007},
$p(k,\theta) \equiv \TFCor(k |\uveck - \uveck'|) = \TFCor\big(2k |\sin(\theta/2)| \big)$, 
where $\theta$ is the angle between the unit vectors $\uveck$ and $\uveck'$ 
and $k \equiv |\veck|=|\veck'|$.
In this case, rotation invariance ensures that the eigenproblem~(\ref{eigeq-os}) is solved by cylindrical (2D) or spherical (3D) harmonics.

\subsection{Two-dimensional case}
In the 2D isotropic case,
inserting the cylindrical harmonics $Z_0= 1$, $Z_l^{+1}= \cos(l \theta)$ and $Z_l^{-1}= \sin(l \theta)$ into Eq.~(\ref{eigeq-os}), we find
\be
\lambda_E^{l,m} = \frac{\int_{0}^{2\pi} \ud \theta \, p(\kE, \theta) \, \cos(l \theta)}
{\int_{0}^{2\pi} \ud \theta \, p(\kE, \theta)},
\label{eq:iso-eigen}
\ee
where $l \geq 0$ and $m\in \{-1,+1\}$ are integer numbers.
In particular, we find $\lambda_E^{l=0}=1$ in agreement with Eq.~(\ref{lambda0}).
They are doubly-degenerated for $l>0$ and the corresponding normalized eigenfunctions are proportional to the orthonormal cylindrical harmonics, with the prefactor determined by the normalization condition~(\ref{comp1}):
\be
\phi_{E,\uveck}^{l=0} = Z_0(\theta) \sqrt{\frac{\int_{0}^{2\pi} \ud \theta' \, p(\kE, \theta')}{\pi}},
\label{eq:iso-phi0}
\ee
and
\be
\phi_{E,\uveck}^{l,\pm 1} = Z_l^{\pm 1}(\theta) \sqrt{\frac{\int_{0}^{2\pi} \ud \theta' \, p(\kE, \theta')}{\pi}}.
\label{eq:iso-phil}
\ee
In the calculation of the diffusion constant, it is actually possible to see that only the first term plus the $l=1$ terms (with $m=-1, +1$) in the summation of the right-hand side of Eq.~(\ref{DBE}), contribute to the diffusion coefficient.
More precisely the on-shell scattering mean free time $\tau_{E,\uveck}$ does not depend on $\uveck$, $\upsilon_{x}$ (respectively $\upsilon_{y}$) is a $2\pi$-periodic and even (resp. odd) function of $\theta$, and $Z_l^{+1}$ (resp. $Z_l^{+1}$) is $2\pi/l$-periodic and even (resp. odd).
Therefore, when performing the angular averaging of the product $\tau_{E,\uveck} \upsilon_i \phi_{E,\uveck}^n$ in Eq.~(\ref{DBE}), one finds that only the term with $l=1$ and $m=+1$ (resp. $m=-1$) couples to $\upsilon_x$ (resp. $\upsilon_y$) and contribute to $\DB^x$ (resp. $\DB^y$).
Then, inserting Eqs.~(\ref{eq:iso-eigen}), (\ref{eq:iso-phi0}) and (\ref{eq:iso-phil}) into Eq.~(\ref{DBE}), we find
\be
D_B(E) = \frac{\hbar E}{m N_0(E)} \frac{1}{\int_{0}^{2\pi} \ud \theta \,  (1 - \cos\theta) \, p(\kE, \theta)}.
\ee
This formula agrees with the result of Ref.~\cite{kuhn2007}, obtained by a different approach.

\subsection{Three-dimensional case}
In the 3D isotropic case, proceeding in a similar way, we find that
the eigenvalues of Eq.~(\ref{eigeq-os}) are given by
\be
\lambda_E^{l,m} = \frac{\int_{0}^{\pi} \ud \theta \, \sin \theta \, p(\kE, \theta) \, P_l( \cos \theta)}
{\int_{0}^{\pi} \ud \theta \, \sin \theta \, p(\kE, \theta)}, 
\label{eq:eigenv-iso}
\ee
with the index $l=0,1,...,+\infty$ and $m=-l,-l+1,...,+l$
and where $P_l( \cos \theta)$ are the Legendre polynomials.
The eigenvalues are 
($2l+1$)-degenerated and 
the corresponding 
normalized eigenfunctions are propotional to orthonormal spherical harmonics, with the prefactor determined by the normalization condition~(\ref{comp1}):
\be
\phi_{E,\uveck}^{l,m} = Y_l^m(\theta,\phi) \, \sqrt{2\pi \int_{0}^{\pi} \ud \theta' \, \sin \theta' \, p(\kE, \theta')},
\label{eq:eigenf-iso}
\ee 
In the calculation of the diffusion constant,
using the same type of symmetry arguments as in the 2D case,
we find that only the $l=1$ (with $m=-1,0,1$) terms couple to $\boldsymbol{\upsilon}$ and contribute in the summation of Eq.~(\ref{DBE}). 
We thus find
\be
\DB(E) = \frac{2}{3\pi} 
\frac{\hbar E}{m N_0(E)} 
\frac{1}{\int_{0}^{\pi} \ud \theta \, \sin \theta \, \big( 1 - \cos \theta \big) \, p(\kE, \theta)},
\label{eq:iso-diff-tens}
\ee
which agrees with the expression found in Ref.~\cite{kuhn2007}.

\subsection{Three-dimensional isotropic speckle \label{ap:iso-test}}
A simple model of 3D speckle with isotropic correlation properties, is found when considering the light pattern obtained inside an integrating sphere lit by a laser beam of wavevector $\kL$.
The real-space correlation function is given in Eq.~(\ref{eq:corr3Diso}) and the
associated power spectrum
\be
\TFCor(\veck)=\frac{\Vr^2 \pi^2 \sigmar^2}{|\veck|} \Theta(2 \sigmar^{-1}-|\veck|)
\label{eq:corr-iso}
\ee
is isotropic.
Although this isotropic model is unrealistic from an experimental point of view, it is useful here in two respects.
First, it bears the same divergence as the anisotropic 3D models of disorder considered in Sec.~\ref{part:correl}: $\TFCor(\veck)\propto 1/|\veck|$ when $|\veck| \rightarrow 0$.
Second, several properties of this model are analytical and known~\cite{kuhn2005,kuhn2007}, and therefore provides a test for our numerical methods.

As done previously, for the diagonalization of the integral operator~(\ref{eigeq-os}) we use $2^7 \times 2^7$ points regularly spaced on the $\veck$-space shell $\vert\veck\vert=\kE$.
\begin{figure}[!t] 
\begin{center}
\includegraphics[width=0.7\textwidth]{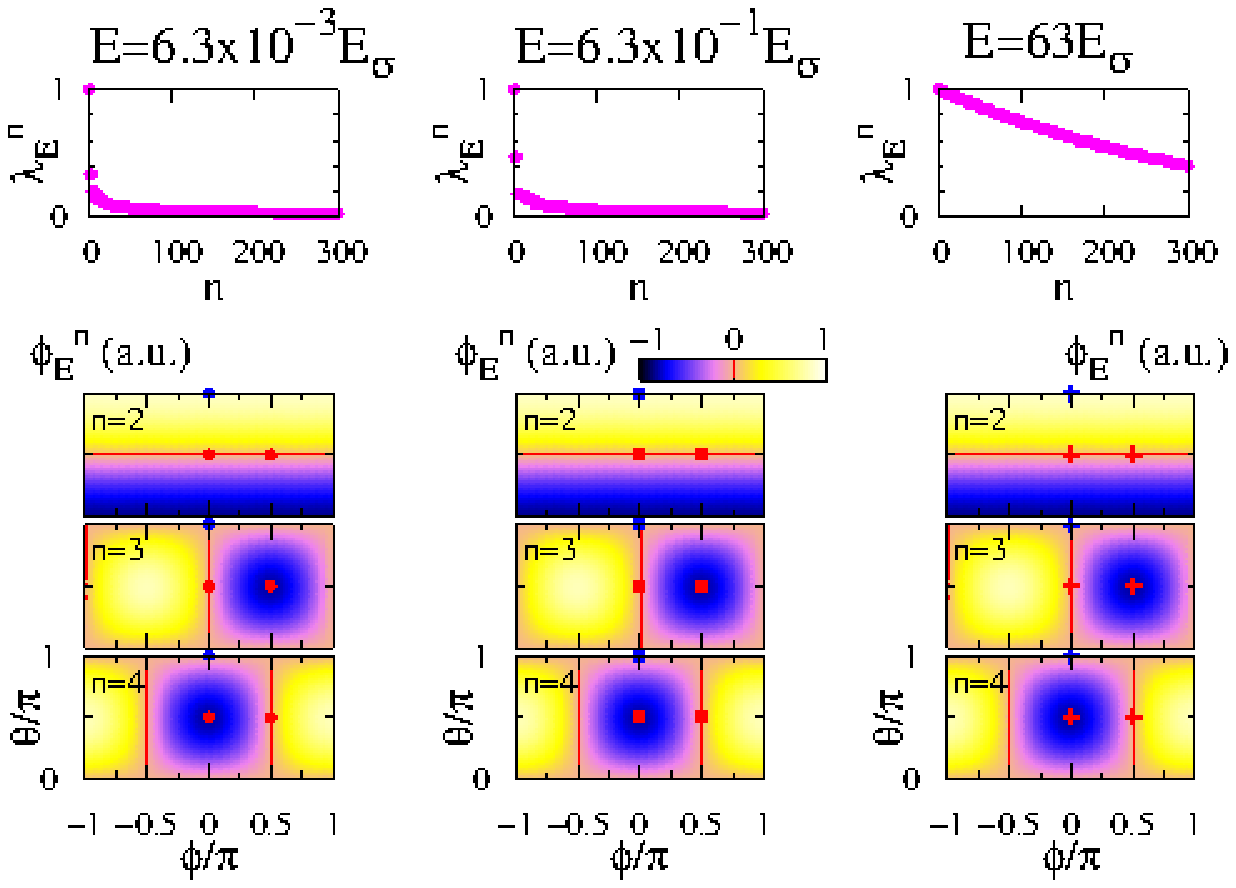}
\end{center} 
\caption{\small{(Color online) \emph{Isotropic 3D speckle}. Eigenvalues of Eq.~(\ref{eigeq-os}) (top row) for the isotropic 3D speckle with power spectrum given by Eq.~(\ref{eq:corr-iso}).
Topology of the main eigenvectors $\phi_{E,\uveck}^n$ contributing to $\DB^x$ (bottom row), $\DB^y$ (2$^{\textrm{nd}}$ row) and $\DB^z$ (3$^{\textrm{rd}}$ row) [with the parametrization $\uveck=(\uveck_x,\uveck_y,\uveck_z) \equiv (\sin\theta \cos\phi, \sin\theta \sin\phi, \cos\theta)$], the red lines locate the nodal lines.
From left to right $E=6.3\times10^{-3}E_{\sigmar}$, $E=6.3\times10^{-1}E_{\sigmar}$ and $E=63E_{\sigmar}$.}}
\label{Figure:spectr3DIso}
\end{figure}
Some eigenfunctions and eigenvalues of Eq.~(\ref{eigeq-os}) are presented in Fig.~\ref{Figure:spectr3DIso}.
We indeed find spherical harmonics [see Eq.~(\ref{eq:eigenf-iso})], and the eigenvalues $\lambda_E^n$ agree well with theory [Eq.~(\ref{eq:eigenv-iso}) with $\TFCor$ given by Eq.~(\ref{eq:corr-iso}), not shown on the figure].
\begin{figure}[!t] 
\begin{center}
\includegraphics[width=0.7\textwidth]{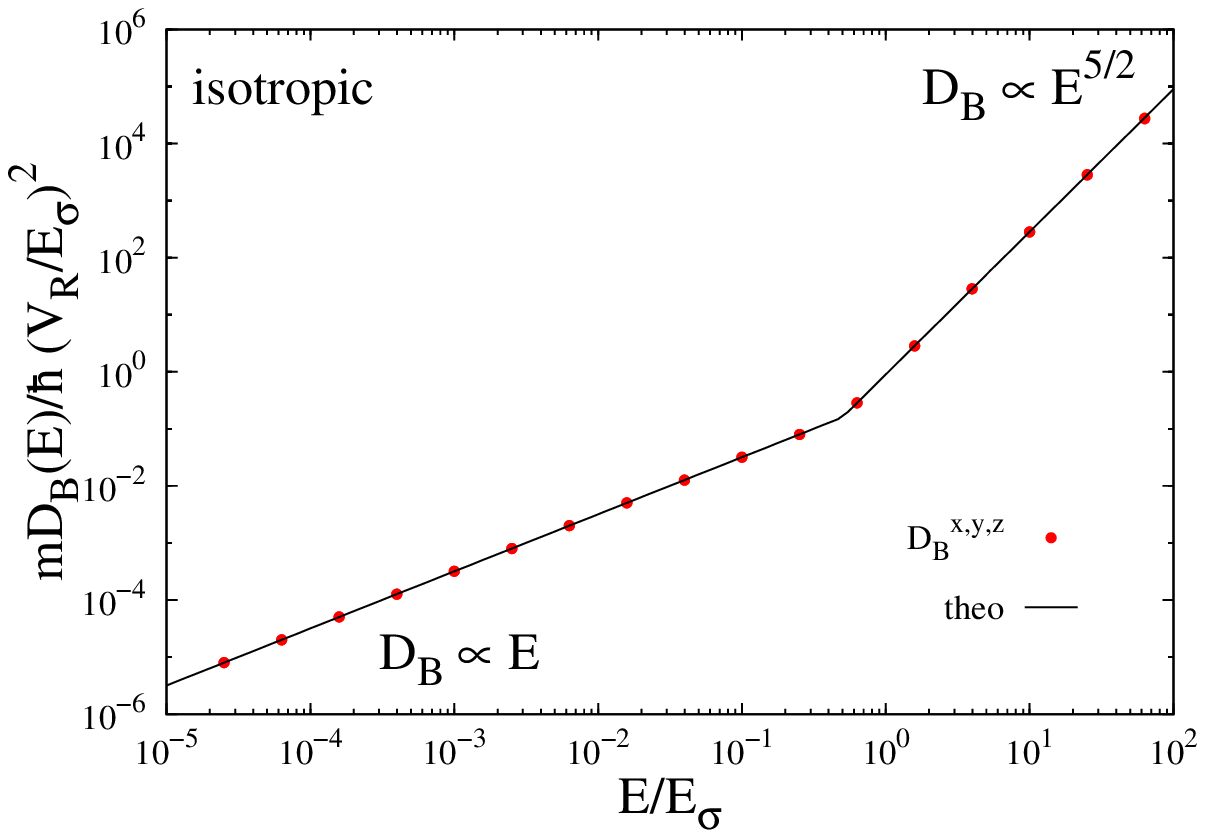}
\end{center} 
\caption{\small{(Color online) Boltzmann diffusion coefficient for the \emph{isotropic 3D speckle} configuration of power spectrum given by Eq.~(\ref{eq:corr-iso}).
The solid black line is the theoretical prediction, red dots are numerical results.
}}
\label{Figure:Dx_Dy_iso}
\end{figure}
We further incoporate these results in Eq.~(\ref{DBE}).
Figure~\ref{Figure:Dx_Dy_iso} presents the numerical results for the Boltzmann diffusion constant (red dots) which agree very well with the analytic formula (solid black line) found when incorporating Eq.~(\ref{eq:corr-iso}) into Eq.~(\ref{eq:iso-diff-tens}).
Note that we recover the same asymptotic behaviours as for our anisotropic cases: $\DB(E)\propto E$ for $E/E_{\sigmar}<1/2$ and $\DB(E)\propto E^{5/2}$ for $E/E_{\sigmar} \geq 1/2$.
In particular, those tests show that the discretization used here correctly treats the $|\veck| \rightarrow 0$ divergence.

%% file: TexFiles/Appendix-Einstein.02.tex
\subsection{Einstein relation \label{ap:Einstein}}
As presented in Sec.~\ref{part:Einst-rel}, we expect 
$\sigtens(\omega=0) \propto \Diff$
in the linear response regime.
Here we calculate $\sigtensB(\omega=0)$ in the Boltzmann approximation and verify this relation explicitly, which enables us to find the proportionality factor in Eq.~(\ref{eq:releinstein}).

Let us first rewrite the Boltzmann diffusion tensor as
\be
\DB^{i,j}(E)= \frac{1}{\hbar N_0(E)}  
\left\langle \tau_{E,\uveck} v_i J_{\veck,j} \right\rangle,
\ee
where $\vv{J}_{\veck}$ is the renormalized current vertex~:
\be
\frac{\vv{J}_{\veck}}{\hbar}= {\boldsymbol \upsilon}+
 \frac{2 \pi}{\hbar}  \sum_{\lambda^n_E\neq 1} \frac{\lambda^n_E}{1-\lambda^n_E}
\left\langle \tau_{E,\uveck'} {\boldsymbol \upsilon}'  \phi_{E,\uveck'}^n \right\rangle \phi_{E,\uveck}^n.
\label{eq:renorm-v}
\ee
We want to calculate the conductivity $\sigtensB$ in the ladder approximation.
We have to evaluate
\vspace{0.2cm}
\input{Diagrammes/diag-cond-bol}
where $\Gamma$ is defined in Eq.~(\ref{Diag:gamma}).
It reads
\beq
\sig_{\textrm{\tiny B}}^{i,j}&&(E) = \int \frac{\ud \veck}{(2 \pi)^d} \, v_i f_{E,\veck} v_j +\\
&& \int \frac{\ud \veck}{(2 \pi)^d} \frac{\ud \veck'}{(2 \pi)^d} \, v_i f_{E,\veck} \Gamma_{\veck,\veck'}(0,0,E) f_{E,\veck'} v'_j. \nonumber
\eeq
As $\Gamma_{\veck,\veck'}(0,0,E) 
= \sum_{\lambda^n_E\neq 1} \frac{\lambda^n_E}{1-\lambda^n_E}
\phi_{E,\uveck}^n \, \phi_{E,\uveck'}^n$~\footnote{Equation~(\ref{Diag:gamma}) gives $\Gamma=\mathrm{U}_{\mathrm{B}} [1 - \av{\Gr} \otimes \av{\Ga} \,\U]^{-1}$. The components $\Gamma_{\veck,\veck'}(0,0,E)$ can be found from the results of appendix~\ref{part:ap-BSE}.},
and $f_{E,\veck} \simeq \tau_{E,\uveck}\, A_0(E,\veck)/\hbar$,
one easily finds
\beq
\sig_{\textrm{\tiny B}}^{i,j}&&(E) = \frac{2\pi}{\hbar} \bigg\{ 
\left\langle \tau_{E,\uveck} v_i v_j \right\rangle 
+ \frac{2 \pi}{\hbar} \\
&& \times \sum_{\lambda^n_E \neq 1} \frac{\lambda^n_E}{1-\lambda^n_E} 
\left\langle \tau_{E,\uveck} v_i \phi_{E,\uveck}^n \right\rangle \,
\left\langle \tau_{E,\uveck} v_j \phi_{E,\uveck}^n \right\rangle \bigg\}.
\nonumber
\eeq
Therefore, we have $\sigtensB=2 \pi N_0(E) \DiffTensB /\hbar$.
We have thus verified Einstein's relation for the classical dc conductivity in anistropic disorder.

%% file: Diagrammes/diag-cond-bol.tex
\begin{fmffile}{diag-cond-bol-fmf}
\be
\sigtensB = 
\parbox{0.35\linewidth}{
	    \begin{fmfgraph*}(35,14)
		\fmfleft{i1}
		\fmfright{o1}
		\fmfpoly{phantom,tension=0.}{v1,v2,v3,v4,v5,v6}
		\fmf{wiggly,tension=5.,label=${\boldsymbol \upsilon}$,l.s=right}{i1,v4}
		\fmf{plain,left=0.25,tension=0.25,width=2}{v4,v3}
		\fmf{plain_arrow,left=0.25,tension=0.,width=2}{v3,v2}
		\fmf{plain,left=0.25,tension=0.25,width=2}{v2,v1}
		\fmf{plain,left=0.25,tension=0.25,width=2}{v1,v6}
		\fmf{plain_arrow,left=0.25,tension=0.,width=2}{v6,v5}
		\fmf{plain,left=0.25,tension=0.25,width=2}{v5,v4}
		\fmf{wiggly,tension=5.,label=${\boldsymbol \upsilon}$,l.s=right}{v1,o1}
		\fmf{phantom}{v3,v5}
		\fmf{phantom}{v2,v6}
	    \end{fmfgraph*}
}
\, + \,
\parbox{0.35\linewidth}{
	    \begin{fmfgraph*}(35,14)
		\fmfleft{i1}
		\fmfright{o1}
		\fmfpoly{phantom,tension=0.,label=$\Gamma$}{v1,v2,v3,v4,v5,v6}
		\fmf{wiggly,tension=5.,label=${\boldsymbol \upsilon}$,l.s=right}{i1,v4}
		\fmf{plain_arrow,left=0.25,tension=0.,width=2}{v4,v3}
		\fmf{plain,left=0.25,tension=0.}{v3,v2}
		\fmf{plain_arrow,left=0.25,tension=0.,width=2}{v2,v1}
		\fmf{plain_arrow,left=0.25,tension=0.,width=2}{v1,v6}
		\fmf{plain,left=0.25,tension=0.}{v6,v5}
		\fmf{plain_arrow,left=0.25,tension=0.,width=2}{v5,v4}
		\fmf{wiggly,tension=5.,label=${\boldsymbol \upsilon}'$,l.s=right,l.d=3.5}{v1,o1}
		\fmf{plain}{v3,v5}
		\fmf{plain}{v2,v6}
	    \end{fmfgraph*}
}
\ee
\end{fmffile}\\

%% file: TexFiles/Appendix-Vrenorm.01.tex
\subsection{Current vertex renormalization \label{ap:Vrenorm}}
The DC conductivity $\sigtensB$ in the Boltzmann approximation reads (see appendix~\ref{ap:Einstein})
\be
\sig_{\textrm{\tiny B}}^{i,j}(E) = \frac{2\pi}{\hbar}  
\left\langle \tau_{E,\uveck} {\upsilon}_i \frac{J_{\veck,j}}{\hbar} \right\rangle,
\label{sig_renorm}
\ee
where $\vv{J}_{\veck}$, the renormalized vertex function, is given by Eq.~(\ref{eq:renorm-v}).
Diagrammatically we can absorb this renormalization in one of the vertices as shown in Eq.~(\ref{Diag:renorm_vertex}).
This is a standard procedure for anisotropic scattering, which is presented for example in Ref.~\cite{akkermans2006}.
\medskip
\input{Diagrammes/diag-renorm-vertex}

%% file: Diagrammes/diag-renorm-vertex.tex
\begin{fmffile}{diag-renorm-vertex-fmf}
\be
\parbox{0.195\linewidth}{
	     \begin{fmfgraph*}(30,15)
		\fmfleft{i1}
		\fmfright{o1}
		\fmfpoly{phantom,tension=0.}{v1,x1,v2,x3,v4,x4,v5,x6}
		\fmf{wiggly,tension=6.,label=${\boldsymbol \upsilon}$,l.s=right}{i1,v4}
		\fmf{plain,left=0.2,tension=0.,width=2}{v4,x3}
		\fmf{plain_arrow,left=0.2,tension=0.,width=2}{x3,v2}
		\fmf{plain_arrow,left=0.2,tension=0.,width=2}{v5,x4}
		\fmf{plain,left=0.2,tension=0.,width=2}{x4,v4}
		\fmf{phantom,tension=100.}{v1,o1}
		\fmf{phantom}{v2,v5}
		\fmf{phantom}{v2,v5}
	    \end{fmfgraph*}
}
\, + \,
\parbox{0.18\linewidth}{
	     \begin{fmfgraph*}(27,15)
		\fmfleft{i1}
		\fmfright{o1}
		\fmfpoly{phantom,tension=0.}{v1,x1,v2,x2,v3,x3,v4,x4,v5,x5,v6,x6}
		\fmf{wiggly,tension=5.,label=${\boldsymbol \upsilon}$,l.s=right}{i1,v4}
		\fmf{plain,left=0.175,tension=0.,width=2}{v4,x3}
		\fmf{plain,left=0.175,tension=0.}{x3,v3}
		\fmf{plain_arrow,left=0.175,tension=0.,width=2}{v3,x2}
		\fmf{plain_arrow,left=0.175,tension=0.,width=2}{x5,v5}
		\fmf{plain,left=0.175,tension=0.}{v5,x4}
		\fmf{plain,left=0.175,tension=0.,width=2}{x4,v4}
		\fmf{phantom,tension=100.}{v1,o1}
		\fmf{phantom}{v3,v5}
		\fmf{phantom}{v2,v6}
		\fmf{plain,tension=0.}{v3,v5}
		\fmf{plain,tension=0.,label=$\Gamma$,l.s=left,l.d=0.9}{x3,x4}
	    \end{fmfgraph*}
}
\, = \,
\parbox{0.3\linewidth}{
	     \begin{fmfgraph*}(33,15)
		\fmfleft{i1}
		\fmfright{o1}
		\fmfpoly{phantom,tension=0.}{v1,x1,v2,x3,v4,x4,v5,x6}
		\fmf{wiggly,tension=4.,label=$\vecJ_{\veck}/\hbar$,l.s=right}{i1,v4}
		\fmf{plain,left=0.2,tension=0.,width=2}{v4,x3}
		\fmf{plain_arrow,left=0.2,tension=0.,width=2}{x3,v2}
		\fmf{plain_arrow,left=0.2,tension=0.,width=2}{v5,x4}
		\fmf{plain,left=0.2,tension=0.,width=2}{x4,v4}
		\fmf{phantom,tension=100.}{v1,o1}
		\fmf{phantom}{v2,v5}
		\fmf{phantom}{v2,v5}
		\fmf{dots,tension=0.}{x3,x4}
	    \end{fmfgraph*}
}
\medskip
\label{Diag:renorm_vertex}
\ee
\end{fmffile}\\

%% file: TexFiles/Appendix-CorrCond.02.tex
\subsection{Weak-localization correction\label{ap:CorrCond}}

\subsubsection{The cooperon}

We calculate the bare cooperon correction, with renormalized current vertices, Diag.~(\ref{Diag:cooperon}) translates into
\be
\Delta \sig_{(X)}^{i,j}(\omega,E)=
\int \frac{\ud \veck}{(2 \pi)^d} \frac{\ud \veck'}{(2 \pi)^d} \, \frac{J_{\veck,i}}{\hbar} f_{E,\veck} X_{\veck,\veck'}(0,\omega,E) f_{E,\veck'} \frac{J_{\veck',j}}{\hbar}.
\ee
Considering that the dominant contribution in the integral comes from $\vecQ\simeq\veck+\veck'\sim 0$ [see Eq~(\ref{eq:coop})], and that $f_{E,\veck}^2\sim 2 (\tau_{E,\uveck}/\hbar)^3 A_0(E,\veck)$ in the on-shell approximation\footnote{The same procedure as described in \new{Sec.~\ref{app:o-s}} is used to obtain those expressions in the on-shell approximation.\label{note:on-shell}}, we get
\be
\Delta \sig_{(X)}^{i,j}(\omega,E)=- \frac{2}{\hbar N_0(E)}
\left\langle \frac{J_{\veck,i} J_{\veck,j}}{\hbar^2} \tau_{E,\uveck} \right\rangle \int \frac{\ud \vecQ}{(2 \pi)^d} \, \frac{1}{-i\hbar \omega + \hbar \vecQ \cdot \DiffTensB(E) \cdot \vecQ}.
\ee

\subsubsection{Hikami contributions}

We now calculate the Hikami corrections [see Diags.~(\ref{Diag:hikami}) and (\ref{Diag:hikami2})]
\beq
\Delta \sig_{(H_1)}^{i,j}(\omega,E)=
\int &\frac{\ud \veck}{(2 \pi)^d} \frac{\ud \veck'}{(2 \pi)^d} \frac{\ud \veck''}{(2 \pi)^d}  \, \frac{J_{\veck,i}}{\hbar} f_{E,\veck} {\UB}_{\veck,\veck'} \Gr(E,\veck'')\\
&\times X_{\frac{\veck+\veck''}{2},\veck'+\frac{\veck''-\veck}{2}}(\veck''-\veck,\omega,E)
\Gr(E,\veck'+\veck''-\veck) f_{E,\veck'} \frac{J_{\veck',j}}{\hbar}. \nonumber
\eeq
In the same way as before, and using the on-shell approximation formulas 
$\Gr(E,\veck) f_{E,\veck} \sim -i (\tau_{E,\uveck}/\hbar)^2 A_0(E,\veck) 
$ and $\Ga(E,\veck) f_{E,\veck} \sim i (\tau_{E,\uveck}/\hbar)^2 A_0(E,\veck)$, we get $\Delta \sigtens_{(H_1)} \simeq \Delta \sigtens_{(H_2)}$ and
\beq
\Delta \sig_{(H)}^{i,j}(\omega,E)&=&\Delta \sig_{(H_1)}^{i,j}(\omega,E)+\Delta \sig_{(H_2)}^{i,j}(\omega,E)\\
&=& \frac{2}{\hbar N_0(E)}
\big\langle \frac{J_{\veck,i}}{\hbar} \tau_{E,\uveck} \int \frac{\ud \veck'}{(2 \pi)^d}\, {\UB}_{\veck,\veck'} f_{E,\veck'} \frac{J_{\veck',j}}{\hbar} \big\rangle \nonumber\\
&\times&\int \frac{\ud \vecQ}{(2 \pi)^d} \, \frac{1}{-i\hbar \omega + \hbar \vecQ \cdot \DiffTensB(E) \cdot \vecQ}. \nonumber
\eeq

\subsubsection{Corrected conductivity tensor}

We now consider the quantity $\vecJ_{\veck} -\int \frac{\ud \veck'}{(2 \pi)^d}\, {\UB}_{\veck,\veck'} f_{E,\uveck'} \vecJ_{\uveck'}$.
Using the relation ${\UB}_{\veck,\veck'}=\sum_{\lambda^n_E \neq 1} \lambda^n_E \phi_{E,\veck}^n \phi_{E,\veck'}^n$, and the parities of the functions $\phi_{E,\uveck}^n$ [see Eqs.~(\ref{parity0}) and (\ref{parityn})], one can show that
\be
\vecJ_{\veck} -\int \frac{\ud \veck'}{(2 \pi)^d}\, {\UB}_{\veck,\veck'} f_{E,\veck'} \vecJ_{\veck'}=\hbar {\boldsymbol \upsilon}.
\ee
Therefore the Hikami contributions renormalize one of the $\vecJ_\veck/\hbar$ back to the bare vertex ${\boldsymbol \upsilon}$, and we have
\beq
\Delta \sig^{i,j}(\omega,E)&=&\Delta \sig_{(X)}^{i,j}(\omega,E)+\Delta \sig_{(H)}^{i,j}(\omega,E)\\
&=& - \frac{2}{\hbar N_0(E)}
\big\langle \frac{J_{\veck,i}}{\hbar} v_{j} \tau_{E,\uveck} \big\rangle \int \frac{\ud \vecQ}{(2 \pi)^d} \, \frac{1}{-i\hbar \omega + \hbar \vecQ \cdot \DiffTensB(E) \cdot \vecQ}, \nonumber
\eeq
which gives the final expression~(\ref{eq:corr-cond}).